\documentstyle[12pt,epsfig]{article}
\def\la{\langle}\def\ra{\rangle}
\def\be{\begin{eqnarray}}\def\bea{\begin{eqnarray}}
\def\ba{\begin{eqnarray}}
\def\ee{\end{eqnarray}}\def\eea{\end{eqnarray}}
\def\ea{\end{eqnarray}}
\def\ben{\begin{enumerate}}\def\bitem{\begin{itemize}}
\def\een{\end{enumerate}}\def\eitem{\end{itemize}}
\def\no{\nonumber}\def\del{\partial}
\def\G0p{$G_0^\prime$}
\def\bi{\bibitem}

\def\Kp{$K^+$}\def\Km{$K^-$}\def\N1520{$N^\star (1520)$}
\def\calL{{\cal L}}\def\calF{{\cal F}}\def\calA{{\cal A}}
\def\calO{{\cal O}}\def\calR{{\cal R}}
\def\calM{{\cal M}}\def\wkm{\omega_{\tiny{K^-}}^\star}
\def\prl{Phys. Rev. Lett.}\def\pr{Phys. Rev.}\def\np{Nucl. Phys.}
\def\pl{Phys. Lett.}
\def\B#1{{}^{#1}\mbox{B}}

\def\Tr{{\mbox{Tr}}}
\newcommand{\e}{{\mbox{e}}}\def\del{\partial}
\def\vr{{\vec r}}
\def\vP{{\vec P}}

\def\hatk{{\hat k}}

\def\mut{{\mbox{M1S}}}
\def\Qt{{\mbox{E2S}}}
\def\rM{{\cal R}_{\rm M1}}\def\rE{{\cal R}_{\rm E2}}
\def\MS{{\mbox{M1V}}}
\def\abs#1{{\left| #1 \right|}}\def\nlo#1{{\mbox{N$^{#1}$LO}}}
\def\roughly#1{\mathrel{\raise.3ex\hbox{$#1$\kern-.75em%
\lower1ex\hbox{$\sim$}}}}\def\lsim{\roughly<}
\def\gsim{\roughly>}\def\fm{{\mbox{fm}}}

\def\ve{\varepsilon}

\def\B#1{{}^{#1}\mbox{B}}

\def\nlo#1{\mbox{N$^{#1}$LO}}

\def\calR{{\cal R}}

\def\itt{\indent\indent}
\def\E{{\cal E}}

\def\Jb{\mbox{\boldmath$ J$}}

\def\jb{\mbox{\boldmath$ j$}}
\def\pb{\mbox{\boldmath$ p$}}\def\qb{\mbox{\boldmath $q$}}

 \def\vb{\mbox{\boldmath$ v$}}

\def\Bp{\mbox{\boldmath$ p$}}\def\Bq{\mbox{\boldmath $q$}}
\def\bfq{\mbox{\boldmath $q$}}

\def\Bk{\mbox{\boldmath$ k$}}\def\Br{\mbox{\boldmath $r$}}

\def\qbhat{\mbox{\boldmath $\hat{q}$}}\def\taub{\mbox{\boldmath $\tau$}}

\def\epsilonb{\mbox{\boldmath $\epsilon$}}

\def\gammab{\mbox{\boldmath $\gamma$}}

\def\sigmab{\mbox{\boldmath $\sigma$}}

\def\Bsigma{\mbox{\boldmath $\sigma$}}

\def\B0{\mbox{\boldmath $0$}}
\def\A0{A_0}
\def\bq{\begin{equation}}
\def\eq{\end{equation}}

\renewcommand{\thefootnote}{\fnsymbol{footnote}}
\begin{document}
\begin{titlepage}
\begin{center}

\vskip 1.0cm {\Large \bf On the Manifestation of Chiral Symmetry\\
in Nuclei and Dense Nuclear Matter}
  \vskip 1.2cm
   {{\large G.E. Brown$^{(a)}$ and Mannque Rho$^{(b,c,d)}$} }
 \vskip 0.2cm

{\it (a) Department of Physics and Astronomy, State University of
New York}

{\it Stony Brook, NY 11794, USA}

 {\it (b) Service de Physique Th\'eorique, CE Saclay, 91191
Gif-sur-Yvette, France}

{\it (c) School of Physics, Seoul National University, Seoul,
Korea}

{\it (d) Institute of Physics, Yonsei University, Seoul, Korea}

\end{center}

\vskip 0.2cm

\centerline{(\today)}
 \vskip 1cm

\centerline{\bf Abstract}
 \vskip 0.5cm
This article reviews our view on how chiral symmetry, its pattern
of breaking and restoration under extreme conditions manifest
themselves in the nucleon, nuclei, nuclear matter and dense
hadronic matter. We discuss how first-principle (QCD) calculations
of the properties of finite nuclei can be effectuated by embedding
the ``standard nuclear physics approach (SNPA)" into the framework
of effective field theories of nuclei that incorporate chiral
dynamics and then exploit the predictive power of the theory to
accurately compute such solar neutrino processes as the
proton-proton fusion and the ``hep" process and such cosmological
nucleosynthesis process as thermal neutron-proton capture etc. The
Brown-Rho (BR) scaling that implements chiral symmetry property of
baryon-rich medium is re-interpreted in terms of ``vector
manifestation" of hidden local symmetry \`a la Harada-Yamawaki. We
present a clear {\it direct} evidence and a variety of {\it
indirect} evidences for BR scaling in nuclear processes at normal
nuclear matter density probed by weak and electromagnetic fields
and at higher density probed by heavy-ion collisions and
compact-star observables. We develop the notion of ``broadband
equilibration" in heavy-ion processes and sharpen the role of
strangeness in the formation of compact stars and their collapse
into black-holes. We revisit the Cheshire-Cat phenomenon first
discovered in the skyrmion structure of baryons and more recently
revived in the form of ``quark-hadron continuity" in mapping
low-density structure of hadrons to high-density structure of
quarks and gluons and argue once more for the usefulness and power
of effective field theories based on chiral symmetry under extreme
conditions. It is shown how color-flavor locking in terms of QCD
variables and hidden local symmetry in terms of hadronic variables
can be connected and how BR scaling fits into this ``continuity"
scheme exhibiting a novel aspect of the Cheshire Cat phenomenon.

\end{titlepage}
\newpage
\tableofcontents
\newpage
\renewcommand{\thefootnote}{\arabic{footnote}}
\setcounter{footnote}{0}

\section{INTRODUCTION}\label{intro}
\setcounter{equation}{0} 
\renewcommand{\theequation}{\mbox{\ref{intro}.\arabic{equation}}}
\itt In a recent paper, Morgenstern and Meziani~\cite{morgenstern}
reported a remarkable result of their extensive analysis of the
longitudinal and transverse response functions in medium-weight
and heavy nuclei and concluded that the available world data
showed unequivocally the quenching of $\sim 20\%$ in the
longitudinal response function and the only viable way to explain
this quenching is to invoke the Brown-Rho (BR) scaling~\cite{BR91}
formulated to account for the change of the strong-interaction
vacuum induced by matter density. Partly motivated by this
development and other indirectly related developments including
the most recent one on color-flavor-locking and color
superconductivity, we shall make in this review {\it our} overview
of evidences that indicate both directly and indirectly that the
BR scaling is indeed operative in finite nuclei and dense (and
superdense) nuclear matter. We shall develop our arguments
starting with the basic structure of the nucleon, then go to that
of nuclei and of nuclear matter and finally to that of dense
matter that one expects to create in relativistic heavy-ion
collisions and find in the interior of highly compact stars. The
principal theme will be that the chiral symmetry of quantum
chromodynamics -- its spontaneous breaking in free space and
partial or full restoration in medium -- underlies the common
feature from elementary hadrons to complex dense systems. We admit
that our views are not necessarily shared by others in all details
and that some points may not be fully correct. We are however
confident that the general theme that we have developed and we
shall review here will survive the experimental test that is to
come.

It was recognized a long time ago that chiral symmetry, now
identified as an essential ingredient of QCD associated with the
light-mass up (u), down (d) and strange (s) quarks, with its
``hidden" realization in Nambu-Goldstone mode plays an important
role in nuclear physics. In fact, the emphasis on its role in a
variety of nuclear processes predates even the advent of QCD
proper and subsequent acceptance as {\it the correct} theory of
strong interactions~\cite{comments-BR}. Among the time-tested and
established is the role of chiral symmetry in exchange currents in
electro-weak processes in nuclei~\cite{chemrho,Rbnp} and its
subsequent importance in probing nuclear structure at various
electron machines, now out of operation, throughout the
world~\cite{mathiot-frois}. Since our early review on the
subject~\cite{comments-BR}, there has been continuous and
significant evolution in the field specifically associated with
the role of QCD in nuclear physics, more or less unnoticed by
workers in other areas of physics. The evolution touched on the
structure of the nucleon, the proton and the neutron, starting
with the notion of a ``little bag"~\cite{littlebag}, the
recognition of the preeminent importance of Goldstone pion clouds
in both the nucleon structure and nuclear
forces~\cite{littlebag,cloudybag}, followed by the resurrection of
the Skyrme soliton model that contains baryons -- which are
fermions -- in a bosonic field theory~\cite{skyrme,witten} and
then the emergence of the notion of ``Cheshire Cat phenomenon."
At present, the issue of both ``spontaneous" breaking and
restoration of chiral symmetry occupies one of the central
positions in current activities in both nuclear and hadron
physics communities. The development up to 1995 has been
summarized in a recent monograph by Nowak, Rho and
Zahed~\cite{NRZ}.

It is now becoming clearer in which direction the chiral symmetry
of QCD will steer the {\it next} generation of nuclear and
hadronic physicists, with the advent of new accelerators such as
the Jefferson Laboratory (JL) and the RHIC at Brookhaven -- both
of which are already operating -- and ALICE/LHC at CERN -- which
is to come in a few years. On one hand, the electron machine at
the JL will probe deeper into shorter distance properties of the
nucleon-nucleon interactions, exploring how the chiral structure
of the nucleon in medium changes from hadronic over to
quark/gluon picture as shorter distances are probed. As will be
repeatedly stressed, one does not expect any abrupt phase
transitions along the way; as the change will most likely be
continuous (for a multitude of reasons we will discuss), it will
then be a matter of economy which language will be more efficient
for given processes with given conditions, implying that there
will be a continuous map between the various descriptions. This
continuity will play a central role in our discussion and will
later be referred, quite broadly, to as ``Cheshire Cat
Phenomenon."

On the other hand, the heavy-ion machines at RHIC and ALICE/LHC
will create systems at high temperature and density, mimicking
the early Universe. At high temperatures, QCD predicts a phase
transition from chiral symmetry in the Goldstone mode to that in
the Wigner mode with a consequent change in
confinement/deconfinement. This is more or less confirmed on
lattice in QCD. Indications are coming out already from available
experimental data that such a transition has been seen. Matter at
high density is a completely different matter. Up to date, it has
not been feasible to put density on lattice, so there is no
model-independent theoretical indication for a similar phase
change as density is increased. Theoretical models however do
predict that there can be a series of phase changes including the
chiral symmetry restoration. As will be discussed in this
article, there are a variety of processes involving nuclei and
nuclear matter that provide, albeit indirect, evidence for such
phase changes, most intriguingly of which are possible signals
from compact (neutron) stars.

Going to the physics of matter under extreme conditions --- high
temperature, high density or both -- encompasses a multitude of
scales. At low density $\rho <\rho_0$ (where $\rho_0$ is nuclear
matter density), one can rely on the rich phenomenology available
in the guise of standard nuclear physics to work with an effective
Lagrangian field theory that is anchored on the premise of quantum
chromodynamics (QCD), e.g., chiral perturbation theory. We may
call this ``phenomenological perturbation approach (PPA)." However
as one approaches the density $\rho_0$, an effective field theory
constructed in the matter-free vacuum encounters difficulty and
becomes more or less unpredictive: Presently there is no way to
derive a nucleus starting from a first-principle Lagrangian, not
to mention denser hadronic matter. The reason why this is so will
be explained in the course of this review. The attitude we will
adopt here is then to develop a generic description of dense
matter based on global characteristics  of hadrons in the
environment of surrounding dense medium. For this, we exploit the
chiral structure of the QCD vacuum and the excitations thereon
(i.e., hadrons) following the notion of ``vector manifestation"
recently proposed and developed by Harada and Yamawaki~\cite{VM}.
In the Harada-Yamawaki scenario, chiral restoration from the
spontaneously broken mode to an unbroken mode takes place with the
massless pions (in the chiral limit) coming together with the
would-be scalar Goldstone bosons that are the longitudinal
components of massive vector mesons, with the massless vector
mesons consequently decoupling. This scheme requires that near the
chiral phase transition --- independently of whatever form it may
take, the vector ($\rho$ and $\omega$) meson masses drop as the
chiral transition point is approached bottom-up. This, we will
see, provides a compelling theoretical support to BR scaling. We
will also see that most remarkably, this scenario receives further
support from the QCD structure of quark condensates in medium
associated with color-flavor locking.

BR scaling as originally formulated in \cite{BR91} corresponded
to a mean-field approximation in a large $N_c$ effective chiral
Lagrangian theory (where $N_c$ is the number of colors in QCD)
with the strong interaction vacuum~\footnote{Throughout this
article, the terminology vacuum will be used in the generalized
sense of a ground state in the presence of matter.} ``sliding"
with density or temperature of the system. For a given density,
say, quantum fluctuations or loop corrections are to be
calculated for correlation functions etc with a chiral Lagrangian
whose parameters are suitably defined at the sliding vacuum
appropriate to physical processes of given kinematical
conditions. How this works out has been explained in the
literature but will be summarized in the review at pertinent
places. We propose that the way the BR-scaling mean-field
quantities vary as a function of the medium condition (density or
temperature) should roughly match with the change of the vacuum
suggested by Harada-Yamawaki's vector manifestation. As mentioned
above, at small external conditions (e.g., at low density), the
PPA using phenomenological Lagrangians fixed in free space will
be in no disagreement with the notion of BR scaling. However as
density increases, fluctuations are to be built on a vacuum
modified from the free-space one. At some point, this effect can
no longer be accessed adequately by the PPA. This can be seen in
what we call ``sobar" description as explained in a later section.
Eventually a correct theory will have to approach Landau
Fermi-liquid fixed point at normal nuclear matter density. We will
show how this comes about with BR scaling chiral Lagrangians: BR
scaling and a Landau parameter can be identified.

One of the major themes of this review is that this picture
combining the vector manifestation of chiral symmetry,
color-flavor locking and BR scaling passes several tests in
nuclear and dense hadronic systems. This, we believe, points to
the validity of the notion that understanding what goes on in
many-body systems of nucleons at high density and understanding
the chiral structure of the nucleon at the QCD level relies on
the same principle, a long standing issue in nuclear and hadronic
physics. We shall summarize the evidences accumulated since our
earlier publication~\cite{comments-BR,BRPR}.

\section{FROM QUARK TO NUCLEON}\label{sec2}
\setcounter{equation}{0} 
\renewcommand{\theequation}{\mbox{\ref{sec2}.\arabic{equation}}}
\subsection{Modeling the Nucleon in QCD}
 \itt For concrete applications of QCD to physical
hadronic processes, the only model that purports to encode QCD and
at the same time is versatile enough for a wide range of strong
interaction physics has been, and still is, the bag model. All
other models of equal versatility can be considered as belonging
to the same class. There is a long history in the development of
this model which, as we shall mention later, continues even today
in the guise of string theory and large $N$ Yang-Mills gauge
theory. The first such model that incorporates confinement and
asymptotic freedom of QCD is the MIT bag model~\cite{MITbag}.
However it was realized early on that since this model by
construction lacks chiral symmetry, it could not be directly
applied to the description of nuclear interactions, the domain of
strong interaction physics most thoroughly and accurately studied.
A simple stability argument showed that the bag for the nucleon
must have a bag size typically of $\sim 1$ fm, essentially the
confinement scale implicit in the model. And this is simply too
big if one naively applies the picture to nuclear systems. As a
possible way to reconcile this simple bag modelling of QCD with
the ``size crisis", it was proposed in 1979~\cite{littlebag} to
implement the spontaneously broken chiral symmetry and bring in
pion cloud to the model. How to introduce pion cloud to the MIT
bag producing a ``chiral bag" was already known and discussed in
the literature~\cite{chodos}. It consisted of adding into the
model Goldstone pion fields outside of the bag and imposing
chiral-invariant boundary conditions that assure both chiral
invariance and confinement. The initial idea in \cite{littlebag}
was to use the pressure generated by the external pion cloud to
squeeze the bag to a smaller size so as to accommodate the
standard nuclear physics picture of meson exchange interactions.
At the time the idea was proposed, we had no idea how to squash
the bag into a ``little bag" in a way consistent with the premise
of QCD. We simply dialed the parameters of the model such that the
size diminished from $\sim 1$ fm to $\sim 0.3$ fm, a size known to
be a spatial cutoff between two nucleons. We now know that this
prescription was not natural and in fact totally unnecessary. How
to reconcile the bag size and nuclei can now be put in the form of
what is known as ``Cheshire Cat Principle"~\cite{cat}\footnote{A
historical note: Simultaneously and independently of \cite{cat}, a
similar idea was proposed based on phenomenology by Brown et
al~\cite{BJRV}.}.

Apart from the concern with the nucleon size in nuclei and {\it
apparent} incompatibility with the successful standard
meson-exchange potential picture, the ``little bag" was not
indispensable for understanding the wide variety of nucleon
dynamics as has been argued by Thomas and his
co-workers~\cite{cloudybag,cloudyrev}. In fact, this latter
approach where the pion field enters only as a fluctuating field
at the bag boundary with its size comparable to that of the
original MIT one has been reasonably successful in a wide range
of applications to those nucleon and nuclear processes that are
dominated by pionic perturbative effects and in which
nonpertubative chiral effects play a subdominant
role~\cite{cloudyrev}.

It is now well understood how both the ``big cloudy bag" of
\cite{cloudybag} and the ``little bag" of \cite{littlebag} can be
accommodated in a unified framework of the chiral bag model. In
fact, we now know how to shrink the bag even further from the
``little bag" and eventually obtain a ``point bag,"  the
skymion~\cite{skyrme,witten} which can be considered as the limit
that the bag size is shrunk to zero. What has transpired from the
development since late 1970's/early 1980's up until now is that
the bag size that appears in the simple confinement model is more
like a gauge degree of freedom~\cite{mattis-gauge,DNS} and
depending upon the gauge choice, it can be of any size without
changing basic physics. That physics should not depend upon the
confinement (bag) size is known as ``Cheshire cat phenomenon" and
has been amply reviewed in the literature (for an exhaustive list
of references, see \cite{NRZ,mrPR,tokietal}). Unfortunately there
is no known exact formulation of this ``Cheshire-Cat gauge
theory." This is mainly because there is no exact bosonization of
QCD known in four dimensions, the skyrmion limit corresponding to
the totally bosonized theory, and the chiral bag partially
bosonized. Asking what size of the chiral bag is needed for a
particular phenomenology is like asking what gauge choice is
optimal in gauge theories that are solved only partially as is
the case with QCD. In practice, what bag size is optimal depends
upon what physical processes are concerned.

Thus far our argument has been phrased in the framework of a bag
picture. The notion of Cheshire Cat is however much more general
than in this restricted context. It is a reflection of a variety
of ``duality" in the problem which is a statement that there are
a variety of different ways of describing the same physical
process. In other areas of physics (such as string theories,
condensed matter physics etc), such a notion can even be
formulated in an exact way. In the strong interaction dynamics,
however, the notion is at best approximate and hence its
predictive power semi-quantitative. Even so, it is sufficiently
versatile as to be applied to a variety of problems in nuclear
physics. As we will describe in later sections, a remarkable
recent development in a related context is the ``complementarity"
of two descriptions in terms of hadrons and in terms of
quark/gluon variables for both dilute matter and dense matter as
discussed in a later section of this review. This suggests that
the Cheshire Cat notion can be generalized to a wider class of
phenomena, going beyond the ``confinement size."

 \subsection{The Chiral Bag Lagrangian}
 \itt
The Cheshire cat phenomenon in our view is a generic feature in
{\it modelling} QCD or other gauge theories. One may therefore
formulate it in various different ways. The use of the chiral bag
is not the only way or maybe not even the best way to exploit it.
Since we understand the chiral-bag structure better than others,
however, we write here the explicit form of the Lagrangian that
the chiral bag model takes and that should capture the essential
physics of the nucleon.  We shall do this for the flavor $SU(3)$
(with up, down and strange quarks) although the nucleon structure
is mainly dictated by the non-strange quarks, with the strange
quark playing a minor and as yet obscure role. It can be written
generically in three terms,
 \be
S=S_V+S_{\tilde{V}}+S_{\del V}\label{model}.
 \ee
The first term defined in the volume $V$ is the ``bag" action
that reflects the explicit QCD variables, quarks $\psi$ and gluons
$G_\mu^a$:
 \be
S_V=\int_V d^4x \left(\bar{\psi}i\not\!\!{D}\psi -\frac{1}{2}
{\rm tr}\ G_{\mu\nu}G^{\mu\nu}\right)+\cdots\label{in}
 \ee
where the trace goes over the color. Here $\psi$ is the quark
field with the color, flavor and Poincar\'e indices suppressed,
$G_{\mu\nu}$ the gluon field tensor and $D_\mu$ the covariant
derivative. The ellipsis stands for quark mass terms that we will
leave unspecified. The action for the outside sector occupying the
volume $\tilde{V}$ contains relevant physical hadronic fields of
zero baryon charge representing color-singlet effective degrees
of freedom. It includes the octet Goldstone bosons and
non-Goldstone excitations as well as the singlet $\eta^\prime$
which acquires mass through $U(1)_A$ anomaly and takes the form
 \be
S_{\tilde{V}}=\frac{f^2}{4}\int_{\tilde{V}} d^4x \left(\Tr\
\del_\mu U^\dagger \del^\mu U +\frac{1}{4N_F}
m^2_{\eta^\prime}[\Tr({\rm ln}U-{\rm ln}U^\dagger)]^2 \right)
+\cdots + S_{WZW}\label{out}
 \ee
where $N_F=3$ is the number of flavors, $\Tr$ goes over the flavor
index and
 \be
U=e^{i\eta^\prime/f_0}e^{2i\pi/f},\\
f_0\equiv \sqrt{N_F/2} f. \nonumber
 \ee
Here the ellipsis stands for mass terms and heavy-mass fields or
higher derivative terms etc. The $S_{WZW}$ is the
Wess-Zumino-Witten term~\cite{witten1} that encodes chiral
anomalies present in QCD:
 \be
S_{WZW}=-N_c \frac{i}{240\pi^2}\int_{D_5[\del D_5=V\times [0,1]]}
\epsilon_{\mu\nu\lambda\rho\sigma} \Tr (L^\mu L^\nu L^\lambda
L^\rho L^\sigma)\label{WZW}
 \ee
with $L=g^\dagger (x,s) dg (x,s)$ defined as $g(x,s=0)=1$ and
$g(x,s=1)=U(x)$. The inside QCD sector and the outside hadronic
sector must be connected by an action that ``translates" them.
This is the role of the surface term. It has two terms, \be
 S_{\del V}=S_{\del V}^{(n)}+S_{\del V}^{(an)}
\ee with the ``normal" boundary term
 \be
S_{\del V}^{(n)}=\frac{1}{2}\int_{\del V} d\Sigma^\mu (n_\mu
\bar{\psi} U^{\gamma_5}\psi)+\cdots \label{surface1}
 \ee
with
 \be
U^{\gamma_5}=e^{i\eta^\prime\gamma_5/f_0}e^{2i\pi\gamma_5/f}
 \ee
and the ``anomalous" boundary term
 \be
 S_{\del V}^{(an)}=i\frac{g^2}{32\pi^2}
\int_{\del V} d\Sigma^\mu {K_5}_\mu (\Tr\ {\rm ln} U^\dagger-\Tr\
{\rm ln} U)+\cdots\label{surface2}
 \ee
where $K_5^\mu$ is the ``Chern-Simons current"
 \be
K_5^\mu=\epsilon^{\mu\nu\alpha\beta} (G_\nu^a
G_{\alpha\beta}^a-\frac{2}{3} gf^{abc} G_\nu^a G_\alpha^b
G_\beta^c).\label{CS}
 \ee
Again irrelevant terms are subsumed in the ellipses. The ``normal"
term (\ref{surface1}) assures chiral invariance of the model as
well as color confinement at the classical level. The ``anomalous"
term -- which is not gauge invariant -- takes care of the quantum
anomaly (Casimir) induced inside the bag that would violate color
confinement~\cite{nrwz,NRZ} if left un-eliminated~\footnote{This
form of bulk action plus a surface Chern-Simons term for the bag
structure has recently been seen to arise from non-perturbative
string theory~\cite{castro}. It turns out that in four dimensions,
the interior of the hadronic bag can be identified as the 3-brane
and the boundary of the bag as the world-volume of the
Chern-Simons 2-brane in which the dynamics is entirely lodged.
This theory does not contain quarks as in the model we are
considering but the topological term that is added carries the
information on the anomaly present in the theory.

We should also point out that the chiral bag structure of the
sort we are discussing arises from QCD proper in the large $N_c$
limit~\cite{manohar}.}.

Up to date, nobody has been able to extract the full content of
this model. There are some unresolved technical difficulties that
frustrate its solvability. Even so, whatever we have been able to
learn from it so far has been found to be fully consistent with
Nature. We believe that physics of any other viable chiral model,
soltitonic or non-solitonic (e.g., \cite{diakonov}), is
essentially -- though perhaps not in detail -- captured in the
model (\ref{model}).

Since there have been extensive reviews on the matter~\cite{NRZ},
we shall simply summarize the salient features obtained from the
model:
 \bitem
\item Due to a Casimir effect in the bag controlled by the surface
term (\ref{surface1}), the baryon charge is fractionized as a
function of the bag radius into the $V$ and $\tilde{V}$ sectors
such that the conserved integer baryon charge is recovered. The
baryon charge fractionization was first noticed in 1980 by Vento
et al.~\cite{ventoetal} with a reasoning based on an analogy to
the monopole-isodoublet system studied by Jackiw and
Rebbi~\cite{jackiwrebbi} and established later~\cite{rgb} for the
special chiral angle (``magic angle") of the pionic hedgehog field
and then generalized in \cite{goldstone} for arbitrary chiral
angle. The result is that the confinement size or the bag radius
has no physical meaning although the baryon size does. The
``cloudy bag" model~\cite{cloudyrev} ignores the hedgehog
component of the pion field, so in that picture, the baryon charge
will be entirely lodged by fiat inside the bag. This implies, in
the framework of the chiral bag, that the cloudy bag is {\it
forced to be big} to be consistent with Nature. The real question
here is whether or not such a big bag can capture the essential
physics involved. This does not address the question of
consistency. On the other hand, the limit where the bag radius is
shrunk to zero with the totality of the baryon charge ``leaked"
into the ``pion cloud" describes the skyrmion, with the pion cloud
picking up the topological charge. The only issue here is whether
or not the point-bag description captures the essential physics
involved. There is no inconsistency in this description of the
baryon.

That the physics should not depend upon the bag radius is an
aspect of the ``Cheshire Cat Principle" (CCP)~\cite{cat}. It is
the trading between the topological character of the skyrmion
configuration and the explicit fermionic charge of the quarks
that makes it satisfy exactly the CCP. It is possible to
formulate this notion as a gauge symmetry. Formulated as a gauge
symmetry, picking a particular bag radius $R$ is equivalent to
fixing the gauge as formulated by Damgaard, Nielsen and
Sollacher~\cite{DNS}. We will call this ``Cheshire-cat gauge
symmetry."
 \item
 While the topological nature of the baryon charge implies an {\it
exact} Cheshire-Cat, the flavor quantum numbers such as isospin,
strangeness etc. are not directly linked to topology. Therefore
their conservation is not directly connected with topology. They
are quantum variables given in terms of collective coordinates.
This means that while conserved flavor quantum numbers in QCD
remain conserved after collective quantization (unlike the
classically conserved baryon charge in the skyrmion limit),
various static quantities such as axial charges $g_A$ (both
singlet and non-singlet), magnetic moments as well as energies and
form factors may not necessarily manifest the {\it perfect} CC
property. They must depend more or less on the details of the
parameters and dynamics taken into account.

A good illustration of this feature is the $N_c$ dependence of the
theory (where $N_c$ is the number of colors). While the skyrmion
sector is strictly valid in the large $N_c$ limit, the bag sector
containing the QCD variables encompasses all orders of $1/N_c$.
This means that in order to assure the CC for the quantities that
depend on $N_c$, the skyrmion sector would have to have
appropriate $1/N_c$ corrections that map to the quark-gluon
sector. In practice this means that other hadronic (massive)
degrees of freedom than pionic must be introduced into the outside
of the bag. What enters will depend on the processes involved.

\begin{figure}
\centerline{\epsfig{file=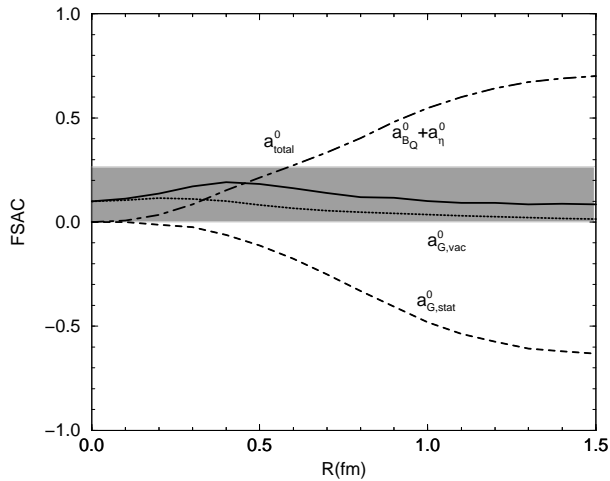, width=9cm}} \caption{Various
contributions to the flavor singlet axial current of the proton
as a function of bag radius and comparison with the experiment:
(a) the ``matter" (quark plus $\eta$) contribution ($a^0_{B_Q} +
a^0_\eta$), (b) the gauge field: the static gluon contribution due
to quark source ($a^0_{G,stat}$) and the gluon vacuum
contribution ($a^0_{G,vac}$), and (c) the total ($a^0_{total})$.
The shaded area corresponds to the range admitted by
experiments.}\label{fsac}
\end{figure}

Most remarkably, however, extensive studies carried out in the
field, chiefly by Park~\cite{BYP}, confirmed that {\it all}
nucleon properties, be they static or non-static, topological or
non-topological, can be satisfactorily and consistently described
in terms of the Cheshire-Cat picture. Even more surprisingly, the
so-called ``proton-spin problem" can also be resolved in a simple
way within the model (\ref{model})~\cite{FSAC}. For this, the
anomalous surface term (\ref{surface2}) associated with the
$U(1)_A$ anomaly and the $\eta^\prime$ mass figures importantly.
As in other Cheshire Cat phenomena (e.g., the baryon charge),
various radius dependent terms conspire to give a more or less
radius independent result: Here the cancellation is between the
``matter" contribution (from both the quark and $\eta^\prime$) and
the ``gauge field" (gluon) contribution, each of which is
individually bag-radius dependent and grows in magnitude with bag
radius. The resulting flavor-singlet axial charge $a^0$ for the
proton shown in Fig. \ref{fsac} comes out to be small as observed
in the EMC/SMC experiments. We should stress that this is not the
only explanation for the small $a^0$. There are several seemingly
alternative explanations but our point here is that the Cheshire
Cat emerges even in a case where QCD aspects are highly
non-trivial and subtle involving quantum anomalies of the chiral
symmetry.
\item So far we have been dealing with long-distance phenomena only.
What about shorter-distance physics? Can Cheshire Cat for instance
address such short-distance problems as nucleon structure
functions seen in deep inelastic lepton scattering where
asymptotic freedom is operative and hence QCD should be
``visible"? The answer to this is that as one probes shorter
distances, the model (\ref{model}) should be modified with more
and more higher-dimension operators implied by the ellipsis in
the action. There is nothing that suggests that the skyrmion
picture when suitably extended should fail if one works hard
enough. It is simply that the calculation will get harder and
harder and it will be more economical to switch over to the ``big
bag" or partonic (quark/gluon) picture. Logically, therefore, a
hybrid description that exploits both regimes would have a more
predictive power. Thus it is easy to understand that the chiral
quark-soliton model, a variation on the model (\ref{model})
formulated in a relativistically invariant form~\cite{diakonov},
has been successfully applied to the problem. We will see later
that even at asymptotic density where the QCD variables are the
correct variables, the notion of effective fields is applicable.
\eitem

\setcounter{equation}{0} 
\renewcommand{\theequation}{\mbox{\ref{EFTNucleus}.\arabic{equation}}}
\section{EFFECTIVE FIELD THEORIES IN NUCLEAR
PHYSICS}\label{EFTNucleus} \subsection{Chiral Symmetry in Nuclear
Processes} \itt
 The next question we wish to address is:
Given the Cheshire Cat freedom, how does one treat nuclei,
namely, two- -or more-nucleon systems? That is, how is nuclear
physics formulated in QCD?

For this, it is clear that the most efficient and economical way
would be to adopt the ``point-bag" description~\footnote{In
accordance with the Cheshire-Cat principle, one could continue
using the hybrid bag-meson description by incorporating relevant
(hadron) degrees of freedom~\cite{saito}. One advantage of this
approach might be that the quark-bag introduction can implement
more readily short-range physics that enters in many-body systems,
albeit at a sub-leading level.}. Once we adopt the zero-radius bag
surrounded by meson cloud as a nucleon, we can represent it as a
local field interacting with pions. How to do nuclear physics in
terms of {\it local} nucleon and meson fields and still do QCD is
encoded in what is called ``Weinberg theorem"~\cite{weinbergth}.
In the present context, the theorem states that doing {\it in a
consistent way} an effective field theory with the nucleon, pion
and other meson fields is no more and no less than doing the gauge
field theory QCD in terms of quarks and gluons.~\footnote{To quote
Weinberg~\cite{weinbergth}: ``{\it We have come to understand that
particles may be described at sufficiently low energies by fields
appearing in so-called effective quantum field theories, whether
or not these particles are truly elementary. For instance, even
though nucleon and pion fields do not appear in the Standard
Model, we can calculate the rates for processes involving
low-energy pions and nucleons by using an effective quantum field
theory of pion and nucleon fields rather than of quark and gluon
fields.... When we use a field theory in this way, we are simply
invoking the general principles of relativistic quantum theories,
together with any relevant symmetries; we are not making any
assumption about the fundamental structure of physics.}"} What is
required is that such an approach preserve the necessary
symmetries -- Lorentz invariance, unitarity, cluster decomposition
etc. -- and is quantum mechanical. We are simply to write down the
most general such local Lagrangian with all possible terms
consistent with the symmetries involved. This strategy is
powerfully illustrated in the description of the baryon (octet and
decaplet baryon) structure in terms of an effective chiral
Lagrangian constructed of baryon and Goldstone boson
fields~\cite{meissner}~\footnote{While the Lagrangian
(\ref{model}) is a {\it model} of QCD, the baryon chiral
Lagrangian used here is an effective {\it theory} of QCD. It is
therefore reasonable to expect that being a theory, the more
judiciously organized and the more terms computed, the more
accurate the calculation will become. There is of course the
question of convergence but applied within appropriate kinematics,
the theory should work out better the more one works. This is
indeed what is being found in baryon chiral perturbation
calculations for which the strategy is fairly well formulated.
This is not the case with ``nuclear chiral perturbation theory"
where no systematic strategy has yet been fully worked out as we
will stress below.}. An initial attempt to implement Weinberg
theorem in nuclear physics was made in 1981~\cite{ericeMR} (it was
incomplete in the chiral counting at the time but was completed in
1991 as mentioned later) but the major development came after
Weinberg's paper in 1990~\cite{wein}.
 \subsection{Objectives of EFTs in Nuclear Physics}
 \itt
There are two major roles of EFT in nuclear physics. One is to
establish that nuclei as strongly interacting systems -- that
have been accurately described in the past in what one would call
``standard nuclear physics approach (SNPA)" or alternatively
``potential model (PM)" based on phenomenological potentials --
can {\it also} be understood from the point of view of a
fundamental theory, QCD, and the other is to be able to make
nontrivial and precise predictions that are important not only for
nuclear physics {\it per se} but also for other areas of physics
such as astrophysics and condensed matter physics. The former
renders the impressively rich variety of nuclear processes a
respectable domain of research. The latter is to provide an
invaluable tool for progress in related areas of physics. In this
review some recent developments that are relevant to the general
theme of this review will be discussed.

The examples of remarkable success are the $n+p\rightarrow
d+\gamma$ at threshold~\cite{np} -- a classic nuclear physics
process -- and the inverse $\gamma+d\rightarrow
n+p$~\cite{chensavage} -- a process relevant to cosmological
baryosynthesis. Going beyond the simplest nuclear systems
involving two nucleons, the method can be applied to processes
that involve $n$ nucleons where $n>2$, such as for instance the
solar ``hep" process~\cite{rocco,hep,tspetal-hep} $^3{\rm
He}+p\rightarrow ^4{\rm He}+e^+ +\nu_e$ which figures in the solar
neutrino problem and the axial charge transitions in heavy nuclei
(to be discussed later) which provide evidence for BR scaling.

An exhaustive and beautiful review on the subject with a somewhat
different emphasis has recently been given by Beane et
al~\cite{seattle} where a comprehensive list of references is
found.

In addressing nuclear EFT, there are currently two complementary
-- and {\it not different} as one might be led to believe -- ways
of organizing the expansion, one based on Weinberg's
strategy~\cite{wein} and the other based on Seattle-Pasadena
strategy~\cite{KSW}. We will loosely refer to the former as
``Weinberg scheme" and to the latter as ``KSW
(Kaplan-Savage-Wise) scheme." The Weinberg scheme contains the
SNPA/PM (standard nuclear approach/potential model) as a
legitimate and essential component of the theory whereas the KSW
scheme by-passes it, although the latter can be effectively
exploited to justify the former in certain kinematic domains that
we are interested in. We will focus on the former as we find it
more natural and straightforward from nuclear physicists' point of
view and for the processes we will discuss, sometimes
considerably more predictive.

The Weinberg scheme~\cite{wein,bira} that we will use here is the
version that has been formulated by Park et al~\cite{MR91,PKMR}
(this will be referred to hereafter as PKMR) and uses the
following basic strategy. It starts by recognizing that the
SNPA/PM (e.g., \cite{pm,argonne}) based on solutions of the
Schr\"odinger equation with the ``most realistic" potential
available in the market fine-tuned to a large number of
experimental data on one and two nucleon systems, captures most
accurately the essential physics of, and hence describes, when
implemented by many-body forces, the bulk of properties of
few-nucleon systems, in some cases (e.g., low-energy scattering)
within $\sim 1$ \% accuracy~\footnote{There is a well-known caveat
in carrying out this program in general and in assessing the
``accuracy" of approximations in particular for $n$-body systems
with $n>2$. While there are useful constraints that help in
organizing, in Weinberg's terminology, the ``reducible" and
``irreducible" terms for two-body systems, this is not usually the
case for many-body systems. Together with many-body forces in the
potential which are difficult to pin down precisely although
numerically small, there can be a variety of organizational
ambiguities in classifying ``reducible" and ``irreducible"
contributions, e.g., off-shell effects, that could contaminate the
calculations resorting to a variety of approximations one is
compelled to make, even though a full consistent field theoretic
approach should in principle be free from such ambiguities. This
is an issue that will require considerable effort to sort out. We
are grateful to Kuniharu Kubodera for stressing this point in the
contex of the PKMR approach.}. We will therefore consider the
SNPA/PM results as {\it input} (e.g., ``counter terms") in lieu of
attempting to compute them from first principles. Our aim is {\it
not} to recalculate them from a {\it more fundamental theory} but
to correctly interpret and incorporate them as a legitimate
leading term in a systematic expansion of a more fundamental
framework~\cite{KDR,MR91}, namely, an effective theory of QCD.
This, we believe, is completely in line with the notion of
effective theory, e.g., the one \`a la Wilson. Our key point is
that {\it it is in how to do embed, and how to systematically
calculate the quantities that are missing from, the SNPA/PM
results into a fundamental theory that EFT has its true power}.
Since the standard nuclear physics approach/potential model
(SNPA/PM) results are to be inputs to the more fundamental theory,
it is clearly more profitable to look at the {\it response
functions} to external fields rather than at scattering amplitudes
(which the SNPA/PM can address accurately) to make meaningful
predictions. Our thesis is that this is where chiral symmetry and
its broken mode can play their primary role in nuclear physics, in
both quantitatively describing and accurately predicting nuclear
phenomena.
 \subsection{Effective Chiral Lagrangians}
 \itt
For very low-energy processes that we are dealing with, the
nucleons can be considered ``heavy," i.e., nonrelativistic while
the pions are relativistic. We shall therefore use the
heavy-baryon formalism although a relativistic formulation can be
made, particularly for the case where the nucleon mass drops as
in BR scaling that we will discuss below. The leading-order
chiral Lagrangian that consists only of local nucleon and pion
fields -- with other heavy fields integrated out -- takes the form
 \be
\calL_0 &=& {\bar N}\left[ i v\cdot D + 2 i g_A S\cdot \Delta
\right] N - \frac{1}{2} \sum_A C_A \left({\bar N} \Gamma_A
N\right)^2
\nonumber \\
&&+\, f_\pi^2 {\rm Tr}\left(i \Delta^\mu i \Delta_\mu\right) +
\frac{f_\pi^2}{4} {\rm Tr}(\chi_+) \label{lag1}
 \ee with
 \be
D_\mu N &=& (\del_\mu + \Gamma_\mu) N ,
\nonumber \\
\Gamma_\mu &=& \frac{1}{2} \left[\xi^\dagger,\, \del_\mu
\xi\right] -\frac{i}{2}\xi^\dagger {\cal R}_\mu \xi - \frac{i}{2}
\xi {\cal L}_\mu\xi^\dagger,
\nonumber \\
\Delta_\mu &=& \frac{1}{2} \left[\xi^\dagger,\, \del_\mu
\xi\right] +\frac{i}{2}\xi^\dagger {\cal R}_\mu \xi - \frac{i}{2}
\xi {\cal L}_\mu\xi^\dagger,
\nonumber \\
\chi_+ &=& \xi^\dagger \chi \xi^\dagger + \xi \chi^\dagger \xi
\label{deltamu}
 \ee
where ${\cal R}_\mu = \frac{\tau^a}{2} \left(
  {\cal V}^a_\mu + {\cal A}^a_\nu\right)$
and ${\cal L}_\mu = \frac{\tau^a}{2} \left(
  {\cal V}^a_\mu - {\cal A}^a_\nu\right)$
denote, respectively, the left and right external gauge fields,
$\chi$ is proportional to the quark mass matrix and if we ignore
the small isospin-symmetry breaking, becomes $\chi=m_\pi^2$ in the
absence of the external scalar and pseudo-scalar gauge fields, and
 \be
\xi = \sqrt{\Sigma} =
  {\rm exp}\left(i\frac{{\vec \tau}\cdot {\vec \pi}}{2 f_\pi}\right)
 \ee
is the chiral field for the Goldstone bosons. The $\calL_0$ is the
leading order {chiral Lagrangian} in the sense that it is leading
order in derivatives, in pion mass and in ``$1/M$" where $M$ is
the free-space nucleon mass. This is a suitable Lagrangian for
(chiral) perturbation near the medium-free vacuum described by the
so-called ``irreducible graphs." However for nuclear processes in
general, certain ``reducible graphs" involve propagators that are
infra-red enhanced and require terms higher order in $1/M$,
namely, nucleon kinetic energy term. For the KSW approach, the
{\it ab initio} account of this ``correction" term is essential.
For the PKMR where the reducible graphs are accounted for via
Schr\"odinger equation or Lippman-Schwinger equation, the chiral
expansion is all that matters, so this Lagrangian plays a key
role.

The second term in (\ref{lag1}) is the leading four-fermion
interaction and contains no derivatives. We will specify the
explicit form later. For convenience, we will work in a reference
frame in which the four velocity $v^{\mu}$ and the spin operator
$S^{\mu}$ are \be v^\mu = (1,\, {\vec 0}) \ \ \ \mbox{and} \ \
S^\mu = \left(0,\, \frac{{\vec \sigma}}{2}\right).
 \ee
The next-to-leading order Lagrangian including four-fermion
contact terms can be written as
 \be
\calL_1 &=& {\bar N} \left(
  \frac{v^\mu v^\nu - g^{\mu\nu}}{2 m_N} D_\mu D_\nu +
\frac{g_A}{m_N}\{S\cdot D,v\cdot D\}+c_1 \Tr\chi_+ +
4\left(c_2-\frac{g_A^2}{8m_N}\right) (v\cdot i\Delta)\right.\nonumber\\
\ \ \ && + \left. 4 c_3 i\Delta\cdot i\Delta
   + \left(2 c_4 + \frac{1}{2m_N}\right)
  \left[S^\mu, \,S^\nu\right] \left[ i \Delta_\nu,\, i\Delta_\nu\right]
\right)N
\nonumber \\
 && -\ 4 i d_1 \,
 {\bar N} S\cdot \Delta N\, {\bar N} N
 + 2 i d_2 \,
  \epsilon^{abc}\,\epsilon_{\mu\nu\lambda\delta} v^\mu \Delta^{\nu,a}
 {\bar N} S^\lambda \tau^b N\, {\bar N} S^\delta \tau^c N
+ \cdots, \label{Lag1}
 \ee
where $\epsilon_{0123}=1$, $\Delta_\mu = \frac{\tau^a}{2}
\Delta^a_\mu$. Terms that are not relevant for the present
discussion are implied in the ellipses.
 \subsection{Nucleon-Nucleon Scattering}\label{nnscattering}
 \itt
Although the genuine predictivity of the EFT is in response
functions, we start with nucleon-nucleon scattering at very low
energies as this issue has been the focus of the recent activity
in the field. If the energy scale involved is much less than the
pion mass, $m_\pi\sim 140$ MeV, then we might integrate out the
pion field and work with an effective theory containing only the
nucleon field. We will reinstate the pions later since the PKMR
strategy relies on the pion degree of freedom explicitly. Since
the pionless theory has interesting aspects on its own that merits
study, we will consider this before putting the pion field into
the picture.

When the pions are integrated out, the Lagrangian (\ref{lag1})
further simplifies to
 \be
\calL =&& N^\dagger\left(i\del_t
+\frac{\vec{\nabla}^2}{2M}\right)N
+\cdots\no\\
&-& \frac 12 [C^s (N^\dagger N)^2 + C^t (N^\dagger \Bsigma N)^2
+\cdots]\label{lag2}
 \ee
where the nucleon kinetic energy term (which is formally higher
order from the heavy-baryon chiral counting) is reinstated. With
the pion fields removed from the theory, this Lagrangian has no
remnant of chiral symmetry. The nucleon field is just a matter
field and does not know anything about the symmetry. This does
not mean that the theory is inconsistent with the symmetry of
QCD. We shall consider the S-wave channels $^1S_0$ and $^3S_1$
(in the notation of $^{2S+1}S_J$). The $C$ coefficients for these
channels are related to the $C$ coefficients in (\ref{lag2}) by
 \be
C^{(^1S_0)}=C^s -3C^t, \ \ \ \ C^{(^3S_1)}=C^s+C^t.
 \ee

\begin{figure}[ht]
\hskip 1.5in \centerline{\epsfig{figure=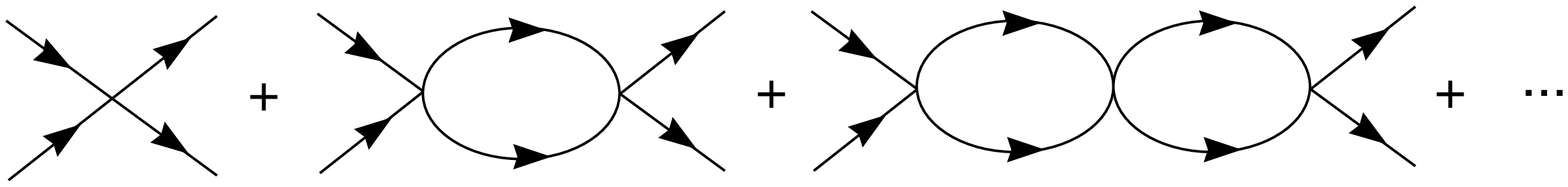,height=1.0in}}
\caption{Graphs contributing to the scattering
amplitude}\label{bubble}
\end{figure}

The scattering amplitude is $\calA$ given by the sum of the
Feynman diagrams given in Fig.\ref{bubble}:
 \be
 \calA=-\frac{C}{1-C(GG)}\label{amp}
\ee where $C$ is the four-fermion interaction constant in
(\ref{lag2}) and $(GG)$ is the two-nucleon propagator connecting
the two vertices, both written symbolically. In the
center-of-mass (CM) frame, this propagator has a linear divergence
in dimension $D=4$ as one can see from a naive dimensional
counting of
 \be
 (GG)&=& -i\int \frac{d^4q}{(2\pi)^4}
\left(\frac{i}{E+q_0-\Bq^2/2M+i\epsilon}\right)
\left(\frac{i}{-q_0-\Bq^2/2M+i\epsilon}\right)\no\\
&=&\int \frac{d^3q}{(2\pi)^3}
\left(\frac{1}{E-\Bq^2/M+i\epsilon}\right)\label{GG} \ee where
$|\Bk|=\sqrt{ME}$ is the CM momentum. To make the integral
meaningful, let us put a momentum cutoff at say at $\Lambda$.
Then (\ref{GG}) is given by
 \be
 (GG)_\Lambda= -\frac{M}{4\pi}\left(\frac{\Lambda}{\pi}
 +i|\Bk|\right).\label{gglambda}
\ee This linear divergence poses no fundamental problem here since
the theory is an effective one with the degrees of freedom above
a certain cutoff having been integrated out. On the contrary, in
the Wilsonian sense, this cutoff dependence has a physical
meaning: It delineates the onset of ``new physics". Substituting
this into the amplitude formula (\ref{amp}), we have
 \be
\calA=
\frac{-C}{1+\frac{CM}{4\pi}(\frac{\Lambda}{\pi}+i|\Bk|)}.\label{amp1}
 \ee
Next, we trade in the theoretical constant $C$ for a physical
quantity. To do this, the scattering amplitude (for $S$ wave) is
written in terms of the phase shift $\delta\equiv\delta_S$ as
 \be
 \calA=\frac{4\pi}{M}\frac{1}{|\Bk| \cot\delta -i |\Bk|}.
\ee Now using the effective range formula
 \be
|\Bk|\cot\delta=-\frac{1}{a} +\frac 12 r_0 |\Bk|^2 +{\cal O}
(|\Bk|^4)\label{er}
 \ee
where $a$ is the scattering length and $r_0$ the effective range,
we have, to the lowest order in momentum,
 \be
\calA=-\frac{4\pi}{M} \frac{1}{1/a +i|\Bk|}.\label{amp2}
 \ee
So from (\ref{amp1}) and (\ref{amp2}), we find
 \be
C(\Lambda)=\frac{4\pi}{M}\left(\frac{1}{-\frac{\Lambda}{\pi}+1/a}\right)
\label{c0lambda}
 \ee
where we have explicited the cutoff dependence of the effective
Lagrangian which shows how an effective theory carries
the``memory" of the degrees of freedom that have been integrated
out. Obviously physical observables should not depend upon what
value one takes for the cutoff $\Lambda$, since the cutoff can be
chosen arbitrarily. This is a statement of renormalization group
(RG) invariance of the physical quantities, which reads for the
amplitude as
 \be
 \Lambda\frac{d\calA}{d\Lambda}=0.
\ee One can easily verify that (\ref{c0lambda}) is the solution
of this renormalization group equation (RGE).

In standard field theory calculations in particle physics, one
uses dimensional regularization (DR) instead of cutoff. This is
because the DR has the advantage that it preserves symmetries,
e.g., chiral symmetry in our case, order by order. In the case we
are concerned with, with no pions, no symmetry is spoiled by the
cutoff regularization at least up to the order considered. In
general when one does higher order calculations, however, it is
not a straightforward matter to preserve chiral symmetry with the
cutoff regularization since it requires judicious inclusion of
symmetry breaking counter terms so as to restore symmetry. This is
why some authors (e.g., the KSW collaboration) prefer using DR.
Now what does the above linear divergence mean in DR?

For this we go back to the propagator $(GG)$, Eq.(\ref{GG}), and
work out the integral using DR: \be
 (GG)_\mu &=& -i(\frac{\mu}{2})^{4-D}\int \frac{d^Dq}{(2\pi)^D}
\left(\frac{i}{E+q_0-\Bq^2/2M+i\epsilon}\right)
\left(\frac{i}{-q_0-\Bq^2/2M+i\epsilon}\right)\no\\
&=&(\frac{\mu}{2})^{4-D}\int \frac{d^{D-1}q}{(2\pi)^{D-1}}
\left(\frac{1}{E-\Bq^2/M+i\epsilon}\right)\no\\
&=&M(-|\Bk|^2)^{(D-3)/2}\Gamma(\frac{3-D}{2})\frac{(\frac{\mu}{2})^{4-D}}
{(4\pi)^{(\frac{D-1}{2})}}. \ee
 This is regular for $D\neq 3$ but singular for $D=3$. In
renormalizable field theories where power divergences are absent
in principle, it is legitimate to simply set $D=4$, thereby
dropping power divergences. But in non-renormalizable theories,
one cannot sweep under the rug the singularities that lurk behind
the physical dimension. This is because in the Wilsonian
renormalization scheme to which our EFT belongs, there are no true
singularities and as mentioned, the power divergence structure
instead carries physical information. As mentioned, the situation
here is quite analogous to the quadratic divergence that needs to
be included in hidden local symmetry (HLS) theory discussed
above~\footnote{Taking into account the power divergence (i.e.,
quadratic divergence) in HLS is crucial in arriving at the the
vector manifestation of chiral symmetry of Harada and
Yamawaki~\cite{VM} discussed in Section \ref{how} and also in
Section \ref{complementarity}.}. The physics behind may be the
same. In fact, this must be a generic feature of all effective
field theories of the sort we are dealing with here, that is, the
theories that have fixed points nearby. To do a correct
regularization with DR in the present case, the linear power
divergence -- the only power divergence there is in this theory at
one loop -- has to be subtracted. In this sense, the ``power
divergence subtraction (PDS)" scheme used by KSW is a proper
implementation of Wilsonian effective field theory in
nucleon-nucleon scattering, the unnatural length scale associated
with the large scattering length signaling the presence nearby of
an infrared fixed point.

Subtracting the linear divergence, we have the
power-divergence-subtracted (PDS) propagator
 \be
(GG)_\mu^{PDS}=(GG)_\mu -\frac{M\mu}{4\pi (D-3)} =
-\frac{M}{4\pi}(\mu  +i|\Bk|).
 \ee
Comparing with (\ref{gglambda}), we see that the scale parameter
$\mu$ in DR is equal to the cutoff $\Lambda/\pi$.

As noted by \cite{wisecft}, the large scattering length for the
$S$-wave scattering, i.e., $a^{(^1S_0)}=-23.714$fm $\sim (1/(8 \,
{\rm MeV}))$ reflects that nature is close to a fixed point and
conformal invariance. This is seen in (\ref{c0lambda}) in the
limit $1/a\rightarrow 0$:
 \be
\Lambda \frac{d}{d\Lambda} (\Lambda C)|_{1/a\rightarrow 0}=0.
 \ee
In view of the length scale involved, $a\rightarrow \infty$ is
close to nature, signaling a quasi-bound state. Note that the
linear divergence subtraction plays a crucial role here, a
feature analogous to the conformal invariance of hidden local
symmetry (HLS) theory at chiral restoration with the quadratic
divergence playing a crucial role~\cite{VM}.

The discussion made up to this point is general, clarifying the
essential feature of effective field theories in two-body nuclear
systems. The KSW approach has been developed further, including
pions as perturbation~\cite{seattle}. As we will show below, the
PKMR strategy, while faithful to the EFT strategy, departs from
this point in that the rigorous counting rule is sacrificed {\it
somewhat} in favor of predictivity.
 \subsection{The PKMR Approach: {\it More Effective} EFT}
\itt
 The faithful adherence to
the counting rule has the short-coming in that while post-dictions
are feasible provided enough experimental information is
available, it is hard if not impossible to make accurate
predictions. The reason is that as one goes to higher orders to
achieve accuracy, one encounters increasing number of counter
terms that remain as unknown parameters and these parameters can
at best be fixed by the {\it very process} one would like to
predict. This is just a fitting, and not a prediction. Thus
instead of adhering strictly to the counting rule that requires
by-passing wave functions, we shall develop a scheme along the
line conceived by Weinberg that would allow us to exploit the wave
functions that are ``accurately" determined from the standard
nuclear physics/PM procedure.

The key observation that we will exploit is the following. All
nuclear processes involve two sorts of graphs: ``reducible" and
``irreducible." In the case of two-body systems, they will be
two-particle reducible and two-particle irreducible (2PI). Instead
of summing the two classes of graphs order-by-order in strict
accordance with a given counting scheme as in KSW, one computes
irreducible graphs to a given order in chiral perturbation theory
($\chi$PT) and then account for reducible graphs by solving
Schr\"odinger equation or Lippman-Schwinger equation with the
irreducible vertex injected into the equation. This procedure
accords to potentials and wave functions a special role as in the
standard nuclear physics calculations. What we hope to do is then
to combine the accuracy of the standard nuclear physics/potential
model approach with the power of $\chi$PT. We might call this
{\it more effective} effective field theory (MEEFT). The
advantage of working with wavefunctions is that one can make {\it
predictions} in nuclear electro-weak response functions that are
difficult to make in an order-by-order calculation because of
un-fixed counter terms. The disadvantage is of course that one is
forced to sacrifice the strict adherence to a counting rule. It
turns out however that in the processes so far studied with
controlled approximations, this sacrifice is minor numerically. A
clear explanation as to how this comes about in two-body systems
was given by Cohen and Phillips~\cite{cohen-phillips}.
 \subsubsection{\it NN scattering}
 \itt
To justify the method that we will apply to response functions,
we first discuss how the method works out in nucleon-nucleon
scattering which is well understood in various different ways,
field theoretical or non-field theoretical. Compared with the KSW
approach, the present scheme is a bit less elegant but the point
of our exercise is to show that it can be at least as accurate as
the KSW scheme~\cite{cohen-phillips}.

For simplicity consider proton-neutron scattering in the $^1S_0$
channel~\footnote{The proton-proton interaction as in the solar
proton fusion process requires the Coulomb interaction in addition
to the QCD interaction. Although technically delicate, this part
is known, so we will not go into the detail. We shall also
restrict to the $^1S_0$ channel. The $^3S_1$ channel is slightly
complicated because of the coupling to the $D$-wave but involves
no new ingredient.}. Since we are to account for the reducible
graphs by Lippman-Schwinger equation, we only need to compute the
potential as the sum of irreducible graphs to a certain chiral
order in the Weinberg counting. For this we can work directly with
the heavy-baryon chiral Lagrangian (\ref{lag1}) with the pion
exchange put on the same footing as the contact four-fermion
interaction. The potential calculated to the next-to-leading
order (NLO) is of the form
 \be
{\cal V}(\Bq) &=& - \tau_1\cdot\tau_2 \,\frac{g_A^2}{4 f_\pi^2}\,
\frac{ \sigma_1\cdot \Bq\,\sigma_2\cdot \bfq}{\Bq^2+m_\pi^2} +
\frac{4\pi}{M} \left[C_0 + (C_2 \delta ^{ij} + D_2 \sigma^{ij})
q^i q^j \right]+\cdots \label{Vq}
 \ee
with
\begin{equation}
\sigma^{ij} = \frac{3}{\sqrt{8}} \left( \frac{\sigma_1^i
\sigma_2^j + \sigma_1^j \sigma_2^i}{2} - \frac{\delta^{ij}}{3}
\sigma_1 \cdot \sigma_2 \right),
\end{equation}
where $\Bq$ is the momentum transferred. The $C_2$ and $D_2$
terms containing quadratic derivatives are NLO contributions that
are added to (\ref{lag2}) since we will work to that order. There
are one-loop corrections involving two-pion exchange that appear
at the NLO and that can be taken into account but as shown by
Hyun, Park and Min~\cite{hyun}, in the kinematic regime we are
considering here, they are not important for the key point we want
to discuss here. As it stands, the potential (\ref{Vq}) is not
regularized, so it has to be regularized just as the Feynman
propagator $(GG)$, Eq.(\ref{GG}), had to be. One can do this in
various different ways; they should all give the same result. A
particularly convenient one for solving differential equations in
coordinate space is to take the following form when one goes to
coordinate space \be V(\Br) &\equiv& \int\!
\frac{d^3\Bq}{(2\pi)^3} \, \e^{i\Bq\cdot \Br}\, S_\Lambda(\Bq^2)\,
 {\cal V}(\Bq),
\label{regV}\ee with a smooth regulator $S_\Lambda(\bfq^2)$ with
a cutoff $\Lambda$. For our purpose it is convenient to take the
Gaussian regulator,
 \be
S_\Lambda(\Bq^2) = \exp\left( - \frac{\Bq^2}{ \Lambda^2}
\right).\label{Sdef}\ee

In the PKMR formalism, pions play a special role due to what is
called ``chiral filter mechanism" (to be defined below) but we can
illustrate our essential point without the pions, so let us drop
the pion exchange term for the moment. Given the
Fourier-transformed potential properly regulated as mentioned,
the Lippman-Schwinger equation for the $S$-wave can be solved in
a standard way. The resulting wave function can be explicitly
written down:
\begin{eqnarray}
\psi({\bf r}) &=& \varphi({\bf r})
 + \frac{S(\frac{M E}{\Lambda^2})\,C_E}{1- \Gamma_E C_E} \,
  \left[
1 - \frac{\sqrt{Z} C_2}{C_E} (\nabla^2 + ME) \right.
\nonumber \\
&&\left. - \frac{\sqrt{Z} D_2}{C_E}
 \frac{S_{12}({\hat r})}{\sqrt{8}}
 r \frac{\partial}{\partial r} \frac{1}{r} \frac{\partial}{\partial r}\right]
 {\tilde \Gamma}_\Lambda({\bf r})
\label{fullwave}
\end{eqnarray}
where $\varphi$ is the free wave function and
\begin{eqnarray}
\Gamma_E &=& 4\pi \int \frac{d^3{\bf p}}{(2\pi)^3}
\frac{S_\Lambda^2({\bf p})}{ME - {\bf p}^2 + i0^+}, \label{GE}
\\
{\tilde \Gamma}_\Lambda({\bf r}) &=& 4\pi \int \frac{d^3{\bf
p}}{(2\pi)^3} \frac{S_\Lambda({\bf p})}{ME - {\bf p}^2
 + i0^+} \mbox{e}^{i {\bf p}\cdot {\bf r}}\label{GLambda} ,
\\
Z &=& (1 - C_2 I_2)^{-2},
\label{Z}\\
C_E &=& a_\Lambda \left(1 + \frac12 a_\Lambda r_\Lambda ME\right)
 + (\sqrt{Z} D_2 ME)^2 \Gamma_E ,
\end{eqnarray}
with
\begin{eqnarray}
a_\Lambda &\equiv& Z\left[C_0 + (C_2^2 + \delta_{S,1} D_2^2)
I_4)\right],
\\
r_\Lambda  &\equiv& \frac{2 Z}{a_\Lambda^2} \left[
   2 C_2 - (C_2^2 - \delta_{S,1} D_2^2) I_2 \right]
\label{arC}\end{eqnarray} where $I_{n}$ ($n=2,\,4$) are defined by
\begin{equation}
I_{n} \equiv
 - \frac{\Lambda^{n+1}}{\pi} \int_{-\infty}^\infty
 dx\, x^{n} S^2(x^2).
\end{equation}

The phase shift can be calculated by looking at the large-$r$
behavior of the wavefunction, \begin{equation} p\cot \delta =
\frac{1}{S^2(\frac{ME}{\Lambda^2})} \left[
 I_\Lambda(E)
 - \frac{1-\eta^2(E)}{a_\Lambda (1 + \frac12 a_\Lambda r_\Lambda ME)}
\right], \label{pCot1S0}\end{equation} where the $\eta(E)$ is the
$D/S$ ratio, which vanishes for the $^1S_0$ channel and
$I_\Lambda(E)= \Gamma_E+i\sqrt{ME} S^2 (ME/\Lambda^2)\equiv\Lambda
I(\frac{ME}{\Lambda^2})$. The expression for the phase shift
contains a lot more than what was considered above even at LO
since the wavefunction is computed to all orders in the potential
and we are using a Gaussian cutoff. As such it is not obvious
that the renormalization group invariance, i.e., the $\Lambda$
independence, is preserved. This may appear to be a shortcoming
of the approach and would constitute a defect if the $\Lambda$
dependence were non-negligible.

\begin{figure}[t]
\centerline{\epsfig{file=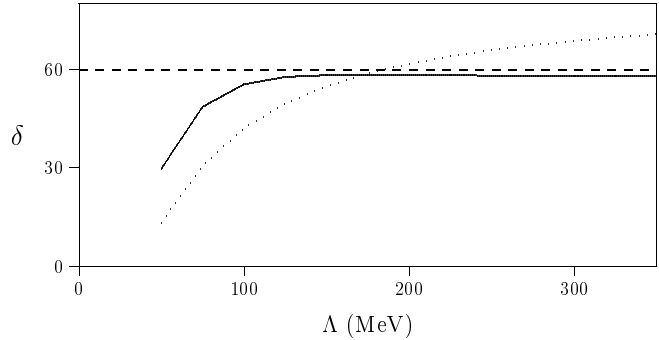,width=3.375in}}
\caption[deviate]{\protect \small Cutoff dependence/independence
in pionless theory: $np$ $^1 S_0$ phase shift (degrees) vs. the
cut-off $\Lambda$ for a fixed CM momentum $p= 68.5$ MeV. The
solid curve represents the NLO result, the dotted curve
the LO
result and the horizontal dashed line represents the experimental
value.}\label{lamdep}
\end{figure}

Let us see how this works out with the present pionless example.
As before, we use the effective range formula (\ref{er}) to fix
the coefficients $C_{0,2}$. We have \begin{eqnarray}
\frac{1}{a_\Lambda} &=& \frac{1}{a} + \Lambda I(0)
 = \frac{1}{a} - \frac{\Lambda}{\sqrt{\pi}},
\\
r_\Lambda &=& r_e - \frac{2 I'(0)}{\Lambda} - \frac{4 S'(0)}{a
\Lambda^2} = r_e - \frac{4}{\sqrt{\pi}\Lambda} + \frac{2}{a
\Lambda^2}.
\end{eqnarray}
At threshold, it can be easily verified that the
renormalization-group (RG) invariant form for $C_0$,
(\ref{c0lambda}), is recovered. What happens away from the
threshold is less trivial. For illustration we take the CM
momentum $p=68.5$ MeV corresponding to $\sim m_\pi/2$. The result
for the $^1S_0$ channel is given in Fig.\ref{lamdep}; here the
phase shift $\delta$ (in degree) is plotted vs. $\Lambda$.
We learn from this exercise that within the present scheme it is
clearly inconsistent with the renormalization group invariance if
the potential is calculated only up to LO. It is at NLO and
$\Lambda\gsim m_\pi$ that the cutoff independence is recovered.
We see that $\Lambda\sim m_\pi$ is the scale at which a ``new
degree of freedom" enters into the pionless theory, an obvious
but important fact.

\begin{figure}[htbp]
\centerline{\protect \epsfig{file=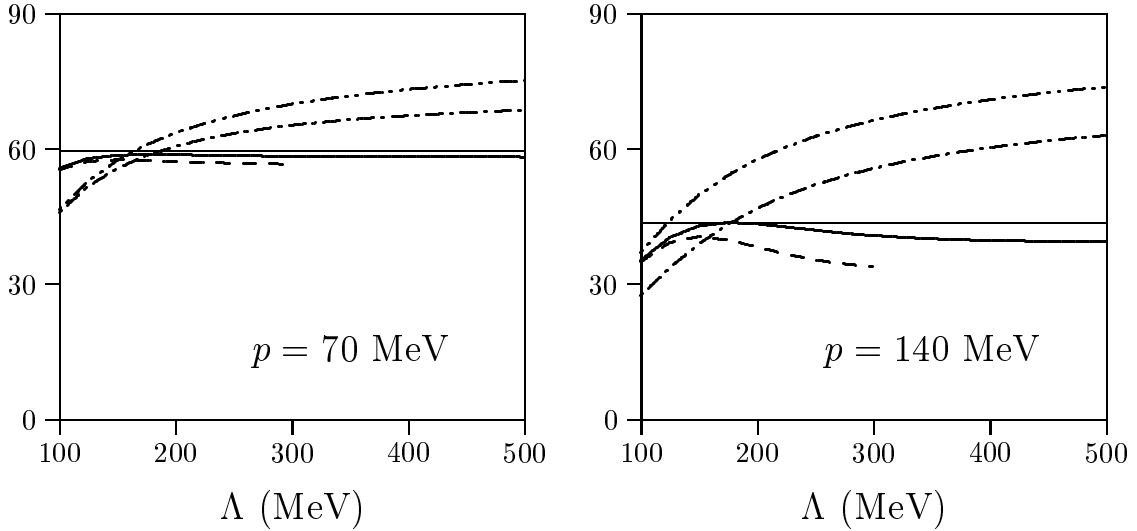}}
\caption[deviate]{\protect \small The $np$ $^1 S_0$ phase shift
(degrees) vs. the cutoff $\Lambda$ for fixed CM momenta $p=70$
MeV (left) and $p= 140$ MeV (right). The next-to-leading order
(NLO) results are given by the solid (with pions) and dashed
(without pions) curves, and the leading-order (LO) results by the
dot-dashed (with pion) and dot-dot-dashed (without pion) curves.
The horizontal line represents the experimental data obtained
from the Nijmegen multi-energy analysis. The NLO result without
pions is drawn only up to $\Lambda \gsim 300$ MeV, above which
the theory becomes meaningless, that is, unnatural as well as
inconsistent.\label{eft-figure1}}
\end{figure}

Next we incorporate pions. Once pions are included, the cutoff
should be put above $m_\pi$. Furthermore the pion presence should
reduce the $\Lambda$ dependence -- if any --  for given CM
momenta. This feature is seen in Fig.\ref{eft-figure1}.
We see clearly that the pion presence improves markedly the
ability to describe scattering. The interplay between the probe
momentum, the pion presence, the cutoff and the chiral order
becomes transparent: The more refined the potential and the
higher the chiral order, the more consistent and more accurate
becomes the prediction. This will be the key point of our next
consideration when we look at response functions.
 \subsubsection{\it Electroweak response functions: Accurate
post-dictions and predictions} \vskip 0.3cm

We now turn to our principal assertion that it is in calculating
nuclear response functions to the electroweak (EW) external fields
that the PKMR approach has its power. We will first discuss
two-body systems and then make predictions for $n$-body systems
with $n\geq 2$. In doing this, we will put pions to start with
since we will exploit what is known as ``chiral filter mechanism"
first introduced in \cite{KDR} and elaborated further in
\cite{PKMR}. The chiral filter mechanism which has been verified
in $\chi$PT states that whenever allowed by symmetry and
kinematics, one-soft-pion exchange in the electroweak currents
dominates, with the corrections to which are in principle
controllable by $\chi$PT and, conversely, whenever forbidden by
symmetry or suppressed by kinematics, the corrections to the
leading order current are uncontrolled by chiral expansion and
require going beyond standard low-order $\chi$PT. We will refer
to these two phenomena as ``two sides of the same coin." In this
formalism, therefore, pions play a prominent role. At a momentum
scale $p\ll m_\pi$, one can work with a pionless EFT as described
in \cite{seattle} in which the role of the chiral filter would of
course be moot. However in this case, one loses the predictive
power since one would be left with one or more counter terms that
are left undetermined by theory or experiments. The PKMR strategy
circumvents this difficulty.

The PKMR strategy is quite simple. We take the most sophisticated
wavefunctions from the SNP (standard nuclear physics)/PM
(potential model) approach with the potential fit to an ensemble
of empirical data on two-body systems and compute the EW currents
using the chiral Lagrangian (\ref{lag1}) to as high an order as
possible in the chiral counting. Such realistic potentials --
which have been extensively studied and confirmed empirically --
can then be interpreted as resulting from high-order $\chi$PT
calculations. For two-body systems, we use as the input the most
``sophisticated" potential available in the literature, i.e., the
Argonne $v18$ potential -- and the resulting
wavefunctions~\cite{argonne}. For making predictions in $n$-body
systems with $n>2$, we rely on Ref.\cite{pm}. The power of these
wavefunctions in confronting few-body nuclei is well documented
in the literature. In all cases, the scattering and static
properties of the nuclei involved are extremely accurately given
and more or less justified within $\chi$PT {\it whenever}
many-body corrections are unimportant. Whenever many-body
corrections are important, on the other hand, they can in turn be
calculated by $\chi$PT as we shall show using the chiral filter
argument. In all cases studied up to date, it is possible to make
error estimates by assessing the deviation from the RG
invariance, that is, the $\Lambda$ dependence.

\begin{figure}[hbtp]
\centerline{\protect \epsfig{file=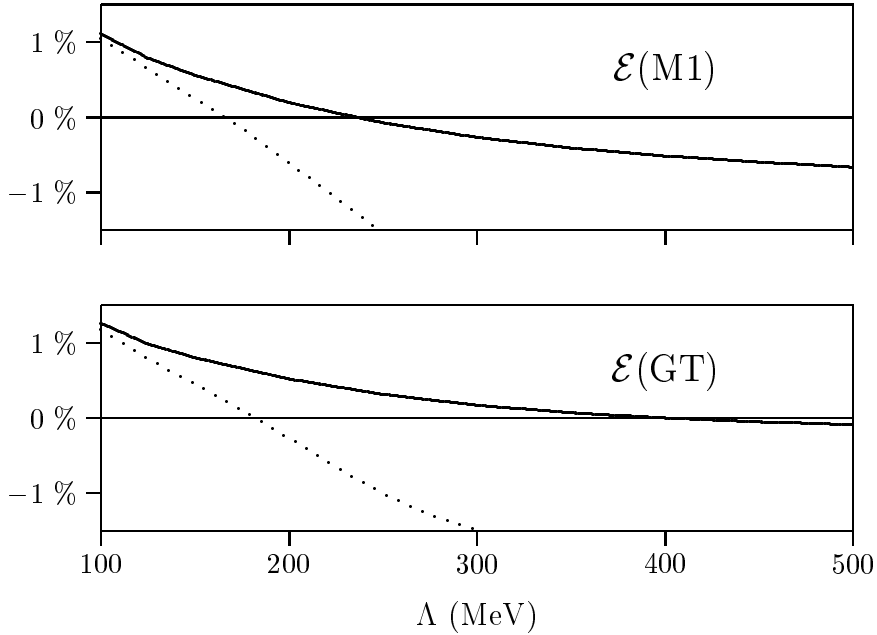}}
\caption{\protect \small ${\cal E}(\mbox{M1})$ (upper) and ${\cal
E}(\mbox{GT})$ (lower) vs. the cutoff $\Lambda$. The solid curves
represent the NLO results with pions and the dotted curves
without pions. When pions are included, the physically relevant
cutoff must lie between the pion mass $m_\pi$ and the lowest
resonance, say, $m_\sigma\sim 450$ MeV. \label{E-MGT}}
\end{figure}

Before calculating the full amplitudes to confront nature, we
first verify that using the ``realistic wavefunction" is {\it
indeed} consistent with actually calculating the wavefunctions
systematically in the chiral counting starting with (\ref{lag1})
or (\ref{lag2}). For this purpose, it suffices to look at the
{\it single-particle} $M1$ matrix elements for the process
 \be
n+p\rightarrow d+\gamma\label{np}
 \ee
and the {\it single-particle} Gamow-Teller matrix element in the
solar proton fusion process
 \be
p+p\rightarrow d+e^+ +\nu_e\label{pp}
 \ee
and verify that there is a matching between a bona-fide EFT and
the hybrid scheme we are using. We are interested~\footnote{These
are of course not the whole story since there are two-body
corrections known as ``meson-exchange currents" that we will
consider shortly.} in ${\cal E}(\mbox{M1})$ and ${\cal
E}(\mbox{GT})$ defined by
 \be {\cal E}(\mbox{M1}) \equiv
\frac{\calM_{\rm M1}^{\rm th}
 - \calM_{\rm M1}^{v18}}{\calM_{\rm M1}^{v18}},
\ \ \ {\cal E}(\mbox{GT}) \equiv \frac{\calM_{\rm GT}^{\rm th}
 - \calM_{\rm GT}^{v18}}{\calM_{\rm GT}^{v18}},
 \ee
where $\calM_{\rm M1}^{\rm th}$ and $\calM_{\rm M1}^{v18}$
denote, respectively, the {\it single-particle} $M1$ transition
matrix element of our NLO calculation considered above for the
$np$ scattering and that of the Argonne potential, and similarly
for ${\cal E}(\mbox{GT})$. Here we are taking the Argonne
potential as the most accurate and closest to ``experiment." In
Fig.\ref{E-MGT} are plotted the deviations from the Argonne
result as a function of the cutoff. We see that as in the case of
the phase shift for $p\sim m_\pi$, the inclusion of the pion
degree of freedom markedly reduces the cutoff dependence, in
conformity with the stated requirement for a viable effective
theory.

Leaving the details to the literature~\cite{PKMR,np}, we merely
summarize a few concrete results, both post-dictions and
predictions, of full calculations to NLO or NNLO including pions.
Some technical details including precise definitions of the
quantities involved can be found in the next subsection
immediately following this where we address the question of
``hard-core correlations" invoked extensively in standard
calculations dealing with short-distance nuclear interactions.
 \bitem
\item {\small\it POST-DICTION}

The most celebrated case of post-diction is the thermal $np$
capture process (\ref{np}). The unpolarized cross section is
dominated by the $M1$ operator -- defined below -- that is
protected by the ``chiral filter" and has been calculated in the
PKMR approach with the Argonne $v18$ to the accuracy of $\sim
1\%$~\cite{np},
 \be
\sigma_{th}=334\pm 3 \ {\rm mb}
 \ee
to be compared with the experimental value $\sigma_{exp}=334.2\pm
0.5 \ {\rm mb}$. The error bar on the theoretical value represents
the range of uncertainty in the short-distance physics unaccounted
for in the NNLO chiral perturbation series that is reflected in
the cutoff dependence. Here the correction to the single-particle
$M1$ matrix element is dominated by the soft-pion contribution
with the remaining corrections amounting to less than 10\% of the
{\it unsuppressed} soft-pion term in conformity with the chiral
filter argument. Since this calculation has {\it no free
parameters}, it is actually a prediction whereas the KSW
calculation without pions has one unknown counter term as the
leading correction term~\cite{ksw-np}~\footnote{In order to
explain the experimental value in this calculation, the counter
term turns out to require to be considerably bigger in magnitude
than, and to have a sign opposite to that of, the soft-pion
correction of \cite{np}. This seems to imply that the counter term
that appears in this pionless theory describes a different physics
than the chiral-filter protected mechanism.}. The upshot of the
$\chi$PT result \`a la PKMR is that it confirms the result of
Riska and Brown~\cite{Rbnp} {\it and in addition} supplies the
corrections to their result with an error estimate, which is the
ultimate power of the approach as far as post-dictions are
concerned. Now having explained accurately the threshold $np$
capture, one can then proceed to make predictions for the inverse
process $d+\gamma\rightarrow n+p$ for which cross section data are
available as well as for the process $e+d\rightarrow n+p+e$ as a
function of momentum transfer. A good agreement with experiments
is obtained~\cite{dgamma} for the photodisintegration of the
deuteron $d+\gamma\rightarrow n+p$. Remarkably the simple theory
seems to work also for the latter to a momentum transfer much
greater than justified by the cutoff
involved~\cite{mathiot-frois}.
\item {\small\it PREDICTIONS}

To bring our point home, we cite a few cases of predictions that
precede experiments. For these predictions, the $\chi$PT
Lagrangian (\ref{lag1}) is used for calculating the irreducible
graphs entering in the electroweak currents.
 \ben
\item While the unpolarized cross section for the $np$ capture
process is dominated by the isovector $M1$ operator that is
protected by the chiral filter mechanism, the process
 \be
\vec{n}+\vec{p}\rightarrow d+\gamma\label{polarized}
 \ee
with polarized neutron and proton can provide information of the
suppressed isoscalar $M1$ and $E2$ matrix elements. The isoscalar
$M1$ matrix element turns out to be particularly interesting
because a precise calculation of this matrix element requires
implementing ``hard-core correlations" of standard nuclear
physics in terms of regularization in field theory. Although the
understanding is still poor, the issue presents a first
indication of how the short-distance prescription encoded in the
hard core may be understood in a more systematic way.

As noted more precisely below, the matrix elements involved are
suppressed by three orders of magnitude with respect to the
isovector $M1$ matrix element -- due to the fact that these are
unprotected by the chiral filter mechanism, so cannot be ``seen"
directly in the cross section but can be singled out by
polarization and anisotropy measurements (see below for
definitions). The ratios with respect to the isovector $M1$ are
predicted to be \be
 \calR_{M1}=(-0.49\pm 0.01)\times 10^{-3}, \ \ \calR_{E2}=(0.24\pm
 0.01)\times 10^{-3}.\label{supp}
\ee There are no data available at the moment; in this sense
these are true predictions. The error bars represent the possible
uncertainty in the cutoff dependence, indicating the uncertainty
in the short-distance physics. We should mention that the
identical results were obtained in the KSW scheme, with however,
somewhat bigger error estimates~\cite{seattle}.
\item The next prediction is on the proton fusion process (\ref{pp}).
Here the dominant matrix element involved, i.e., Gamow-Teller, is
unprotected by the chiral filter but it turns out that the
single-particle matrix element which can be precisely calculated
in the standard nuclear physics/PM approach with accurate wave
two-nucleon wave functions dominates. The remaining correction
terms are not protected by the chiral-filter mechanism and hence
can be subject to some uncertainty although numerically
small~\cite{PKMR-pp}. It turns out however that to
next-to-next-to-leading order (NNLO) it is possible to make an
extremely accurate totally parameter-free estimate of the
correction~\cite{tspetal-pp}. The quantity that carries
information on nuclear dynamics is the astrophysical $S$ factor,
$S_{pp}=\sigma (E) E E^{2\pi\eta}$ where $\sigma$ is the cross
section for the process (\ref{pp}) and $\eta=M\alpha/(2p)$. The
prediction for $S$ is~\cite{tspetal-pp}
 \be
S_{pp} (0)=(3.94\times 10^{-25})(1\pm 0.0015\pm 0.001) \ {\rm MeV\
\ barn}
 \ee
where the first error is due to the uncertainty in the one-body
matrix element and the second due to the uncertainty in the
two-body matrix element that is truncated at the NNLO.
Higher-order corrections are expected to be completely negligible.
This probably is the most accurate prediction anchored on the
``fundamental principle" achieved in nuclear physics.

In the solar neutrino problem, the $S$ factor is an input as a
Standard Model property. In view of the important role that the
$pp$ fusion process will play in the future measurements of the
solar neutrinos, this accurate Standard Model prediction will
prove valuable for the issue.
\item So far we have been concerned with two-nucleon systems that
can be more or less equally well treated by both the
Kaplan-Savage-Wise (KSW) approach and the Park-Kubodera-Min-Rho
(PKMR) approach. We now turn to a case where the PKMR can make a
possible prediction~\cite{hep,tspetal-hep} whereas a strict
adherence to the counting rule would frustrate the feat due to too
many unknown parameters. This is the solar ``hep" problem
 \be
p+^3{\rm He}\rightarrow ^4{\rm He} + e^+ +\nu_e.
 \ee
This process produces neutrinos of the maximum energy
$E_\nu\approx 19.8$ MeV; for this lepton momentum, not only the
$S$ wave but also the $P$ wave need to be taken into account.
Computing the neutrino flux as needed for the solar neutrino
problem requires both vector and axial-vector matrix elements.
While the vector matrix element involves no subtleties -- and
hence is straightforwardly calculated (including two-body
currents), the principal matrix element of the axial-vector
current turns out to be extremely intricate. Since the $P$ wave
enters non-negligibly, the time component of the axial current
cannot be ignored. However this part of the current is under
theoretical control since the axial charge operator is
chiral-filter protected~\cite{KDR,MR91}: the corrections to the
known single-particle axial-charge matrix element are dominated by
one-soft-pion-exchange two-body matrix elements (with three-body
operators absent to the order considered) and are accurately
calculable, given the ``accurate" wave functions available in the
literature. The principal difficulty is in the matrix element of
the space component of the axial current, the most important of
which is the Gamow-Teller operator. (Three-body contributions are
higher-order than those coming from one-body and two-body terms
and can be ignored). Normally the matrix element of the
single-particle Gamow-Teller operator $g_A\sum_i
\tau^{\pm}(i)\sigma^a (i)$ is of order 1 if the symmetries of the
initial and final nuclear states are normal since the operator
just flips spin and isospin. However in the ``hep" case, the
dominant components of the initial state $|p^3$He$\ra$ and the
final state $|^4$He$\ra$ have different spatial symmetries so the
overlap is largely suppressed. One finds, numerically, that to the
chiral order that is free of unknown parameters (e.g.,
next-to-next-to leading order, NNLO, or ${\cal O} (Q^3)$), the
single-particle GT matrix element comes out to be suppressed by a
factor of $\sim$3 relative to the normal matrix element. Now main
corrections to the one-body Gamow-Teller are expected to be from
two-body terms with three-body terms down by a chiral order.
However these corrections are chiral-filter-unprotected as we have
explained and hence cannot in general be calculated accurately in
chiral perturbation theory unlike the corrections to the
single-particle axial charge matrix elements. Furthermore, the
two-body corrections typically come with a sign opposite to the
sing-particle one with a size comparable to the main term, thus
causing a drastic further suppression. This shows why it is {\it
essential} to be able to calculate in a controlled fashion higher
chiral order corrections to the GT matrix element.

Such a calculation turns out to be feasible given accurate wave
functions for three- and four-nucleon systems~\cite{rocco}. The
key point is that the calculation of this
chiral-filter-unprotected term parallels closely the calculation
of the suppressed isoscalar $M1$ and $E2$ matrix elements in the
polarized $np$ capture mentioned above, thereby enabling the PKMR
approach to pin down the most problematic part of the ``hep"
process within a narrow range of uncertainty dictated by the range
of the cutoff involved~\cite{hep}. Referring for details to the
literature~\cite{hep,tspetal-hep}, we simply summarize what enters
in the calculation. The first is that as in the case of the $pp$
case, there are no unknown parameters once the accurate data on
triton beta decay are used. The second is that the strategy for
controlling the chiral-filter-unprotected two-body corrections to
the Gamow-Teller matrix element parallels that for the $pp$ fusion
process, i.e., the renormalization-cutoff interpretation of what
is known in nuclear physics as ``hard-core correlation" discussed
below.

The $S$ factor so obtained~\cite{tspetal-hep} is
 \be
S_{hep} (0)=(8.6\times 10^{-20})(1\pm 0.15) \ {\rm MeV\ \
barn}.\label{heptsp}
 \ee
The error bar stands for uncertainty due to the cutoff dependence
which signals the uncertainty inherent in terminating the chiral
series at NNLO. The uncertainty is bigger here than in the $pp$
case since one is dealing with four-nucleon systems. Higher-order
corrections are expected to considerably reduce the error bar.

A similar calculation in a standard nuclear physics approach based
on the traditional treatment of exchange currents~\cite{chemrho}
has been performed by Marcucci et al~\cite{rocco}. This approach
uses the same accurate wave functions as in \cite{tspetal-hep},
without however resorting to a systematic chiral counting. That
the standard nuclear physics approach of \cite{rocco} agrees with
(\ref{heptsp}) within the error bar suggests that the standard
approach of \cite{rocco} may well be consistent -- to the order
considered -- with the EFT approach to the ``short-range
correlation."

 \end{enumerate}
\eitem
 \subsubsection{\it The chiral filter: Two sides of the same coin}
 \itt
While the processes protected by the chiral filter, such as the
$M1$ operator and the axial charge operator~\cite{KDR}, are
accurately calculable since one-soft-pion exchange dominates with
correction terms suppressed typically by an order of magnitude
relative to the dominant soft-pion exchange, the situation is not
the same for those processes that have no chiral-filter
protection. Even so, the PKMR approach can still make a meaningful
statement on such processes. That is the story of the other side
of the same coin.

If the chiral filter does not apply, then the role of the pion is
considerably diminished and, compounded with higher order
effects, clean and reliable low-order calculations will no longer
be feasible. A prime example of this situation is the process
$p+p\rightarrow p+p+\pi^0$. Here one-soft-pion exchange is
suppressed by symmetry and a variety of correction terms of
comparable importance compete in such a way that no simple
$\chi$PT description can be obtained. Both the PKMR and the KSW
are unsuccessful for this process.

The proton fusion process (\ref{pp}) presents a different
situation at least in the PKMR approach~\cite{PKMR-pp}. Since the
process is essentially dominated by the Gamow-Teller operator
with no accidental suppression, the single particle matrix
element in the standard nuclear physics (SNP)/potential model (PM)
approach simply dominates. On the other hand, since one is
dealing with low-energy (or momentum) probes with the relevant
scale much less than the pion mass, one may simply integrate out
the pion and work with a pionless effective Lagrangian as in
\cite{butler}. However the price to pay here is to face
higher-order counter terms. We learn from the chiral filter that
these corrections cannot be computed in a low-order expansion.
This is because those terms reflect the physics of a distance
scale shorter than that that can be captured in a few perturbative
expansion. While in the PKMR approach, the counter terms of high
order that are required are numerically small compared with the
leading single-particle term given in the SNP/PM approach, with
no help of the chiral filter, there is nothing that says that
subleading corrections are necessarily small compared with the
leading-order correction that one may be able to calculate. For
example in high-order KSW calculations in the ``pionless EFT" for
the proton fusion (see \cite{butler}), the corrections occur at
fifth order. Now even if the counter terms that appear at that
order are -- eventually -- fixed by the inverse neutrino
processes that may be measurable in the laboratories, the error
incurred in this fifth order calculation will not be any smaller
than what the PKMR approach can obtain at a lower order.
 \subsubsection{\it An interpretation of hard-core correlation in an
effective field theory}
 \itt
The isoscalar matrix elements of the polarized neutron and proton
capture process (\ref{polarized}),
 \be
\vec{n}+\vec{p}\rightarrow d+\gamma\label{polnp}
 \ee
are {\it not} chiral-filter-protected. Nonetheless it was possible
to make the predictions (\ref{supp}). How this comes about is
interesting from the point of view of learning something about
short-range correlations in nuclei. We elaborate on this point in
the rest of this section.

At threshold, the initial nuclear state in (\ref{polnp}) is in
either the ${}^1S_0 (T=1)$ or ${}^3S_1(T=0)$ channel, where $T$
is the isospin. The process therefore receives contributions from
the isovector M1 matrix element (M1V) between the initial
${}^1S_0$ ($T=1$) and the final deuteron ($T=0$) state, along
with the isoscalar M1 matrix element (\mut) and the isoscalar E2
(\Qt) matrix element between the initial ${}^3S_1$ ($T=0$) and
the final deuteron ${}^3S_1\!-\!{}^3D_1$ states. Since M1V is by
far the largest amplitude, the spin-averaged cross section
$\sigma_{unpol}(np\rightarrow d\gamma)$ is totally dominated by
M1V. Meanwhile, since the initial ${}^1S_0$ state has $J=0$, the
M1V cannot yield spin-dependent effects, whereas \mut{} and \Qt{}
can.

The $T$ matrix for the process can be written as \bea \langle
\psi_d, \gamma(\hatk, \lambda) | {\cal T}|
    \psi_{np}\rangle
 = \chi^\dagger_d\, {\cal M}(\hatk, \lambda)\, \chi_{np}
\eea with \bea {\cal M}(\hatk, \lambda) =
 \sqrt{4\pi} \frac{\sqrt{v_n}}{2 \sqrt{\omega} A_s}
 \, \left[
  i (\hatk \times {\hat \epsilon}_\lambda^*)\cdot (\sigmab_1-\sigmab_2)\, \MS
\right.
\nonumber \\
 \left.
  - i (\hatk \times {\hat \epsilon}_\lambda^*)\cdot
  (\sigmab_1+\sigmab_2)\,\frac{\mut}{\sqrt{2}}
  + (\sigmab_1\cdot\hatk \sigmab_2\cdot {\hat \epsilon}_\lambda^*
   + \sigmab_2\cdot\hatk \sigmab_1\cdot {\hat \epsilon}_\lambda^*)
    \frac{\Qt}{\sqrt{2}}
\right]\label{ampp} \eea where $v_n$ is the velocity of the
projectile neutron, $A_s$ is the deuteron normalization factor
$A_s\simeq 0.8850\ {\rm fm}^{-1/2}$, and $\chi_d$ ($\chi_{np}$)
denotes the spin wave function of the final deuteron (initial
$np$) state. The emitted photon is characterized by the unit
momentum vector $\hatk$, the energy $\omega$ and the helicity
$\lambda$, and its polarization vector is denoted by ${\hat
\epsilon}_\lambda \equiv {\hat \epsilon}_\lambda({\hat k})$. The
amplitudes, $\MS$, $\mut$ and $\Qt$, represent the isovector M1,
isoscalar M1 and isoscalar E2 contributions, respectively, all of
which are real at threshold. These quantities are defined in such
a manner that they all have the dimension of length, and the
cross section for the unpolarized $np$ system takes the form
 \bea
\sigma_{unpol}= \abs{\MS}^2 + \abs{\mut}^2 + \abs{\Qt}^2\,.
\label{xsection} \eea
 The isoscalar terms
($\abs{\mut}^2$ and $\abs{\Qt}^2$) are strongly suppressed
relative to $\abs{\MS}^2$
--- approximately by a factor of $\sim O(10^{-6})$ ---
so the unpolarized cross section is practically unaffected by the
isoscalar terms. As mentioned above, the isovector M1 amplitude
was calculated~\cite{np} very accurately up to ${\cal O} (Q^3)$
relative to the single-particle operator. The result expressed in
terms of $\MS$ is: $\MS{}=5.78\pm 0.03$ fm, which should be
compared to the empirical value $\sqrt{\sigma_{unpol}^{exp}}=
5.781\pm 0.004$ fm. Here, we will focus on the isoscalar
amplitudes.

Now the isoscalar matrix elements can be isolated by
spin-dependent observables, namely by the photon circular
polarization $P_\gamma \equiv \frac{I_{+1}(0^\circ) -
I_{-1}(0^\circ)} {I_{+1}(0^\circ) + I_{-1}(0^\circ)}$ and the
anisotrophy $\eta_\gamma \equiv \frac{I(90^\circ) - I(0^\circ)}
{I(90^\circ) + I(0^\circ)}$ where $I_\lambda(\theta)$ is the
angular distribution of photons with helicity $\lambda=\pm 1$,
with $\theta$ the angle between $\hatk$ (direction of photon
emission) and a quantization axis of nucleon
polarization~\footnote{Explicit forms are not needed for our
discussion but we write them down for completeness:
 \be
P_\gamma &=& |\vP_n|\, \frac{\sqrt{2} (\rM - \rE) +
\frac12(\rM+\rE)^2}{1 + \rM^2 + \rE^2},\nonumber\\
\eta_\gamma &=& pP\, \frac{\rM^2 + \rE^2 - 6 \rM \rE}
     {4(1 - pP) + (4 + pP) (\rM^2 + \rE^2) + 2 pP\, \rM
     \rE}\nonumber
\ee where $pP\equiv \vP_p\!\cdot\!\vP_n$ and $\vec{P}$ is the
polarization three-vector.}. Measurement of $P_\gamma$ and
$\eta_\gamma$ can determine the empirical values of the ratios
 \bea
{\cal R}_{\rm M1} \equiv \frac{\mut}{\MS}\;, \ \ \ {\cal R}_{\rm
E2} \equiv \frac{\Qt}{\MS}\;.\label{ratios}
 \eea

Both $\mut$ and $\Qt$ are suppressed relative to the isovector M1
matrix element (M1V) due to mismatch in symmetries of the wave
functions etc. We will however take the single-particle matrix
elements of $\mut$ and $\Qt$ as the leading order (LO) in chiral
counting of the higher order terms in calculating the ratios
$\rM$ and $\rE$. Radiative corrections to these leading-order
terms come at \nlo{1} and are calculated easily given the wave
functions~\footnote{Although not in the main line of discussion,
we quote as an aside the numbers to give an idea of the size we
are dealing with. With the realistic A$v$18 wavefunctions, the
numerical results for the sum of the LO and NLO come out to be
\be \mut_{\rm 1B}(\fm) &=& (-4.192 - 0.105) \times 10^{-3}
   = -4.297\times 10^{-3},
   \nonumber \\
\Qt_{\rm 1B}(\fm) &=& (1.401 - 0.007) \times 10^{-3}
     = 1.394 \times 10^{-3}.\nonumber
\ee}. Relative to the leading order (LO) terms, corrections will
then be classified by N$^n$NO for ${\cal O} (Q^n)$. For the
chiral-filter protected M1V, the leading two-body correction
appears at \nlo{1} and the one-loops and the counter-terms appear
at \nlo{3}. For the isoscalar current, not protected by the
chiral filter, the counting rules are quite different. While the
E2S receives only negligible contributions from higher order
(\nlo{3} and \nlo{4}) corrections, the situation with M1S is
quite different. The corrections are quite large even though the
leading two-body corrections for the M1S turn out to appear in
tree order at \nlo{3} while the loop corrections come at \nlo{4}
and are finite.

\begin{figure}[htbp]
\centerline{\psfig{file=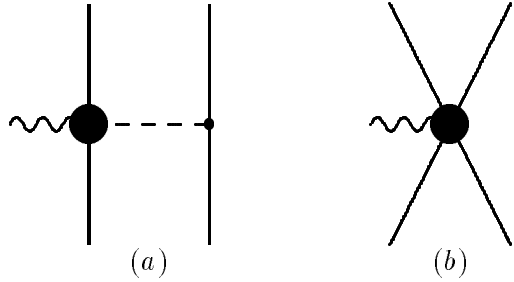}} \caption[gene]{\protect
\small Generalized tree diagrams for the two-body isoscalar
current. The solid circles include counter-term insertions and
(one-particle irreducible) loop corrections. The wiggly line
stands for the external field (current) and the dashed line the
pion. }\label{PKMR-fig1}
\end{figure}

\begin{figure}[htbp]
\centerline{\psfig{file=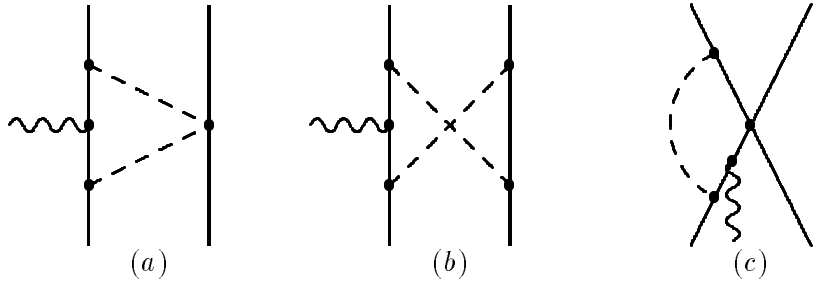}} \caption[deviate]{\protect
\small One-loop graphs contributing to the two-body currents.
They come at $\calO (Q^4)$ and higher orders relative to the LO
one-body term. All possible insertions on the external line are
understood. \label{PKMR-fig3} }
\end{figure}

Given that the higher corrections {\it separately} are quite large
as expected from the chiral-filter argument, there is a problem
which does not arise in chiral-filter-protected processes. It has
to do with the zero-range interactions that come as radiative
corrections involving the four-Fermi contact interactions
(proportional to $C_A$) in Eq.(\ref{lag1}) (see
Fig.\ref{PKMR-fig3}c) and as four-Fermi counter terms (see
Fig.\ref{PKMR-fig1}b). It turns out that there is only one term of
the latter type that contributes, i.e., a term in the Lagrangian
of the form
 \be
\Delta {\cal L} \sim - i g_4 (\del_\mu {\cal B}_\nu - \del_\nu
{\cal B}_\mu)
 {\bar N} [S^\mu,\, S^\nu] N\, {\bar N} N
\ee where ${\cal B}_\mu$ is the isoscalar external field and $g_4$
is an undetermined coefficient. When the chiral-filter mechanism
is operative, these zero-range interactions appear at higher order
and are expected to be suppressed by ``naturalness" conditions and
more importantly by ``hard-core" correlation functions that are
incorporated in the wave functions. In the present case, the
chiral filter is not effective and hence it is not a good
approximation to let hard-core correlation functions ``kill" the
contact interactions. One way to resolve this problem is to
exploit the cutoff regularization that was used above in Section
\ref{nnscattering} for NN scattering. Operationally, we can
achieve this goal by replacing the delta function in coordinate
space attached to zero-range terms by the delta-shell with hard
core $r_c$,
 \be \delta^3(\vr) \rightarrow
\frac{\delta(r-r_c)}{4\pi r_c^2}. \label{deltashell}
 \ee
We then have to assure that the {\it total contribution} we
compute be independent of $r_c$ within a reasonable range. This
requirement constitutes a sort of renormalization-group
invariance. When the chiral filter mechanism is operative,
dropping the zero-range terms is justified since the $r_c$
independence is not spoiled. If this requirement is not satisfied,
the effective theory is not reliable and the result cannot be
trusted. The present case belongs to this class.

It turns out fortunately that for M1S, there enters only one
particular linear combination of four-Fermi terms involving $g_4$
and $C_A$'s and that it is this combination -- call it $X$ -- that
figures in the magnetic moment of the deuteron. Fitting to the
deuteron magnetic moment allows the constant $X$ to be uniquely
determined for any given value of $r_c$. The result for the total
M1S for wide-ranging values of $r_c$ is given for illustration in
Table \ref{m1s-vs-rc}.
\begin{table}[ht]
\begin{center}
\begin{tabular}{|r||r|r|r|r|r|}
\hline $r_c$ (fm) & 0.01 & 0.2 & 0.4 & 0.6 & 0.8
\\ \hline
$\mut$ &
$-2.849$ & $-2.850$ & $-2.852$ & $-2.856$ & $-2.861$ \\
\hline
\end{tabular}
\caption{\label{m1s-vs-rc}\protect \small The total contributions
to \mut\ in unit of $10^{-3}\ \fm$ vs. $r_c$ in fm. Recall the
single-particle contribution: ${\rm M1S}_{1B}\approx -4.30\times
10^{-3}$\ fm.}
\end{center}
\end{table}
The remarkable insensitivity to $r_c$ may be taken as a sign of
RG invariance and a support for the procedure. This is more
remarkable considering that the series in the chiral order does
not appear to be converging. To exhibit this, we quote how each
term of \nlo{n} for $n\geq 3$ contributes to the ratio $\rM$ in
the range of $r_c$
 \be r_c^{min}\equiv
0.01\ {\rm fm} \le r_c \le r_c^{max}\equiv 0.8\ {\rm fm}.
 \ee
Expressed in the chiral order and in the range
$(r_c^{max},r_c^{min})$,  the $\rM$ comes out to be
 \be
\calR_{M1}\times 10^3 &=&``{\rm LO}"  + \ \ \ \ ``{\rm N}^3{\rm
LO}" \ \ + \ \  ``{\rm N}^4{\rm LO}" \ \  + \ ``({\rm N}^3{\rm LO}
+{\rm N}^4{\rm LO})_X"\nonumber\\
 &=& -0.74+(-0.48,-0.74)+(0.23,0.46)+(0.49,0.54)\nonumber\\
 &=& (-0.50,-0.49)
 \ee
where the contribution that depends on the single parameter $X$
fixed  by the deuteron magnetic moment is indicated by the
subscript $X$. This is how the value given in (\ref{supp}) has
been arrived at. The crucial observation which applies directly
also to the {\it hep} process -- and to a somewhat less extent, to
the $hep$ process -- discussed above is that the final result
cannot be arrived at by any partial sum of the terms. Also note
the importance of the $X$-dependent term. Needless to say, an
experimental check of the result obtained here would be highly
desirable.

It should be noted that the scale we are talking about in
connection with the ``hard-core regularization" is of the chiral
scale $\Lambda_\chi\sim 4\pi f_\pi$ that delineates low-energy
degrees of freedom of hadrons from high-energy, and {\it not} that
of nuclei that enters in low-energy two-body interactions
discussed above.
\section{``VECTOR MANIFESTATION" OF CHIRAL\\ SYMMETRY}
 \label{how}
 \setcounter{equation}{0} 
\renewcommand{\theequation}{\mbox{\ref{how}.\arabic{equation}}}
 \subsection{Harada-Yamawaki Scenario}
\itt The major problem that we have to face in going to many-body
systems is how the structure of the ground state, i.e., ``vacuum"
is to be described and how the properties of hadronic degrees of
freedom get modified in the changed background. This question will
be addressed more specifically in the next section in terms of an
effective in-medium field theory. Here we address the generic
aspects based on symmetries involved in the problem. We will
discuss the problem in terms of density but we believe the general
argument applies as well to temperature at least up to the chiral
restoration point. Possible differences that appear after the
critical point will be mentioned in Section \ref{denseqcd}.

To discuss how the vacuum changes as matter density increases and
how the properties of baryonic matter are affected by the vacuum
change, we need to first address how chiral symmetry is affected
by the medium. While it is generally accepted that chiral symmetry
endowed with $N_F$-flavor massless quarks breaks spontaneously in
the (zero-density) vacuum as $ SU(N_F)_L\times
SU(N_F)_R\rightarrow SU(N_F)_{L+R}$, it is not fully known how the
broken symmetry gets restored in medium under extreme conditions,
that is, at high temperature and/or at high density. There are
several different scenarios with which the broken symmetry can get
restored as temperature and/or density are/is dialed. As we will
argue, what happens in nuclei and nuclear matter where matter
density is non-zero depends rather crucially on what scenario the
chiral symmetry restoration adopts.

For concreteness and simplicity, take the two-flavor case with up
and down quarks ($N_F=2$)~\footnote{The Harada-Yamawaki argument
was originally developed with three flavors u, d and s. Although
it has not been verified explicitly, we see no reason why the same
argument should not apply to the two-flavor case with some minor
changes.}. The ``standard scenario" is the $\sigma$ model one
where the broken Nambu-Goldstone mode gets restored to the
Wigner-Weyl mode wherein the triplet of pions $\pi^i$ with $i=\pm,
0$ and the scalar $\sigma$ (in the chiral limit) merge to an
$O(4)$ massless degenerate multiplet at the phase transition. At
the phase transition, other hadrons such as nucleons, vector
mesons etc. could be either massless or massive as long as they
appear in chiral multiplet structure. Now in zero density, the
scalar $\sigma$ need not be the fourth component of $O(4)$ as in
linear $\sigma$ model. It could for instance be a chiral singlet
but what is required is that at the phase transition it joins the
pions into the $O(4)$ as described by Weinberg in his ``mended
symmetry" scenario~\cite{weinberg-salam,beane}. One should note
that the $\sigma$-model scenario is moot on the spin-1 mesons, so,
e.g., the vector meson $\rho$ (which will figure importantly in
later discussions) could become either lighter or heavier as one
approaches the chiral restoration. This model belongs to the
$O(4)$ universality class and has been extensively discussed in
the literature~\cite{pisarski}. At present, it is not inconsistent
with lattice measurements with temperature but it is not the only
viable scheme either.

There is an alternative scenario to the above standard one
recently proposed by Harada and Yamawaki~\cite{VM} which we
believe provides support to the BR scaling to be elaborated below
as well as connects with the color-flavor locking scenario of QCD
discussed in Section \ref{complementarity}. The distinguishing
feature of Harada-Yamawaki's ``vector manifestation" is that
chiral symmetry with, say, two flavors (u and d) is restored by
the triplet of pions merging with a triplet of the longitudinal
components of vector mesons to the representation $(3,1)\oplus
(1,3)$. At the chiral point, the vector coupling $g_V$ flows to
the fixed point $g_V=0, a=1$ corresponding to what Georgi called
vector symmetry limit~\cite{georgi} {\it together with} the pion
decay constant $f_\pi$ going to zero. In this case, the vector
$\rho$ will become massless together with other matter fields,
i.e., constituent quarks and the vector coupling will vanish, a
feature which is an important factor in our discussions to follow.

Limiting ourselves to the two-flavor case (the three-flavor case
can be similarly treated), we consider specifically the hidden
local symmetry (HLS) theory of Bando et al~\cite{bandoetal} with
the symmetry group $[U(2)_L\times U(2)_R]_{global}\times
[U(2)_V]_{local}$ consisting of a triplet of pions, a triplet of
$\rho$-mesons and an $\omega$-meson. Motivated by the observation
that in the vacuum, the $\rho$ and $\omega$ mesons are nearly
degenerate and the quartet symmetry is fairly good
phenomenologically, we put them into a $U(2)$ multiplet. In this
theory, baryons (proton and neutron) do not appear explicitly.
They can be considered as having been integrated out. If needed,
they can be recovered as solitons (skyrmions) of the theory.

The relevant degrees of freedom in the HLS theory are the left and
right chiral fields denoted by $\xi_{L,R}$,
 \ba
\xi_{L,R}=e^{i\sigma/f} e^{\mp i\pi/f}
 \ea
with $\sigma (x)=\sigma^a T^a$ and $\pi=\pi^a T^a$ and the hidden
local gauge fields denoted by
 \ba
V_\mu\equiv V_\mu^\alpha T^\alpha=\frac{\tau^a}{2}\rho_\mu^a
+\frac 12 \omega_\mu
 \ea with $\Tr(T^\alpha T^\beta)=\frac
12\delta^{\alpha\beta}$. If we denote the $[U(2)_L\times
U(2)_R]_{global}\times [U(2)_V]_{local}$ unitary transformations
by $(g_L,g_R,h)$, then the fields transform $\xi_{L,R}\mapsto
h(x)\xi_{L,R}g^\dagger_{\small{L,R}}$ and $V_\mu\mapsto
h(x)(V_\mu-i\del_\mu)h^\dagger (x)$. How chiral symmetry
restoration can come about in the HLS framework at large $N_F$ was
discussed by Harada and Yamawaki~\cite{HY:letter} and at high
temperature by Harada and Sasaki~\cite{harada-sasaki}. Since their
arguments are quite general, it is highly plausible that they can
be applied to high density. This is what we shall do here.
\subsection{Vector Manifestation in Hot Matter}
\itt To motivate our argument to be developed below for
density-driven phenomena, we briefly summarize, without going into
specific details which we will do in the next subsection, the work
of Harada and Sasaki~\cite{harada-sasaki} on how the VM manifests
itself as the critical temperature $T_c$ is approached. We do this
because this work brings out a subtlety in the behavior of
vector-meson excitations in medium in general and in the
neighborhood of the critical temperature in particular that has
not been duly taken into account in other works on the matter
available in the literature.

There are two distinctive features in the HLS approach to the
chiral phase transition. One is that in order to match with QCD
{\it while preserving hidden local symmetry} at the chiral scale
$\Lambda_\chi$ in the $T=n=0$ space~\footnote{In this section, the
matter density will be denoted $n$, reserving $\rho$ for the
$\rho$ meson.}, the $\rho$ and $\omega$ mesons must be considered
as ``light" in the sense that the pion mass is light as first
pointed out by Georgi for a consistent power
counting~\cite{georgi}. As noted below, it is in this sense that
the HLS theory can be mapped to chiral perturbation
theory~\cite{HY:matching}. This feature figures importantly in the
hot (and/or dense) matter. The second feature is that in hot
medium, there are two effects that control hadronic properties.
One is the ``intrinsic" dependence on $T$ through the
$T$-dependence of the QCD condensates (both quark and gluon) of
hadronic masses and coupling constants and the other is the
hadronic thermal fluctuations. The former is controlled by the
matching to QCD and hence so is the latter through the parameters.

In \cite{fate}, Harada and Yamawaki studied the renormalization
group structure of HLS in terms of the parameters of the HLS
Lagrangian that enter at higher orders in the chiral expansion and
showed that effective low-energy theories with the given flavor
symmetries can flow in various different directions as scale is
varied but there is only one fixed point that is consistent with
QCD, namely the Georgi vector limit. This implies that the
intrinsic temperature dependence must be constrained by the Georgi
vector limit, which means that the mass parameter $M_V$ and the
gauge coupling constant $g_V$ will approach zero as $T\rightarrow
T_c$. When $T$ is large, i.e., $M_V\ll T$,~\footnote{This
condition makes sense since $M_V\rightarrow 0$ as $T\rightarrow
T_c$.} the pole mass of the vector ($\rho$) meson mass takes the
simple form
 \be
 m_\rho^2 (T)=M_\rho^2 (T) +g_V^2 J(T).
\ee Here $M_\rho^2=af_\pi^2 g_V^2$ where $g_V$ is the (hidden)
gauge coupling and $a\neq 1$ signals deviation from the Georgi
vector limit and $J(T)$ is a finite temperature-dependent
function. As $T\rightarrow T_c$, $M_\rho\rightarrow 0$ and
$g_V\rightarrow 0$, so the pole mass goes to zero. This is
consistent with BR scaling.

We should stress that while the mass goes to zero at near $T_c$,
the theory is nonetheless consistent with what one would expect at
low $T$. For $T\ll M_\rho$, the pole mass is found to be of the
form~\cite{harada-sasaki}
 \be
m_\rho^2 (T)=M_\rho^2 (T) + c\frac {T^4}{F_\pi^2}+\cdots
 \ee
where $c$ is a numerical constant and $F_\pi\simeq 93$ MeV is the
pion decay constant in the vacuum. The ellipsis stands for terms
higher order in $T$. Note that while this agrees with the
low-temperature result of \cite{deyetal}, the pole mass need not
increase with temperature as suggested in \cite{deyetal}. A close
look indicates that it can even decrease at about the same rate as
$f_\pi (T)$ since $M_\rho$ decreases {\it faster} than $f_\pi
(T)$, suggesting the following -- BR scaling -- relation:
 \be
m_\rho (T)/m_\rho \sim f_\pi (T)/f_\pi.
 \ee
Note also that an early QCD sum-rule calculation by Adami and
Brown~\cite{adami-brown2} of the thermal behavior of the $\rho$
mass contains features that are similar to those in the vector
manifestation. Although the connection to the Georgi vector limit
is yet to be established, the vanishing of the $\rho$ mass in the
work of \cite{adami-brown2} at the critical temperature $T_c$ due
to the approach to zero of the Wilson coefficient that multiplies
a certain combination of gluon condensates with the vanishing of
quark condensate resembles the way the $\rho$ mass goes to zero in
the vector manifestation. In the Adami-Brown formulation, the
tadpoles which give the quark and antiquark masses in NJL are
jammed together in going to deep Euclidean, so that to begin with,
it appears that there is $\langle \bar{q}\bar{q}qq\rangle$
condensate. Adami and Brown factorized this, since they believed
that before deep Euclidean was reached there were really two
$\langle \bar{q}q\rangle$ condensates. Whereas the question of
``factorization" remained in the QCD sum rule approach, the
present work of Harada and Sasaki shows that this was exactly
right; namely, the scale $\Lambda$ changes smoothly as chiral
restoration is reached, with dynamically generated masses going to
zero with it.

\subsection{Vector Manifestation in Dense Matter}
\itt
 In addressing dense matter, we follow the reasoning of
\cite{HY:letter,harada-sasaki,HY:matching}. Here we supply some
details left out in the preceding subsection. We take the HLS
Lagrangian as an effective Lagrangian that results when
high-energy degrees of freedom above the chiral scale $\Lambda$
taken to be higher than the vector meson mass $m_V$ are integrated
out. Now the scale $\Lambda$ will in general depend on the number
of flavors $N_F$, density $n$ or temperature $T$ depending upon
what system is being considered. This is to be taken as a {\it
bare} Lagrangian in the Wilsonian sense with the parameters $g_V
(\Lambda)$, $a(\Lambda)$ and $f_\pi (\Lambda)$ which plays the
role of order parameter for chiral symmetry with $f_\pi=0$
signaling the onset of the Wigner-Weyl phase. Reference
\cite{HY:matching} describes how these parameters can be
determined in terms of QCD condensates by matching -- \`a la
Wilson -- the vector and axial-vector correlators of HLS theory
with the ones of QCD at the chiral scale $\Lambda$ and how by
following renormalization group (RG) flows to low energy scales,
one can determine low-energy parameters that can be related to
those that figure in chiral perturbation theory (for, e.g.,
$\pi\pi$ scattering). An important observation here is that the
(assumed) equality at chiral restoration (where $\la
\bar{q}q\ra=0$) of the vector and axial-vector correlators, i.e.,
 \ba
\Pi_V|_{\la \bar{q}q\ra=0}=\Pi_A|_{\la \bar{q}q\ra=0}
 \ea
where
 \ba
i\int d^4xe^{iq\cdot x} \la 0|T J_\mu^a (x) J_\nu^b (0)|0\ra &=&
\delta^{ab} (q_\mu q_\nu -g_{\mu\nu} q^2)\Pi_V (-q^2),\nonumber\\
i\int d^4xe^{iq\cdot x} \la 0|T J_{5\mu}^a (x) J_{5\nu}^b (0)|0\ra
&=& \delta^{ab} (q_\mu q_\nu -g_{\mu\nu} q^2)\Pi_A (-q^2)
 \ea
implies that at the critical point, {\it independently of how the
symmetry change is driven}, the HLS theory approaches the Georgi
vector limit, namely, $g_V=0$ and $a=1$, in addition to the
vanishing of $f_\pi$. That chiral symmetry is restored with
$f_\pi=0$ at the Georgi vector limit follows from a proper account
of quadratic divergence in the renormalization group flow equation
for $f_\pi$ and $a$.~\footnote{The quadratic divergence is present
when the cutoff regularization is used whereas it is absent at
four dimensions $D=4$ when the dimensional regularization is used.
The quadratic divergence present in the cutoff regularization
corresponds to a singularity at $D=2$ in the dimensionally
regularized integral from one-loop graphs which comes out to be
proportional to the $\Gamma$ function
 \ba
\sim \Gamma (1-D/2)=\frac{\Gamma (3-D/2)}{1-D/2} -\Gamma
(2-D/3).\nonumber
 \ea
In a renormalizable theory, this singularity (or any other power
divergence) does not figure at $D=4$ but in an effective
(nonrenormalizable) theory like HLS, this singularity present at
$D=2$ -- which is absent at $D=4$ -- has to be subtracted. This
procedure is quite similar to the subtraction of the singularity
at $D=3$ that was required in the KSW scheme for two-nucleon
scattering discussed in Section \ref{nnscattering} and we suspect
that a similar physics is at work here. Note that were it not for
this quadratic divergence, the $f_\pi$ would not run and hence
would not go to zero at the point where the Georgi vector limit is
reached. Thus the quadratic term plays a crucial role in
specifying the symmetry structure of the chiral restoration phase
transition.} In terms of baryon density $n$, we interpret this to
imply that at the critical density $n=n_c$, we must have
 \ba
g_V(\Lambda(n_c); n_c)=0, \ \ \ a(\Lambda(n_c); n_c)=1
 \ea
where we have indicated the density dependence of the cutoff
$\Lambda$. The reasoning used here is identical to that for the
temperature-driven transition.

In HLS theory, the vector mass is given by the Higgs mechanism. In
free space, it is of the form
 \be
m_V\equiv m_\rho=m_\omega =\sqrt{a(m_V)} g_V (m_V) f_\pi
(m_V)\label{ksrf}
 \ee
where the cutoff dependence is understood. Here the parameter
$a(m_V)$ etc means that it is the value at the scale $m_V$
determined by an RG flow from the {\it bare} quantity
$a(\Lambda)$.
Note that (\ref{ksrf}) is similar, but not identical, to the KSRF
relation $m_{\rho {\tiny (KSRF)}}=\sqrt{2}g_{\rho\pi\pi} f_\pi
(0)$~\footnote{However numerically they are very close. We explain
as a side remark how this comes about. While Eq.(\ref{ksrf}) for
$m_V=M_V(Q=m_V)$ is approximate with numerically small corrections
of order $g_V^2/(4\pi)^2$ ignored there, the corresponding formula
for $Q=0$, i.e., $M^2_V (0)=a(0) g^2(0) f^2_\pi (0)$ is an exact
low-energy theorem~\cite{HKY}. The small corrections arise in
going from the off-shell point $Q=0$ to the on-shell point
$Q=m_\rho$. Noting that the KSRF relation is given in terms of the
$\rho$ on-shell constant $g_{\rho\pi\pi}$ and the pion on-shell
constant $f_\pi (0)$, we first extract the former from HLS
Lagrangian given in terms of on-shell quantities:
 \ba
g_{\rho\pi\pi}=\frac{g(m_\rho)}{2}\frac{f^2_\sigma
(m_\rho)}{f^2_\pi (0)}\nonumber
 \ea
where $f_\sigma$ is the decay constant of the scalar that goes
into the longitudinal component of the $\rho$ by Higgs mechanism
(which becomes equal to $f_\pi$ at the Georgi vector limit). The
pion decay constant $f_\pi$ runs with pion loops below the $\rho$
scale but $f_\sigma$ does not run below $m_\rho$ since the $\rho$
decouples. Using $f^2_\sigma (m_\rho)=a(m_\rho)f^2_\pi (m_\rho)$,
we can rewrite the KSRF mass formula as
 \ba
m^2_{\rho{\tiny (KSRF)}}=M^2_\rho
(m_\rho)\frac{a(m_\rho)}{2}\frac{f^2_\pi (m_\rho)}{f^2_\pi
(0)}.\nonumber
 \ea
By the renormalization group equations given by Harada and
Yamawaki, we have $a(m_\rho)\simeq 1$ and $\frac{f^2_\pi
(m_\rho)}{f^2_\pi (0)}\simeq 2$. Thus, we get
 \ba
m^2_{\rho {\tiny (KSRF)}}\simeq M_\rho (m_\rho)\equiv
m^2_\rho.\nonumber \ea}. Now in medium with $n\neq 0$, this mass
formula will remain the same except that it will depend upon
density~\footnote{Note that the mass in this formula is a {\it
parameter} in the Lagrangian and not necessarily a pole mass
except at the phase transition point. The pole mass will
generically have additional density-dependent terms which will go
to zero as the coupling $g_V$ reaches the Georgi vector limit.},
 \be
m^\star_V\equiv m^\star_\rho=m^\star_\omega =\sqrt{a(m^\star_V)}
g_V (m^\star_V) f_\pi (m^\star_V).\label{ksrfn}
 \ee The density
dependence is indicated by the star. As in the case of $N_F$
discussed in \cite{VM}, the cutoff $\Lambda$ will depend upon
density, say, $\Lambda^\star$ implicit in (\ref{ksrfn}).

The Harada-Yamawaki argument (or theorem) suggests that at $n=n_c$
where $\la\bar{q}q\ra=0$ and hence $\Pi^\star_V (n_c)=\Pi^\star_A
(n_c)$,  the Georgi's vector limit $g_V=0$, $a=1$ is reached
together with $f_\pi=0$. This means that both the mass parameter
(\ref{ksrfn}) and the pole mass vanish,
 \ba
m_V^\star (n_c)=0.
 \ea

At the point where the vector meson mass vanishes, the quartet
scalars will be ``de-Higgsed" from the vector mesons and form a
degenerate multiplet with the triplet of massless pions (and
``$\eta$"~\footnote{We assume that the fourth pseudoscalar
``$\eta$" gets a mass by $U(1)_A$ anomaly, so can be excluded from
our consideration.}) with the massless vectors decoupled. This
assures that the vector correlator is equal to the axial-vector
correlator in the HLS sector matching with the QCD sector. In this
scenario, {\it dictated by the renormalization group equations,
the vector meson masses drop as density (or T) increases}. Note
that this scenario is distinct from the ``standard" (as yet
to-be-established) picture in which the $\rho$ and $a_1$ come
together as do the pions and a scalar $\sigma$. In the standard
scenario, there is nothing which forces the vector mesons to
become massless and decouple. They can even become more massive at
chiral restoration than in the vacuum as discussed by
Pisarksi~\cite{pisarski2}~\footnote{An important point to note
here is that Pisarski's argument hinges on the vector dominance
picture. However Harada and Yamawaki show that at the vector
manifestation, the vector dominance is strongly
violated~\cite{fate}. This is one of the basic differences in
treating vector mesons.}. Thus the vanishing of the vector-meson
mass is a {\it prima facie} signal for the phase transition in the
Harada-Yamawaki picture.

Here using the same reasoning as for large $N_F$ and high $T$, we
have arrived at the conclusion that the same vector manifestation
can take place at the critical density~\footnote{There is a puzzle
here. The analysis of \cite{appelquist} for the $N_F$-driven phase
transition is that the top-down phase transition from $N_F
>N_{Fc}$ to $N_F<N_{Fc}$ is neither first-order nor second-order.
The order parameter is found to be continuous at the phase
transition, so it is not first-order. However the correlation
length does not diverge at the phase transition, so it is not
second-order either. If correct, this phase transition may appear
to be different from that expected in temperature or density.}. As
in the case of temperature, the ``running" in density of the
parameters $g_V$, $a$ and $f_\pi$ implies that away from the
critical point, the vector meson (pole) masses drop as density is
increased. In the next section we will interpret this phenomenon
in terms of BR scaling.

In this section as well as in what follows, we are interpreting
the vector-meson ``mass" in the sense of the BR-scaling mass. Of
course in the presence of a medium with nonzero temperature and/or
density, the lack of Lorentz invariance gives rise to different
components and it will be necessary to specify which component
gets ``de-Higgsed" at the phase transition. The meaning of the
vector mass as used here is not the same as the standard
definition as we will try to clarify in the following sections.

\section{LANDAU FERMI LIQUID FROM CHIRAL\\ LAGRANGIANS}\label{fermiliquid}
\setcounter{equation}{0} 
\renewcommand{\theequation}{\mbox{\ref{fermiliquid}.\arabic{equation}}}
\subsection{Fluctuating Around Zero-Density Vacuum}
 \itt
We have developed the thesis that chiral symmetry can be suitably
implemented into nuclear problems involving a few nucleons.
Furthermore EFT can be formulated in such a way that the
traditional nuclear physics approach with realistic potentials
(i.e., PM) can be identified as a {\it legitimate} part of a
consistent EFT supplying the leading term in the expansion. This
means that what nuclear physicists have been doing up to today --
and their highly successful results -- can be considered as an
essential part of the modern structure of strong interactions
based on QCD. For the few-nucleon processes we have been dealing
with, e.g., NN scattering, $n+p\leftrightarrow d+\gamma$, solar
$p+p\rightarrow d+e^+ +\nu_e$ etc., what enters into the theory
is the chiral symmetric Lagrangian determined in free space, that
is, at zero density. The starting point there is an effective
chiral Lagrangian that describes QCD in the matter-free vacuum,
with masses and coupling constants all determined in particle
experiments in medium-free space. The few-nucleon problems are
then treated by fluctuating around the matter-free vacuum near a
``fixed point" (at infinite scattering length) discussed above.
The machinery for doing this is the well-established chiral
perturbation theory ($\chi$PT) involving baryons. Since the
processes involved occur near the fixed point of the theory, one
can successfully do the perturbation.

Now how do we go to heavier nuclei, nuclear matter and denser
matter which must live near a different fixed point, say, the
Fermi-liquid fixed point to be described below which is not
necessarily near the above fixed point? This is the principal
question that we wish to address in this review. If there are no
phase changes along the way, one may attempt to build many-nucleon
systems in a perturbative scheme starting with the free-space
Lagrangian mentioned above. Indeed there have been several such
attempts in $\chi$PT to reach at nuclear
matter~\cite{lynn,friman}. It is not obvious though that this can
be done beyond the normal matter density since a strong-coupling
expansion is involved and large anomalous dimensions arise. This
means that it may be simply too hard if not impossible to handle
various phase changes (kaon condensation, chiral restoration,
color superconductivity etc.) that are conjectured to occur as one
increases density toward chiral restoration as demanded for
relativistic heavy-ion processes and for describing compact stars.
A different but perhaps more precise way of saying the same thing
is in terms of the skyrmion picture for dense matter which is
closest to QCD if $N_c$ is taken to be large. In the skyrmion
description, a lump of baryonic matter with a given baryon number
$B$ is characterized by a ``topological vacuum" associated with a
conserved winding number $B$. A system with density $\rho_B^\prime
\gg\rho_B$ means that $B^\prime\gg B$ for a given volume. Now the
point is that the $B^\prime$ vacuum may not be simply connected to
the $B$ vacuum by a small perturbation. In what follows,
therefore, we shall take a different route and formulate a chiral
Lagrangian field theory using a ``sliding" vacuum with parameters
of the Lagrangian running with matter density. This leads to the
notion of Brown-Rho scaling and in-medium effective field theory
for a system with Fermi seas, i.e., Landau Fermi-liquid theory.
\subsection{Skyrmion vs. Q-Ball}
\itt
 Consider bound many-nucleon systems corresponding to nuclei.
If the number of colors can be taken to be very large, then one
may approach them starting with a skyrmion-type Lagrangian
constructed of Goldstone boson fields (and heavy mesons) and
looking for solutions of the winding number $W$ equal to the
baryon number $B$ or the mass number $A$ in a nucleus. The Skyrme
Lagrangian which consists of the current algebra term and the
quartic term
 \be
 \calL_{sk}=\frac{f^2}{4} \Tr (\del_\mu U \del^\mu U^\dagger)
 +\frac{1}{32e^2}\Tr [U\del_\mu U^\dagger, U\del_\nu
 U^\dagger]^2+\cdots\label{skyrme}
  \ee
which may be viewed as an approximate zero bag size limit of the
Lagrangian (\ref{model}) has been extensively studied
mathematically for baryon number up to $B=22$ with a fascinating
result~\cite{skyrmenuc}. (The ellipsis stands for higher
derivative and/or heavy-meson and chiral-symmetry breaking terms
that are to be added as needed.) The Lagrangian of the type
(\ref{skyrme}) suitably extended with higher derivatives and/or
massive boson fields is accurate only at large-$N_c$ limit, so it
is not clear what the classical solution of this Lagrangian for
$B\geq 2$ represents since in nature where $N_c\ll\infty$.
Nonetheless this development is quite exciting since it is closest
to QCD at least for $N_c=\infty$. As stated above, the ground
states with baryons numbers $B$ and $B^\prime$ with $B\gg
B^\prime$ cannot be connected by small perturbation because they
represent different topological ``vacua."

A direction we will adopt here is not as ambitious as the above
but we will attempt to account for the different vacuum structure
for dense systems (heavy nuclei) than for dilute systems (light
nuclei). We shall start with a theory for a system with a Fermi
sea. It is possible that such a system emerges as a sort of
nontopological soliton or ``Q-ball" from a theory based on chiral
perturbation theory defined at zero density as suggested first by
Lynn~\cite{lynn} and recently refined by Lutz et
al~\cite{friman}. Although there is no proof, we believe that the
Q-ball system which is non-topological is essentially equivalent
to the skyrmion system which is topological. This equivalence is
somewhat like in the nucleon structure where the two pictures
work equally well. The essential point we shall accept here is
that a non-topological soliton solution exists possessing the
liquid structure that is supposed to be present in nuclear
matter. We will later identify this with what we do know but for
the moment let us assume that we have a Fermi sea filled with
nucleons. It is fortunate for our purpose that such a system is
accessible to an elegant and powerful effective field theory
treatment as shown by various authors~\cite{shankar} that leads
to Landau's Fermi liquid theory applied to nuclear
systems~\cite{migdal}. \vskip 0.3cm
\subsection{Effective Chiral Lagrangian for Many-Nucleon
 Systems}
\itt What we wish to do in our case is to arrive, following the
developments in \cite{frsall,songPR,mr-migdal}, at the Landau
Fermi liquid theory starting from an effective chiral Lagrangian
that we have been discussing above. Treating the nucleon in terms
of a local field denoted by $N$ and the Goldstone boson field by
$\xi=\sqrt{U}=e^{i{\bf \pi}\cdot{\bf \tau} /2f_\pi^\star}$, we
can write a simple Lagrangian of the form
 \be \calL=\bar{N}[i\gamma_{\mu}(\del^{\mu}+iv^{\mu}
+g_A^\star\gamma_5 a^{\mu}) -M^\star]N -\sum_i C_i^\star
(\bar{N}\Gamma_i N)^2 +\cdots \label{leff} \ee where the ellipsis
stands for higher dimension nucleon operators and the
$\Gamma_i$'s Dirac and flavor matrices as well as derivatives
consistent with chiral symmetry. The star affixed on the masses
and coupling constants will be defined precisely below. The
induced vector and axial vector ``fields" are given by $v_\mu =
-\frac{i}{2}(\xi^{\dagger}\del_\mu\xi +\xi\del_\mu\xi^\dagger )$
and $a_\mu = -\frac{i}{2}(\xi^\dagger\del_\mu\xi
-\xi\del_\mu\xi^\dagger ).$~\footnote{The $\xi$ field appearing
here is the ``gauge-fixed" one, $\xi_L=\xi_R^\dagger=\xi$. Here
we are dealing with the matter far away from the chiral phase
transition point, so we are taking the unitary gauge.} In
(\ref{leff}) only the pion ($\pi$) and nucleon ($N$) fields
appear explicitly: all other fields have been integrated out. The
effect of massive degrees of freedom will be lodged in
higher-dimension and/or higher-derivative interactions. The
external electro-weak fields if needed are straightforwardly
incorporated by suitable gauging.

When applied to symmetric nuclear matter in the mean field
approximation, the Lagrangian (\ref{leff}) is known to be
equivalent~\cite{gelmini,BR96} to the Lagrangian that contains
just the degrees of freedom that figure in a linear model of the
Walecka-type ~\cite{waleckamodel}~\footnote{For asymmetric
nuclear matter, isovector fields (e.g., $\rho$, $a_1$ etc.) must
be included in a chirally symmetric way.}
 \be \calL &=&
\bar{N}(i\gamma_{\mu}(\del^\mu+ig_v^\star\omega^\mu
)-M^\star+h^\star\sigma )N \nonumber\\ & &-\frac 14 F_{\mu\nu}^2
+\frac 12 (\partial_\mu \sigma)^2
+\frac{{m^\star_\omega}^2}{2}\omega^2
-\frac{{m^\star_\sigma}^2}{2}\sigma^2+\cdots\label{leff2} \ee
 where the ellipsis denotes higher-dimension operators. We should
stress that (\ref{leff2}) is consistent with chiral symmetry
since here both the $\omega$ and $\sigma$ fields are {\it chiral
singlets}. In fact the $\sigma$ here has nothing to do with the
chiral fourth-component scalar field of the linear sigma model
except near the chiral phase transition density; it is a
``dilaton" connected with the trace anomaly of QCD and is
supposed to approach the chiral fourth component in the ``mended
symmetry" way \`a la Weinberg~\cite{weinberg-salam,beane} near
chiral restoration. In practice, depending upon the problem, one
form is more convenient than the other. In what follows, both
(\ref{leff}) and (\ref{leff2}) will be used interchangeably.

 In (\ref{leff}) and (\ref{leff2}), the mass
and coupling parameters affixed with stars depend upon the
``sliding" vacuum defined at a given ``density." (Putting density
into the parameters of a Lagrangian is subtle because of chiral
symmetry and thermodynamic consistency. We will specify more
precisely how the ``density" is to be defined.) We are thinking of
the situation where an external pressure (e.g., gravity for
compact stars) is exerted and hence we can think of the
parameters of the Lagrangian varying as a function of density.
They need not be associated with quantities defined at a {\it
minimum} of an effective potential of the theory without the
external pressure.

The next important observation we shall exploit is that by
Matsui~\cite{matsui} that a Lagrangian of the type (\ref{leff})
can give in the mean field the result of Landau Fermi liquid.
This means that in the same approximation, the chiral Lagrangian
(\ref{leff}) can be mapped to Landau Fermi liquid. From the works
in \cite{shankar} which tell us that Landau theory is a
fixed-point theory with a {\it Fermi-liquid fixed point} which is
distinct from that of the effective Lagrangian used for
two-nucleon systems discussed above, we learn that the theory
(\ref{leff}) in the mean field has two fixed-point quantities,
viz, the effective quasiparticle (or Landau) mass $m^\star_N$ and
the Landau parameters $\calF$ for quasiparticle interactions. The
full interaction between two quasiparticles ${\pb_1}$ and
${\pb_2}$ at the Fermi surface of symmetric nuclear matter
written in terms of spin and isospin invariants is defined by
 \be \label{qpint}
{\calF}_{\pb_1\sigma_1\tau_1,\pb_2\sigma_2\tau_2}&=&
\frac{1}{N(0)}\left[F(\cos \theta_{12})+F^\prime(\cos
\theta_{12})\taub_1\cdot \taub_2+G(\cos \theta_{12})\sigmab_1\cdot
\sigmab_2\phantom{\frac{\qb^{\,2}}{k_F^2}}\right.\nonumber\\
&+&\left. G^\prime(\cos \theta_{12})\sigmab_1\cdot
\sigmab_2\taub_1\cdot\taub_2 +\frac{\qb^{\, 2}}{k_F^2}H(\cos
\theta_{12})S_{12}(\qbhat)\right.\nonumber\\
&+&\left.\frac{\qb^{\, 2}}{k_F^2}H^\prime(\cos \theta_{12})S_{12}
(\qbhat)\taub_1\cdot\taub_2\right]\label{landauF}
 \ee where $\theta_{12}$ is the
angle between ${\pb_1}$ and ${\pb_2}$ and $N(0)=\frac{4
k_F^2}{(2\pi^2)}\left(\frac{dp}{d\ve}\right)_F$ is the density of
states at the Fermi surface. Also, $\qb=\pb_1-\pb_2$ and $
S_{12}(\qbhat) = 3\sigmab_1\cdot\qbhat\sigmab_2\cdot\qbhat -
\sigmab_1\cdot\sigmab_2$,  where $\qbhat = \qb/|\qb |$. The
tensor interactions $H$ and $H^\prime$ are important for the axial
charge to be considered later. The functions $F, F^\prime, \dots$
are expanded in Legendre polynomials, $F(\cos
\theta_{12})=\sum_\ell F_\ell P_\ell(\cos \theta_{12})$,  with
analogous expansion for the spin- and isospin-dependent
interactions.

The fixed point $m^\star_N$ is associated with a given Fermi
momentum $k_F$ or density $\rho$ of the given system, so that
there will be a set of fixed points $\calF$ for each (given)
$k_F$ or $\rho$. Thus we will have the Landau parameters that
depend upon density once we accept that the Landau mass depends
on density as is obvious from the Landau mass formula
 \be
\frac{m_N^\star}{M}=1+\frac 13 F_1=(1-\frac 13
\tilde{F}_1)^{-1}\label{landaumass} \ee where $\tilde{F_1}\equiv
(M/m_N^\star) F_1$ and $F_1$ is the $l=1$ component of the
interaction $F$ in (\ref{landauF}).
 \subsection{Brown-Rho Scaling}
\itt How the fixed point quantities vary with density cannot be
determined from the effective Lagrangian alone. It must be tied
to the vacuum property characterized by the chiral condensate
$\la\bar{\psi}\psi\ra$ where $\psi$ is the quark field. The
dependence must therefore be derived from the fundamental theory,
QCD. In the past QCD sum rules have been used to extract
information on this matter. For our purpose, it is more
convenient to take the Skyrme Lagrangian (\ref{skyrme}) as a
starting point and try to deduce the necessary information. We
expect it to give the leading behavior corresponding to a mean
field theory in the sense to be defined more precisely later.
This is the argument presented in the original BR
paper~\cite{BR91}. Implementing the scalar dilaton field $\chi$
associated with the trace anomaly of QCD to the Lagrangian
extended to take explicitly into account the vector mesons
$V_\mu=\rho_\mu, \omega_\mu$, it has been found that in the mean
field the following scaling holds, e.g., in the Lagrangian
(\ref{leff2})~\footnote{The scalar $\sigma$ as well as the vector
$V$ are absent in (\ref{leff}) since they have been integrated
out. Their scaling will appear therefore in the coefficients
representing those degrees of freedom, namely, $C_i^\star$'s. The
connections between the two are given in
\cite{frsall,songPR,mr-migdal}.}:
 \be
\Phi (\rho) \approx \frac{f^\star_\pi (\rho )}{f_\pi} \approx
\frac{m^\star_\sigma (\rho )}{m_\sigma}\approx\frac{m_V^\star
(\rho )}{m_V} \approx \frac{M^\star (\rho )}{M}. \label{BRscaling}
 \ee
Here $f_\pi$ is the pion decay constant, the $f^\star_\pi$ being
the in-medium one~\footnote{Since Lorentz invariance is lost in
medium, the constants $f_\pi^\star$, $g_A^\star$ etc. contain two
components, space part and time part, that are in general
different. The parameters that appear in BR scaling are
mean-field quantities and are defined more precisely in the sense
described above for a generalized Walecka model (\ref{leff2}). The
same remark applies to the in-medium vector coupling we will
discuss in later sections.}. We caution that the
$M^\star$ is the
nucleon in-medium mass (or BR scaling mass) which is {\it not}
the same as the Landau mass $m_N^\star$ as we will see shortly.
The quantity $\Phi(\rho)$, related to the quark condensate
$\Phi\approx (\la\bar{q}q\ra^\star/\la\bar{q}q\ra)^a$ where $a$
is a constant which we will find in Section \ref{PT} equals $\sim
1$, is the scaling factor that needs to be determined from
(fundamental) theory or experiments.

Let us now examine more closely what the ``density dependence" in
general (not necessarily in the context of BR scaling) means in
the Lagrangian (\ref{leff}) or (\ref{leff2}). For this, let us
take (\ref{leff2}) and introduce the chirally invariant operator
 \be
\check\rho u^\mu \equiv \bar{N}\gamma^\mu N \ee where $u^\mu
=(\sqrt{1-\vb^2})^{-1}(1,\vb ) =(\sqrt{\rho^2-\jb^2})^{-1}(\rho
,\jb )$ is the fluid 4-velocity. Here $\jb = \langle
\bar{N}\gammab N \rangle $ is the baryon current density and $
\rho =\langle N^\dagger N \rangle =\sum_in_i$  the baryon number
density. The expectation value of $\check\rho$ yields the baryon
density in the rest-frame of the fluid. Using $\check\rho$ it is
easy to construct a Lorentz invariant, chirally invariant
Lagrangian with density dependent parameters.

Now a density dependent mass parameter in the Lagrangian should be
interpreted as $ m^\star=m^\star (\check\rho)$. This means that
the model (\ref{leff2}) is no longer linear. It is highly
non-linear even at the mean field level.

The Euler-Lagrange equations of motion for the bosonic fields are
the usual ones but the nucleon equation of motion is not. This is
because of the functional dependence of the masses and coupling
constants on the nucleon field:
 \be
[i\gamma^\mu (\del_\mu +ig_v^\star\omega_\mu -iu_\mu\check\Sigma
) -M^\star +h^\star\sigma ] N=0\label{fer} \ee
 with
 \be
\check\Sigma &=&\frac{\del\calL}{\del\check\rho}\\ &=&
m_\omega^\star\omega^2\frac{\del m_\omega^\star}{\del\check\rho}
-m_\sigma^\star\sigma^2\frac{\del m_\sigma^\star}{\del\check\rho}
-\bar{N}\omega^\mu\gamma_\mu N\frac{\del
g_v^\star}{\del\check\rho} -\bar{N}N\frac{\del
M^\star}{\del\check\rho}. \nonumber \ee
 Here we are assuming that
$(\partial/\partial\check\rho) h^\star\approx 0$. This can be
justified using NJL model~\cite{BBR} but we will simply assume it
here~\footnote{In fact apart from the phenomenological
requirement, there is no reason to assume this. All our arguments
that follow would go through without difficulty even if the
coupling constant were to scale with density.}. The additional
term $\check\Sigma$, which may be related to what is referred to
in many-body theory as ``rearrangement terms", is essential in
making the theory consistent. This point has been overlooked in
the literature.

The crucial observation is that when one computes the
energy-momentum tensor with (\ref{leff2}), one finds, in addition
to the canonical term (which is obtained when the parameters are
treated as constants), a new term proportional to $\check\Sigma$
\be
T^{\mu\nu}=T_{can}^{\mu\nu}+\check\Sigma(\bar{N}{u\cdot\gamma} N)
g^{\mu\nu}.\label{tmunu} \ee
 The pressure is then given by
$\frac13 \langle T_{ii}\rangle_{\vb =0}$. The additional term in
(\ref{tmunu}) matches precisely the terms that arise when the
derivative with respect to $\rho$ acts on the density-dependent
masses and coupling constants in the formula derived from
$T_{00}$: \be p=-\frac{\del E}{\del V}=\rho^2\frac{\del\E
/\rho}{\del\rho}=\mu\rho -\E \ee
 where
\be \E=\langle T^{00}\rangle. \ee This matching assures
energy-momentum conservation and thermodynamic consistency. This
verifies that interpreting ``density-dependent" parameters in our
Lagrangian as the dependence on $\check\rho$ with $\la
\check\rho\ra$ being the density is consistent with both chiral
symmetry and thermodynamics.
 \subsection{Landau Mass and BR Scaling}
\itt
 We end this section by writing down
the relations between the parameters of the Lagrangian
(\ref{leff}) or equivalently (\ref{leff2}) and Landau parameters.
We will use the former and take (including the isovector channel)
 \be
-\sum_i C_i^\star (\bar{N}\Gamma_i N)^2\approx
-\frac{{C_{\tilde{\omega}}^\star}^2}{2} (\bar{N}\gamma_\mu N)^2
-\frac{{C_{\tilde{\rho}}^\star}^2}{2}(\bar{N}\gamma_\mu\tau N)^2
+\cdots. \ee
 This may be considered  as the leading terms that
result when the vectors $\omega$ and $\rho$ are integrated out.
But it is more than just that. In fact, the subscripts can be
taken to represent not only the vector mesons $\omega$ and $\rho$
that nuclear physicists are familiar with but also {\it all}
vector mesons of the same quantum numbers (which account for the
appearance of the tilde on the vector mesons), so the two
``counter terms" on the right-hand side subsume the {\it full}
short-distance physics of the given chiral order. Since the
Lagrangian (\ref{leff}) in tree order must correspond to the mean
field of (\ref{leff2}), one can simply compute the relevant
Feynman graphs with (\ref{leff}) and match them to the Landau
Fermi liquid formula. One obtains~\cite{frsall}
 \be \frac{m_N^\star}{M}
=\left(\Phi^{-1} -\frac 13 \tilde{F}_1
(\pi)\right)^{-1}\label{mstar2}
 \ee
where $\tilde{F}_1(\pi )$ is the pion contribution to $F_1$ which
is completely fixed by chiral symmetry so the only parameter that
enters in (\ref{mstar2}) is the BR scaling $\Phi\approx
M^\star/M$.

Now using the Landau mass formula (\ref{landaumass}) and \be
\tilde{F}_1=\tilde{F}_1 (\tilde{\omega})+\tilde{F}_1 (\pi) \ee
 we find
 \be \tilde{F}_1
(\tilde{\omega})=3(1-\Phi^{-1})\label{mainrelation}
 \ee
where $\tilde{F}_1 (\tilde{\omega})$ includes contributions from
{\it all} excitations of the $\omega$ meson quantum numbers. This
is the main result of the connection between the chiral Lagrangian
with BR scaling and Landau Fermi liquid theory. This is a highly
non-trivial relation which says that BR scaling -- which in the
context of the QCD vacuum reflects modification due to matter
density -- is tied to short-range quasiparticle interactions. This
means that what may be considered to be a basic feature of QCD and
nuclear interactions via nuclear forces are inter-related,
indicating another level of the Cheshire Catism. \vskip 0.3cm

$\bullet$ {\it Determining $\Phi (\rho)$} \vskip 0.3cm

If one accepts the BR scaling (\ref{BRscaling}), there are several
sources to use for determining the scaling factor $\Phi (\rho)$.
For instance, one can determine it from the QCD sum
rule~\footnote{There is a great deal of controversy in the
application of QCD sum rules in dense and/or hot medium. As it
stands, the situation is totally confused and it is perhaps too
dangerous to take seriously any result so far obtained. For a
recent discussion on this matter, see \cite{leupold}. Our proposal
is that the value (\ref{phinuc}) which will result from the fit to
a variety of nuclear processes discussed below be taken as the
value to be obtained from the QCD sum rule approach.} for the mass
ratio $m_\rho^\star/m_\rho$~\cite{sumrule} or from the ratio
$f_\pi^\star/f_\pi$ using the Gell-Mann-Oakes-Renner (GMOR)
formula for pion mass in medium~\footnote{Using the in-medium GMOR
formula to arrive at the value (\ref{phinuc}), it is assumed that
the pion mass remains unaffected by density up to nuclear matter
density. This assumption may have to be re-examined in view of the
``pion mass" measured in deeply bound pionic states which indicate
a mild increase as a function of density as discussed in Section
\ref{PT}. This issue is not so simple to address since it is not
clear that the mass that appears in the GMOR relation is the same
quantity that has been ``measured."} or a fit to nuclear matter
properties~\cite{frsall}. The result comes out to be about the
same at nuclear matter density
 \be
\Phi (\rho=\rho_0)\approx 0.78. \label{phinuc}\ee
 In terms of the parameterization
of the form
 \be
\Phi (\rho)=(1+y\rho/\rho_0)^{-1},
 \ee
the result corresponds to $y(\rho=\rho_0)\approx 0.28$. This gives
the Landau mass $m_N^\star/M\approx 0.70$ which is to be compared
-- with a due consideration of the caveat mentioned above on the
difficulty with QCD sum rules in nuclear physics -- with the
presently available QCD sum-rule value
$0.69^{+0.14}_{-0.07}$~\cite{FJL}. This value is also consistent
with the properties of heavy nuclei.

\section{INDIRECT EVIDENCES FOR BR SCALING}\label{indirect}
\setcounter{equation}{0} 
\renewcommand{\theequation}{\mbox{\ref{indirect}.\arabic{equation}}}
 \subsection{Indications in Finite Nuclei}\label{finite}
 \itt
Before the publication on BR scaling in \cite{BR91}, Brown and
Rho proposed~\cite{BR89,BR90} that some of the missing strength in
the longitudinal response function in nuclei could be explained
by the dropping masses in medium $m_\rho^\star/m_\rho\approx
m_N^\star/m_N\approx 0.75$. The same dropping masses (of both the
nucleon and the $\rho$) could also explain that within certain
range of kinematics relevant to experiments, the transverse
response function does not scale in medium. This is essentially
the picture that is confirmed by the Morgenstern-Meziani analysis
mentioned in the introduction and described in detail in Section
\ref{electron}.

In addition to the electromagnetic response functions, there were
further predictions in \cite{BR90} that followed from the
dropping masses. One was the anomalous gyromagnetic ratio $\delta
g_l$ in nuclei which gets enhanced by the scaling factor. This
will be discussed below in terms of a more modern formulation with
chiral Lagrangians and Landau Fermi liquid theory although the
physics is essentially the same as in this work. The second is
that the dropping mass implies that as density increases, the
tensor force in nuclei gets weakened. The reason is that the
tensor force is given by two agencies, one-pion exchange and
one-$\rho$ exchange coming with opposite signs. At low density,
the pion exchange dominates, so the net effect is more or less
controlled by the pion tensor. However as density increases, the
$\rho$ tensor gets enhanced by the factor $(m_N/m_N^\star)^2$
apart from the increased range of the $\rho$-exchange force.
Therefore the $\rho$ tensor tends to cancel more and more the
pion tensor as density increases, thereby suppressing the net
tensor force in the outer region where pion exchange dominates.
There are some indications that the suppressed tensor force is
consistent with nuclear spectra as well as with Gamow-Teller
transitions in nuclei to which we shall return shortly.

There is, however, a possible caveat to this as we will see later
and that has to do with the possibility that the vector coupling
$g^\star_{\rho NN}$ will also drop as density increases,
eventually decoupling as dictated by the ``vector manifestation"
discussed above. It is not clear at what density this decoupling
sets in but there are empirical indications that the tensor force
weakening is operative at least up to nuclear matter density. The
phenomenology in nuclei discussed above also indicates that at
least at the low densities encountered in nuclei, the ratio
$g^\star_{\rho NN}/m^\star_\rho$ does increase. It should be noted
that even when the vector decoupling takes place, with the flavor
symmetry in the hadronic sector yielding to the color gauge
symmetry in the quark-gluon sector~\footnote{The way this
change-over takes place is discussed in Section \ref{denseqcd}.},
some vector repulsion remains coming from (Fierzed) gluon
exchange~\cite{BRPR}.

 \subsection{The Anomalous Gyromagnetic Ratio in Nuclei}
 \itt
The development given above in Section \ref{fermiliquid} of the
effective chiral Lagrangians with BR scaling and Fermi liquid
fixed point theory can provide a more sophisticated and rigorous
explanation for the enhanced gyromagnetic ratio $g_l$ in nuclei
mentioned above. We discuss this as a strong, though indirect,
evidence for BR scaling. The nuclear gyromagnetic moment $g_l$ in
the convection current for a particle sitting on top of the Fermi
sea is defined as
 \be
\Jb=\frac{\Bk}{M}g_l.\label{convection}
 \ee
We have defined $g_l\equiv 1+\delta g_l$ such that the bare
nucleon mass $M$, {\it not} the effective (Landau) mass
$m_N^\star$, appears in the formula. This is because we want to
preserve charge conservation associated with the factor 1 in
$g_l$, $\delta g_l$ being purely isovector.

The $\delta g_l$ has been determined from experiments. The
measurement that is most relevant to our theory is the one on
giant dipole resonances on heavy nuclei, in particular in the
lead region. The result of \cite{schumacher} on $^{209}$Bi gives
the proton value
\be
 \delta g_l^p=0.23\pm 0.03.\label{deltaglexp}
\ee
 This can also be determined from magnetic moment measurements but
the analysis is somewhat more complex and hence less precise. We
shall refer to (\ref{deltaglexp}).

Given the effective chiral Lagrangian (\ref{leff}), with the
connections we have established so far, it is easy to compute in
the mean field order the nuclear gyromagnetic ratio $g_l$. It
suffices to calculate all terms of the same order to assure that
charge is conserved or what is equivalent, ``Kohn theorem" is
satisfied. The result is the Migdal formula~\cite{migdal} given
in terms of the parameters of the Lagrangian, that is, the BR
scaling $\Phi (\rho)$,
 \be
g_l=\frac{1+\tau_3}{2}+\delta g_l\label{gl}
 \ee
 with
 \be
\delta g_l=\frac 16 (\tilde{F}_1^\prime -\tilde{F}_1)\tau_3
=\frac 49 \left[\Phi^{-1} -1- \frac 12 \tilde{F}_1
(\pi)\right]\tau_3. \label{chptd} \ee

 As mentioned above, $\tilde{F}_1(\pi)$ is completely fixed by
chiral symmetry for any density -- possibly up to chiral
restoration -- so the only quantity that appears here is the BR
scaling factor. At nuclear matter density, we have $\frac 13
\tilde{F}_1 (\pi)|_{\rho=\rho_0}=-0.153$. Now taking $\Phi
(\rho_0)\approx 0.78$ from (\ref{phinuc}), we predict from
(\ref{chptd})
 \ba
\delta g_l^{th}=0.227\tau_3.
 \ea This agrees with the
experiment (\ref{deltaglexp}) providing a quantitative check of
the BR scaling.
 \subsection{Axial Charge Transitions in Heavy Nuclei}
\itt Another process where the Lagrangian (\ref{leff}) with BR
scaling together with the PKMR's EFT can make a prediction is in
the axial charge beta transition of the type \be
A(J^\pm)\rightarrow A^\prime (J^\mp)+e^-(e^+)+\bar{\nu}(\nu) \ \
\ \Delta T=1.\label{ac}
 \ee
This process was studied by several experimentalists with the
objective of exhibiting strong medium-enhancement of the matrix
element of the axial charge operator $A^a_0$ that governs the
process (\ref{ac}) in heavy nuclei. Such an enhancement was first
observed by Warburton in the lead region~\cite{warburton} and
since then several authors confirmed Warburton's finding in
various other heavy nuclei. Warburton focused on the quantity
(that we shall refer to as ``Warburton ratio")
 \be
\epsilon_{MEC}=M_{exp}/M_{sp} \ee
 where $M_{exp}$ is the {\it
measured} matrix element for the axial charge transition and
$M_{sp}$ is the {\it theoretical} single-particle matrix element
for a nucleon {\it without} BR scaling. There are several
theoretical uncertainties involved in extracting the Warburton
ratio. First, one has to extract from given beta decay data what
corresponds to the axial charge matrix element. This involves
estimating accurately other terms than the axial charge that
contribute. The second is what one means precisely by $M_{sp}$
which is a theoretical entity. For these reasons, an unambiguous
conclusion is difficult to arrive at. However what is significant
is Warburton's observation that in heavy nuclei, $\epsilon_{MEC}$
can be substantially larger than the possible uncertainties
involved:
 \be \epsilon_{MEC}^{HeavyNuclei}=1.9\sim 2.0.
 \ee
Furthermore more recent measurements and their detailed analyses
in different nuclei~\cite{minamisono,baumann,vangeert} as shown
in Table \ref{expepsilon} quantitatively confirm this result of
Warburton.

The prediction from the Lagrangian (\ref{leff}) with BR scaling
in the same approximation as in the gyromagnetic ratio case
is~\cite{KR91,frsall}
 \be \epsilon_{MEC}^{\chi th}=\Phi^{-1}
(1+\tilde{\Delta})\label{epsilonth}
 \ee
where $\tilde{\Delta}$ is, in accordance with the chiral filter
mechanism, dominated by the pionic contribution with small
controllable corrections from vector degrees of
freedom~\footnote{In terms of the Landau-Migdal fixed-point
parameters, $\tilde{\Delta}=G_1^\prime/3-10 H_0^\prime/3+
4H_1^\prime/3-2H_2^\prime/15$ with the dominant contribution
coming from the soft pions~\cite{KDR}.}, the magnitude and the
density dependence of which are again totally controlled by
chiral symmetry, with the BR scaling $\Phi$ appearing as an
overall factor. For $\rho=\rho_0/2$ and $\rho=\rho_0$, we
find~\cite{frsall}
 \be
\epsilon_{MEC}^{\chi th}|_{\rho=\rho_0/2}\approx 1.63, \ \ \
\epsilon_{MEC}^{\chi th}|_{\rho=\rho_0}\approx 2.06.
 \ee
This is in an overall agreement with all the available Warburton
ratios (see Table \ref{expepsilon}).

\begin{table}[h]
\caption{``Empirical values" for
$\epsilon_{MEC}$}\label{expepsilon} \vskip .3cm
\begin{center}
\begin{tabular}{cccc}\hline\hline
Mass number A&$\epsilon_{MEC}$& Reference & \\
 \hline
 12&$1.64\pm 0.05$&\cite{minamisono}& \\
50 &$1.60\pm 0.05$ &\cite{baumann}& \\
205 & $1.95\pm 0.05$ &\cite{vangeert}\\
208&$2.01\pm 0.10$&\cite{warburton}& \\
\hline\hline
\end{tabular}
\end{center}
\end{table}

\subsection{Axial-Vector Coupling Constant $g_A^\star$ in Dense
Matter}\label{gAstar}

\itt
 One can make a rather simple statement on how the axial-vector
coupling constant $g_A^\star$ scales in dense matter on the basis
of BR scaling and the Fermi liquid theory.

The story of $g_A^\star$ in nuclei is a long story and rather
involved with nuclear structure effects compounded with strong
nuclear tensor correlations, excitations of nucleon resonances and
possible effects of the vacuum
change~\cite{MR74,arima-core,arima-more,mr91}. Gamow-Teller
transitions in complex nuclei, particularly giant resonances
thereof, play an important role in presupernova stage of the
collapsing stars and the effective coupling constant $g_A^\star$
figures importantly in their description~\cite{presupernova}.

What we will present here is probably not the unique explanation
of what is going on but it is a version that is immediate from,
and consistent, with the general theme of this
review~\footnote{It is perhaps appropriate at this point to
clarify how the absence of Lorentz symmetry manifests itself in
our treatment of $g_A^\star$, an issue we briefly alluded to
earlier. As the two (preceding and this) subsections clearly
show, the effective axial coupling constant $g_A$ in matter is
different for the space and time components of the axial current.
We see that the time component of the axial charge (measured
through axial-charge transitions) is enhanced in nuclear matter
whereas the space component of the axial charge (measured through
Gamow-Teller transitions as here) is quenched. Although we did
not make the distinction in discussing BR scaling, it is evident
that the $g_A^\star$ that appears in the relation between the BR
scaling mass $M^\star$ and the Landau mass $m_N^\star$ involves
the space component measured by Gamow-Teller transitions. A
similar separation can be made for instance for the pion decay
constant $f_\pi^\star$ which will be discussed in Section
\ref{PT}.}.

We have arrived above at the Landau mass formula (\ref{mstar2})
given in terms of the BR scaling parameter $\Phi$. The Landau
mass differs from the BR scaling mass $M^\star$ due to the pion
contribution. From the Skyrme-type Lagrangian from which we have
deduced the scaling behavior, this means that there must be
contributions from other than the property of $f_\pi$. The only
other factor that enters in the skyrmion description is the
coefficient of the Skyrme quartic term $e$ or in terms of a
physical quantity, $g_A$. When this is taken into account, the
resulting expression is found to be~\cite{frsall}~\footnote{Since
the middle expression of (\ref{quenching}) involves the pionic
contribution $F_1 (\pi)$, one might think that $g_A^\star$
differs from $g_A$ merely due to the presence of the pion. This
interpretation is not correct. The Landau parameter $F_1(\pi)$
contains the density of states $N(0)=2k_F m_N^\star/\pi^2$ which
carries the Landau mass $m_N^\star$ and hence does involve BR
scaling. This is clearly seen in the last equality.}
 \be
\frac{m_N^\star}{M}=\sqrt{\frac{g_A^\star}{g_A}}\Phi \ee from
which we get
 \be \frac{g_A^\star}{g_A}=\left(1+\frac 13 F_1
(\pi)\right)^2 =\left(1-\frac 13 \tilde{F}_1
 (\pi)\Phi\right)^{-2}.\label{quenching} \ee
At nuclear matter density, this predicts
\be
 g_A^\star (\rho_0)\approx 1.\label{fixedpoint}
\ee


What does this mean physically? If we accept that the Landau mass
is a fixed point of the Fermi liquid theory, then the $g_A^\star$
must be a constant {\it defined} at the fixed point. It therefore
must correspond to decimating all the way down to the Fermi
surface, that is to say, sending the infrared cutoff $k_c$ in the
renormalization-group equation to zero. In this limit,
$g_A^\star$ must be proportional to the Landau parameter
$G_0^\prime$. This connection in infinite (translationally
invariant) matter is being investigated in~\cite{KBR2001}. What
this means in practice in finite nuclei in which the transition is
measured is not yet fully understood. But it appears reasonable
to suppose that applied to finite nuclei, what we obtained here
is an effective constant that should be used {\it when the
Gamow-Teller matrix element is computed within the valence shell
with the single-particle states undergoing the Gamow-Teller
transition restricted to the lowest-lying excitations on top of
the Fermi sea}. In other words, the $g_A^\star$ so obtained is a
fixed-point quantity.

An apt illustration of how this reasoning seems to be working is
given by the complete $0f1p$-shell Monte Carlo calculation of
Gamow-Teller response functions by Radha et al~\cite{koonin} and
also big shell-model diagonalization by Caurier et
al~\cite{langanka}. Remarkably, it is found that
$g_A^\star\approx 1$ is universally needed in these
calculations~\cite{koonin,zuker,presupernova} which may be taken
as a support for the fixed-point notion of the $g_A^\star$ .
\subsection{Evidences from ``On-Shell" Vector Mesons}
 \itt
The indirect evidences we have discussed above dealt with the
properties of hadrons that are way off-shell, with
$q_\mu^2\approx 0$. The masses and coupling constants that are
probed in such processes correspond to the {\it parameters} of
the effective Lagrangian and they would be near their physical
parameters only if the tree approximations were valid. One would
of course like to confront with measurements of the ``physical
hardrons" propagating in dense matter. In order to do this
quantitatively with a theory that implements BR scaling, one
would have to compute to higher orders, say, in chiral
perturbation suitably formulated to incorporate the ``sliding"
vacuum structure associated with BR scaling. This would of course
give rise to widths of the hadrons involved and to a further shift
in mass from that of BR scaling which is mean field. Such a
program has not yet been clearly formulated. Hence there are no
predictions in our approach. For low density, there are several
calculations that start from the zero density vacuum~\cite{weise}
that predict that while the $\rho$ meson may or may not be
shifted in mass, the $\omega$ will. The $\rho$ will broaden in
width in matter but the $\omega$ will remain narrow and may even
be bound in nuclear matter~\cite{marco}. The experimental finding
of this property of $\omega$ will confirm unambiguously the
observations made in the preceding and following sections.

Experiments are presently being performed to check all this at
various laboratories, notably at GSI and Jefferson Lab.
Specifically, the signals from heavy-ion collisions to be
discussed below address this issue although with results that are
compounded with temperature effects.
\begin{figure}[th]
\centerline{\epsfig{file=ozawaa.eps,width=2.63in}
\epsfig{file=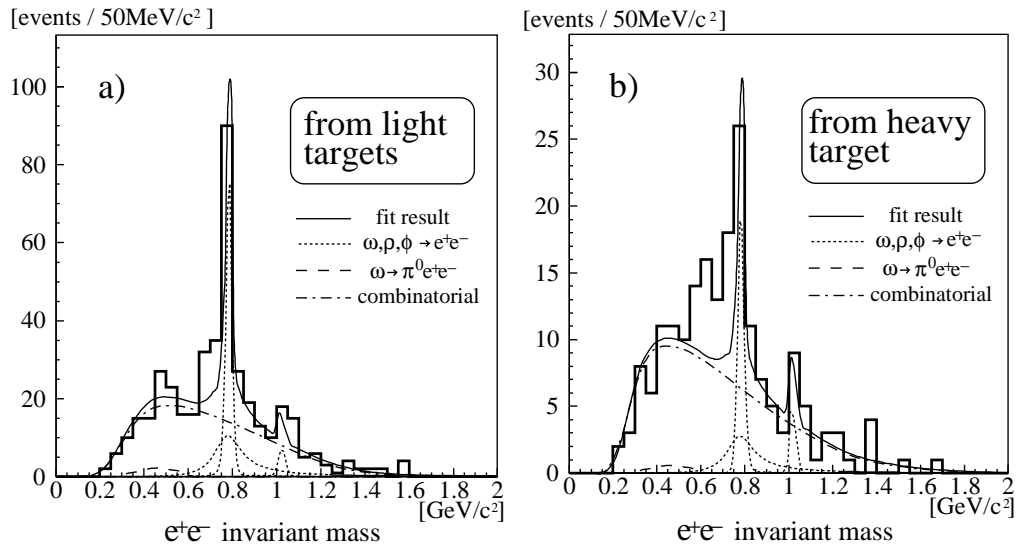,width=2.63in,angle=0}}
\caption{\small Distribution with invariant mass of $e^+ e^-$
pairs~\cite{ozawa}: a) from a carbon target; b) from a copper
target. The difference between histogram and thin solid line
represents the effect of medium (density) on vector-meson mass,
the $e^+e^-$ invariant mass being determined by the in-medium
vector-meson mass. \label{KEK}} \vskip 3mm
\end{figure}

Here we briefly comment on some recent experiments that seem to
indicate that the vector-meson masses do decrease with density in
nuclear matter or that at least the dropping masses are
compatible with the observations. One is the TAGX collaboration
result~\cite{lolos} on the process $^3$He$(\gamma,\rho^0)ppn$
with the tagged photons in the range 800--1120 MeV. The mass
shift of the $\rho^0$ meson reported was $\delta m_\rho=160\pm
35$ MeV. This is nearly the amount predicted for nuclear matter,
i.e., $m_\rho(1-\Phi)\approx 170$ MeV for $\Phi=0.78$ used above.
This seems to be a bit too much a shift for such a small nucleus.
It seems to involve other than just BR scaling. The other
experiment is a more recent measurement~\cite{ozawa} at KEK on
the invariant mass spectra of the $e^+ e^-$ pairs in the target
rapidity region of 12 GeV $p+A$ reactions with the nucleus $A=$ C
and Cu. The experimental result is given in Fig.\ref{KEK}. This
collaboration ``sees" the mass shift of the vectors in the
heavier nucleus indicative of a BR scaling. Although the precise
value of the mass shift is not determined in this work, it is
consistent with the expectation. We learned from private
communication~\cite{Kozawa} that with their kinematic
distribution, about 60\% of the $\rho$ mesons and about 10\% of
the $\omega$ mesons are estimated to decay inside Cu nucleus, so
most of the excess on the low-mass side of the $\omega$ peak is
from the $\rho$ mesons. The authors of \cite{ozawa} considered
the possibility of an in-medium increase in $\omega$ mass, but
this did not give a statistically significant effect. More data
and their analyses are forthcoming and will give a clearer insight
into the properties of the hadrons on the ``mass-shell," thereby
checking the BR scaled effective Lagrangian in more details.

 \section{DIRECT EVIDENCE FROM THE $(e,e^\prime p)$
\\ RESPONSE FUNCTIONS IN NUCLEI}\label{electron}
\setcounter{equation}{0} 
\renewcommand{\theequation}{\mbox{\ref{electron}.\arabic{equation}}}
 \itt
The evidences for BR scaling given so far are more or less
indirect. They are based on consistency with what we already know
of nuclear interactions. In this section, we discuss what we
consider to be the most direct evidence for BR scaling, namely,
the longitudinal and transverse electromagnetic response
functions in medium-heavy and heavy nuclei. Although the process
involves off-shell properties of the vector meson, we believe
that this evidence is cleaner and more direct than what we can
extract from presently available heavy-ion data -- including the
celebrated CERES data -- described below.
\begin{figure}[th]
 \centerline{\epsfig{file=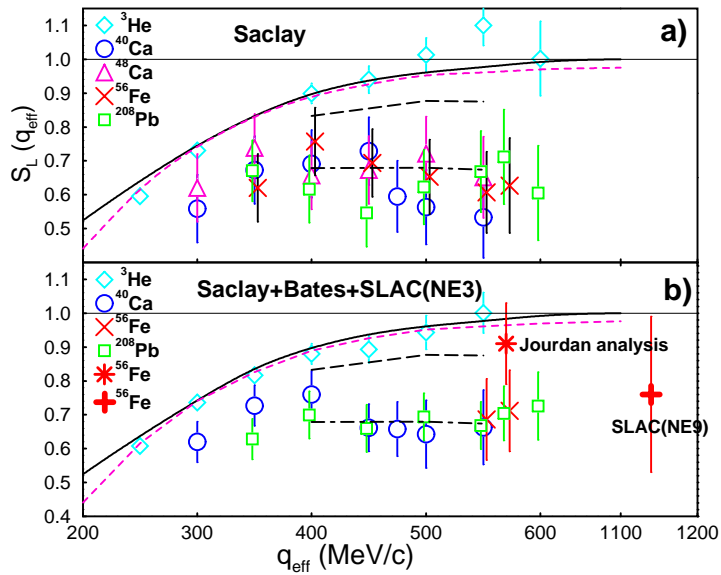,width=9.6cm,angle=0}}
\caption{\small The longitudinal response function $S_L$ as a
function of $q_{eff}$ using (a) Saclay data only and (b) Saclay
data plus SLAC NE3 and Bates data. The SLAC NE9's $^{56}$Fe
result is shown by a cross and that of the Jourdan analysis by a
star in (b). The theoretical curves are: the solid black curve
for nuclear matter without BR scaling and the dashed curve for
$^4$He; the long dashed curve is the same as the nuclear matter
one integrated within the experimental limits of $\omega$. The
dot-dashed curve is with BR scaling. See Section \ref{electron}
for details . \label{morgenstern}} \vskip 3mm
\end{figure}

Electron scattering which knocks nucleons out of nuclei have been
used for many years to tell us about the binding energy of
protons in nuclei and also to pin down the effective mass that the
proton has in the nucleus. To the extent that the electron
couples through vector mesons to the nucleons, these experiments
can also tell us about the {\it in-medium} properties of the
vector mesons. This was worked out first in a paper by Soyeur,
Brown and Rho~\cite{SBR1} following the initial suggestion in
\cite{BR89,BR90}. We follow here a more schematic
treatment~\cite{BRS} which brings out the main ideas more
clearly, and which is in semi-quantitative agreement with the
more detailed calculation.

Before going further, let us remark that the experimental
analysis has been thoroughly confused during the 1990's and only
now is being clarified, chiefly due to efforts of Joseph
Morgenstern.
A global analysis by Jourdan~\cite{jourdan} which relied on
three-dimensional numerical calculation of Onley et
al.~\cite{onley} indicated that medium effects appeared to be
absent, or at least the situation was unclear. The analyses
involved therein were unnecessarily complicated in view of the
fact that Yennie et al~\cite{yennie} had shown that the chief
distortion by the Coulomb field could be reliably taken into
account by shifting the momentum $\Bq$ to an effective momentum
 \be
\Bq_{eff}=\bf (1-V_c/|\Bq|)\label{effk}
 \ee
where $V_c$ is an effective Coulomb potential of the electron.
This is a semiclassical correction and had to be accompanied by
higher order corrections for a quantitative description of the
elastic electron scattering off nuclei~\cite{yennie} since the
wavefunctions oscillate rapidly. Even there the $\Bq_{eff}$ of
eq.(\ref{effk}) gave a semiquantitative description of the
distortion. Quasielastic scattering is very smooth as function of
position in the integrand, and one would expect the introduction
of $\Bq_{eff}$ should be adequate there. For a more recent
discussion on this matter, see Triani~\cite{triani}.

Before the linear accelerator was shut down in Saclay,
quasielastic experiments were performed on $^{12}$C and $^{208}$Pb
with a positron beam. In this case the sign of $V_c$ changes, and
the correction (\ref{effk}) to the momentum is of opposite sign.
Using the effective momentum approximation of eq.(\ref{effk}), it
has been shown~\cite{guye} that both electron and positron
scattering of $^{12}$C and $^{208}$Pb could be well described in a
consistent way. Introduction of $\Bq_{eff}$ handles what is
called the lowest order focusing effect in the theory. Whereas
Onley et al~\cite{onley} found higher-order focusing effects to be
important, introduction of these would ruin the consistency
between the electron and positron scattering. Of course without
substantial numerical work, it is not possible to check the Onley
et al's results.\vskip 0.3cm

$\bullet$ {\bf Experimental Situation}\vskip 0.3cm

Although as theorists we cannot go into the experimental
situation in detail, we can see that the early disagreement
between the MIT-Bates data and Saclay data changed when MIT
replaced their old scattering chamber, where background was
corrected for by simulation, by a new one. The updated MIT-Bates
data and the Saclay data are now in substantial agreement. The
analysis of the positron scattering introduced a new element. It
could be straightforwardly explained by the effective momentum
approximation (EMA). Kim et al~\cite{kimetal} found that
inclusion of higher order screening corrections in an approximate
treatment of electron Coulomb distortion in quasi-elastic
$(e,e^\prime)$ reaction was important. Indeed for forward
electron angles the low-energy side of the DWBA peak looks
similar to the plane-wave result. The authors suggest that the
phase factors effectively cancel the effect of the effective
momentum on the low $\omega$ side of the quasielastic peak.
However, if so, the next correction beyond the EMA would involve
the square of the projectile charge which would be the same for
electron and positron, negating the simple relation (\ref{effk}).
This would also mean that the difference between plane wave and
DWBA for positrons would be large. In any case, these matters can
be only conjectured because the Ohio University collaborations
have not been extended to positrons although it seems obvious that
this should be done.

 It is our belief that the whole situation, especially the
theoretical analysis, has been a real shambles, for which we take
partial responsibility. The Yennie, Boos and Ravenhall work was
so convincing that theorists should have insisted that
experiments were analyzed in the effective momentum
approximation.\vskip 0.3cm

$\bullet$ {\bf Theory with BR Scaling} \vskip 0.3cm

The mechanism for the change in quasielastic scattering can be
quite simply understood by considering the vector dominance model
in which the electron couples to the nucleon through the vector
meson. The vector mass enters into this coupling through
 \be
{\cal D}(m_V)
=\frac{m_V^2}{m_V^2+q^2}=\frac{1}{1+q^2/m_V^2}.\label{prop}
 \ee
Usually the numerator enters into the electron-$\rho$ meson
coupling and the denominator into the propagator, but the terms
should be put together as in eq.(\ref{prop}) because ${\cal
D}(m_V)$ must go to 1 as $q\rightarrow 0$ in order to conserve the
electron charge. If now $m_V\rightarrow m_V^\star\sim 0.8 m_V$
then ${\cal D}(m_V^\star)<{\cal D}(m_V)$.

The longitudinal response measures the charge density for which
the operator is
 \be
{\cal O}^L=\frac 12 +\frac 12 \tau_3.
 \ee
The $\omega$-meson couples to the $\frac 12$ and the $\rho$-meson
to the $\frac 12 \tau_3$. The scattering via the $\omega$-meson
exchange will be chiefly longitudinal in nature, since the
isoscalar magnetic moment, which gives the transverse response,
is small. Thus, the suppression in the longitudinal response is
essentially
 \be
F=\frac{1+q^2/m_\omega^2}{1+q^2/{m_\omega^\star}^2}.
 \ee
Here what figures is the BR scaling in the $\omega$ meson mass.

For the transverse response the main operator is
 \be
{\cal O}^T=\mu_V \frac{\epsilonb\cdot
[\Bsigma\times\Bq]}{2m_N}\tau_3,
 \ee
this response coupling chiefly to the isovector magnetic moment.
Now in medium $m_N\rightarrow m_N^\star$ as well as the
propagator changing, so the relevant factor is
 \be
F_T=\left(\frac{1+q^2/m_\rho^2}{1+q^2/{m_\rho^\star}^2}\right)
\frac{m_N}{m_N^\star}.\label{ftau}
 \ee
What enters here is the BR scaling in the nucleon mass. Now taking
$m_\rho^\star/m_\rho\sim m_N^\star/m_N$ according to the BR
scaling, one finds that the two factors in (\ref{ftau}) nearly
cancel each other in the range of momentum transfers $q\sim$
several hundred MeV in which the longitudinal and transverse
responses have been measured.

In Fig.\ref{morgenstern}, we show comparison of longitudinal data
with the theory of Soyeur, Brown and Rho. Also a point from
Jourdan's analysis using the Onley et al theory is shown. The
drop of the dash-dot line, which included the Soyeur et al medium
effects, below the Fabrocini and Fantoni solid line, which does
not include these effects, is very clear, an $\sim 20$ \% effect.

Both predicted and experimental effects on the transverse
scattering are small. This is an example where BR scaling has a
visible effect. Clearly the separation of longitudinal and
transverse components was essential to expose BR scaling.\vskip
0.3cm

$\bullet$ {\bf Comments}\vskip 0.3cm

The Soyeur-Brown-Rho treatment in ~\cite{SBR1} was more
complicated than our above schematic model because it built in
the quark structure of the nucleon at short distances,
essentially through inclusion of the chiral bag model. Since the
latter was not scaled, the effect of scaling in our schematic
model was somewhat diminished. Nonetheless they were quite large
as seen from the difference between the dashed line and the
stared point ``Jourdan analysis" and Soyeur et al's dash-dot line.

It should be noted that Saito, Tsushima and
Thomas~\cite{saitoetal} obtained within the quark-meson (QMC)
model results for the quasi-elastic electron scattering similar
to those of Soyeur, Brown and Rho.

In Section \ref{heavyion}, we will develop the evidence of
medium-dependent vector-meson masses from the dilepton and photon
experiments carried out at the SPS of CERN. Whereas the excess
dilepton can be as simply produced in the Rapp-Wambach scenario
as in the Brown-Rho scenario, the photons take us to higher
densities and temperature, beyond the chiral restoration
transition. We will develop a notion of complementarity between
the BR dropping mass scenario and quark-gluon plasma (QGP)
description for the photon treatment. As for quasielastic
electron scattering we need not go into QGP since the hadron
language is amply adequate. The dileptons we consider next fit in
also quite simply with the quasielastic electron scattering. Our
treatment in the next section will, however, allow us to unite
the low-energy sector with the chiral restoration region of
energies.
 \setcounter{equation}{0} 
\renewcommand{\theequation}{\mbox{\ref{PT}.\arabic{equation}}}
 \section{BR SCALING IN CHIRAL RESTORATION}\label{PT}
 \itt
One can approach the {\it phenomenology} of the chiral restoration
from a variety of different angles. We define ours by introducing
the role of BR scaling by a simple construction of the chiral
restoration transition as mean field in the Nambu-Jona-Lasinio
model. More precisely we consider the transition as mean field up
to {\it both} $\rho_c$ and $T_c$, the critical density and
temperature, although there may well be a density discontinuity
at the transition, making it first order. Most likely the
transition is second order for $\rho=0$, with a tri-critical point
on the phase boundary going towards finite density. What is
important for our discussion, however, is that masses of mesons
other than the pion go smoothly to zero in the chiral limit. The
simplest possible model for this is mean field NJL. We follow the
construction of Brown, Buballa and Rho~\cite{BBR}. Relevant parts
of this paper were reviewed in Brown and Rho~\cite{BRPR}. The
present description in NJL agrees with the vector manifestation
scenario of Harada-Yamawaki's hidden local symmetry theory (see
Section \ref{how}) and with the quark-hadron complementarity
picture given in Section \ref{denseqcd}.

\subsection{Bag Constant and Scalar Field Energy}
 \itt
Brown, Buballa and Rho showed that the Walecka nuclear mean field
theory at nuclear matter density could be connected with NJL mean
field theory at higher densities; e.g., the $\sigma$-field
coupling the quarks in NJL can be taken to be 1/3 of that to
nucleons in Walecka theory, \be
 g_{\sigma QQ}=g_{\sigma NN}/3\simeq 10/3.
\ee
 Furthermore, the scalar field energy $\frac 12 m_\sigma^2
\sigma^2$ plays the role of the bag constant $B$ in the chiral
restoration transition. On the quark-gluon side of the transition
the appearance of the bag constant corresponds to the
disappearance of the scalar field energy going upwards from below
through the transition. The magnitude of the effective bag
constant is only $\sim 1/2$ of that that would be given by the
trace anomaly: \be
 B=-\frac{\beta(g)}{8g} \la 0|(G_{\mu\nu}^a)^2|0\ra
\ee
 which to one loop (for $N_F=3$) is
\be
 B&=&\frac{9\alpha_s}{8\pi}\la 0|(G_{\mu\nu}^a)^2|0\ra =(245\,{\rm
 MeV})^4\no\\
 &=& 469\ {\rm MeV/fm^3}\simeq 2B_{eff}
\ee
 where $B_{eff}=\frac 12 m_\sigma^2 \sigma^2$. As found by
 Miller~\cite{miller}, only about 50\% of the glue is ``uncovered"
 in the chiral restoration transition. The other 50\% remains as
 ``covered" glue. In Shuryak et al~\cite{shuryak}, the soft glue
 disappears with the dynamically generated quark masses whereas
 the hard glue (``epoxy") is made up of instanton molecules which
 do not break chiral symmetry. The amount of trace anomaly which
 disappears at the transition gives $B_{eff}\simeq 235\, {\rm
 MeV/fm^3}$.

\subsection{``Nambu Scaling" in Temperature}
\begin{figure}[h]
\centerline{\epsfig{file=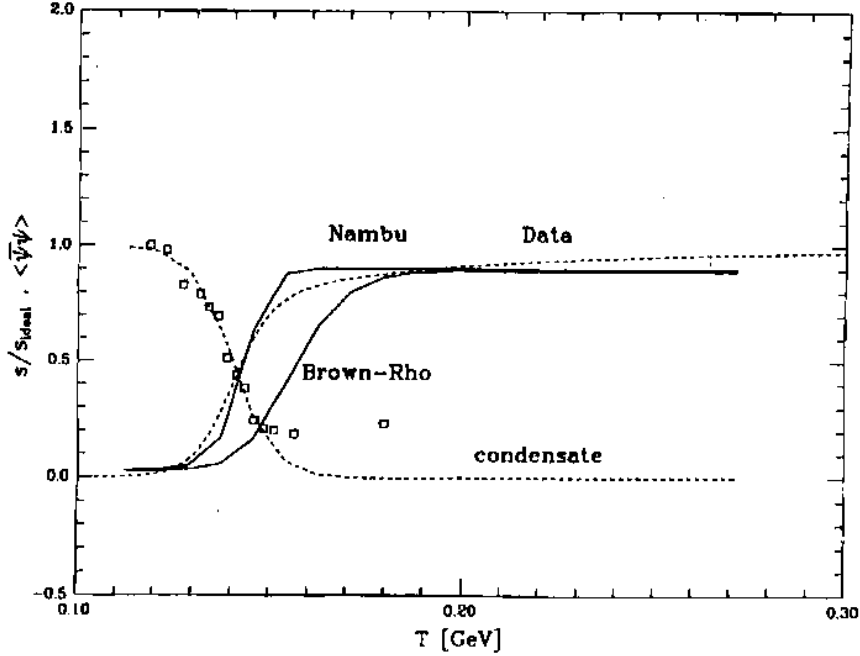,width=10cm,angle=90}}
 \caption{\small Comparison of entropy derived by Koch
and Brown from the lattice data on energy density (dahsed line)
and from dropping masses (full lines) calculated in the Nambu
scaling and in the naive Brown-Rho scaling. In the lower dashed
line for $\la\bar{\psi}\psi\ra$ the bare quark mass has been
taken out. }\label{kochbrown}
\end{figure}

 \itt
From the Gell-Mann-Oakes-Renner relation
 \be
 m_\pi^2=2m_q\la\bar{q}q\ra/f_\pi^2
 \ee
and taking $m_\pi$ not to scale with density, Brown and
Rho~\cite{BRPR} found hadron masses scaling as $f_\pi^\star$ with
$f_\pi^\star\propto |\la\bar{q}q\ra^\star|^{1/2}$.~\footnote{In
\cite{BR91}, using Skyrme's Lagrangian and looking at pions in
medium, the scaling $f_\pi^\star\propto
|\la\bar{q}q\ra^\star|^{1/3}$ was obtained. We now know that
because of chiral symmetry broken only slightly in nature, the
in-medium properties of a pion cannot be given correctly in the
same mean field argument as used for other non-Goldstone bosons.
Therefore it is safe to say that there is no accurate description
of the pionic property based on model considerations. Future
experiments will help in this direction.} It was then with some
surprise that Koch and Brown~\cite{kochbrown} found that the
temperature dependence of the entropy in the many-body system as
calculated in lattice gauge calculations~\cite{kogutL} could be
reproduced in the hadron sector best with the scaling \be
 \frac{m^\star}{m}\approx \frac{\la\bar{q}q\ra^\star}{\la\bar{q}q\ra}
\ee
 for hadron masses other than the pion mass. In Fig.\ref{kochbrown} is
shown how Nambu scaling behaves in the phase transition region.
This scaling follows from the mean field NJL model and is
referred to as ``Nambu scaling" although it is a generic feature
of the linear sigma model. The entropy is particularly suitable to
study, since quark mean fields drop out so it tells us directly
about the number of degrees of freedom.

The hadron language is suitable for low temperatures. As stressed
above and further elaborated later, we believe that our
description is ``dual" to a quark-gluon language as temperature
moves towards the chiral restoration temperature. In particular,
Koch and Brown limited their number of hadrons to 24, the number
of light quarks. Rather than just cutting the number at this
value, the same number can be consistently imposed by introducing
excluded volume effects~\cite{BJBP}. In any case the restriction
that the hadronic description with dropping masses go over, in
terms of the number of degrees of freedom, to the quark picture
at higher temperature, must be built in. Otherwise it would not
be possible to go beyond the Hagedorn temperature. The scale at
which the change-over from hadron to quark language is made is
one of convenience in accordance with the Cheshire Cat principle.
Presumably it is in the region of the chiral restoration
transition. This aspect becomes even more evident with the
``hadron-quark continuity" discussed in Section \ref{denseqcd}.

Finally note that the Nambu scaling that fits the entropy in the
hadronic description comes from mean field, i.e., the coupling
constant $g_{\sigma QQ}$ is not scaled. As long as we deal with
the $\sigma$, the fluctuation field connected with chiral
symmetry restoration, mean field seems to work very well, perhaps
at least approximately up to the chiral restoration transition.
In fact it was noticed in \cite{frsall} that the mean field
description of nuclear matter in terms of chiral Lagrangians
implemented with BR scaling required that the scalar coupling be
unscaled at least up to nuclear matter density. This will also be
observed in the problem of ``strangeness equilibration" discussed
in Section \ref{strangeness}.

\subsection{``Nambu Scaling" in Density}
 \itt
Given that there is nothing which dictates that the scaling in
temperature and that in density be the same, the question is
whether the Nambu scaling also works in density, at variance with
those found in \cite{BR91,BRPR} which are different from the
Nambu one. To answer this question let us first address the
question: How must the pion mass $m_\pi^\star$ change with
density in order to have Nambu scaling; i.e., hadron masses other
than that of the pion to scale as
 \be
\frac{m^\star}{m}=\frac{f_\pi^\star}{f_\pi}=\frac{\la\bar{q}q\ra^\star}
{\la\bar{q}q\ra}\ ?\label{i}
 \ee

From Gell-Mann-Oakes-Renner equation
 \be
f_\pi^2=2m_q\la\bar{q}q\ra/m_\pi^2\label{ii}
 \ee and since the quark mass does not change with density, we
 have
 \be
\frac{f_\pi^\star}{f_\pi}=\left(\frac{\la\bar{q}q\ra^\star}
{\la\bar{q}q\ra}\right)^{1/2}
\frac{m_\pi}{m_\pi^\star}.\label{iii}
 \ee
Now, to linear order in density we have
 \be
\frac{\la\bar{q}q\ra^\star}
{\la\bar{q}q\ra}=1-\frac{\rho\sigma_{\pi N}}{f_\pi^2
m_\pi^2}\approx 0.63 \ \ \ {\rm for}\ \
\rho=\rho_0.\label{nambuvalue}
 \ee
Setting (\ref{i}) equal to (\ref{ii}) at $\rho=\rho_0$ and using
(\ref{iii}), we find that in order for the Nambu scaling to apply
to density, the pion mass should scale up with density by a factor
 \be
\frac{m_\pi^\star}{m_\pi}=\left(\frac{\la\bar{q}q\ra}
{\la\bar{q}q\ra^\star}\right)^{1/2}\approx 1.26 \ \ \ {\rm for} \
\ \rho=\rho_0.
 \ee
It turns out that this increase seems to be supported empirically
although it is not clear whether the ``measured" mass is the same
quantity that appears in the in-medium GMOR relation. Experiments
which pinned down the increase in pion mass from bare mass by a
factor of $1.20 \sim 1.24$ were performed recently at
GSI~\cite{KEK} in the deeply bound states of $^{207}$Pb. Since
the outer part of the pion wave functions is at the densities
rather less than nuclear matter density, these factors may
represent the lower limits and a factor of 1.26 for $\rho=\rho_0$
seems quite reasonable. Thus the Nambu scaling {\it may} hold at
least near nuclear matter density.

We should mention however two caveats to this. One is that
two-loop chiral perturbation calculations find smaller values for
the mass shift. The calculation by Park et al~\cite{PJM} predicts
the pion mass at nuclear matter density to be at most $6\sim 7$
\% higher than that at zero density. This is a calculation where
the off-shell ambiguity associated with the pion in medium being
an off-shell quantity is minimized. For a pion field that
contains substantial off-shell ambiguity, one can easily obtain
the mass factor $1.20\sim 1.24$ but this seems to be
unreasonable. Another two-loop calculation by Kaiser and
Weise~\cite{kaiserweise} for the pion self-energy in asymmetric
nuclear matter obtains about 10 \% increase in mass of the pion
which is consistent with the result of Park et al. The other
caveat is that the scaling (\ref{i}) with the value
(\ref{nambuvalue}) at nuclear matter density would disagree with
the value expected from the experimental value (\ref{deltaglexp})
for $\delta g_l$, Eq.(\ref{chptd}), for which we have $\Phi
(\rho_0)=0.78$.

 In the above we have
established that NJL theory at mean field~\cite{BBR} works well
in describing the scaling of masses. In ref.\cite{BR96} it was
shown that the Walecka mean field theory in terms of nucleons can
be carried over with smooth change of parameters to a mean field
theory of constituent quarks at higher densities. The constituent
quark goes massless in the chiral limit with chiral restoration.
In Section \ref{complementarity}, we will suggest how this can be
understood in terms of color-isospin locking and hidden local
symmetry.

In low-energy nuclear physics we are used to introducing form
factors with couplings. These decrease with increasing scale. In
modern language, the form factors are there as a sort of
regularization that plays an important role in effective field
theories as discussed in Section \ref{EFTNucleus}. Why do we do
so well here without form factors ? We believe that the reason
lies in the special role that the scalar field $\sigma$ has in
chiral restoration. {\it A priori} it would seem reasonable to
continue the linear approximation for $\sigma$-exchange between
quarks
 \be
 \overline{Q}(q^2-{m^\star}_\sigma^2)^{-1} Q
 \ee
where $Q$ is the constituent quark, down to small $m_\sigma^\star$
where higher-order or nonlinear effects must enter. We believe
that these effects must be counterbalanced, to a large extent, by
form factors. As we develop in the next section, this does not
seem to be so with vector mesons, where form factors really do
not cut down the exchanges. As formulated more precisely in
Section \ref{complementarity}, the vector meson exchange at low
densities (i.e., flavor symmetry) changes into gluon exchange at
higher density (QCD symmetry) and asymptotic freedom assures that
their effects go to zero with increasing scale.

\setcounter{equation}{0} 
\renewcommand{\theequation}{\mbox{\ref{heavyion}.\arabic{equation}}}
 \section{SIGNALS FROM HEAVY-ION COLLISIONS}\label{heavyion}
 \itt
Extreme conditions of high temperature and/or high density are
expected to be created in relativistic heavy-ion collisions being
probed in RHIC at Brookhaven and to be probed in CERN's LHC. In
this section, we discuss how BR scaling manifests itself in
heavy-ion processes, leaving high-density zero-temperature
situation to later sections. Before going into the details, we
first define precisely the meaning of BR scaling in the context
that we shall use in this section.
 \subsection{Top-Down and Bottom-Up and How They Connect}
 \itt
There are effectively two ways of approaching BR scaling. The BR
scaling as first suggested in \cite{BR91} is a ``top-down" version
starting, using quark mean field, from a high temperature/density
regime wherein constituent quarks (or quasiquarks) are relevant
degrees of freedom, i.e., near the chiral transition point and
then extrapolating down to a low temperature/density regime
wherein hadrons are appropriate degrees of freedom. In contrast,
Rapp and Wambach and others~\cite{rapp} calculated medium
dependent hadron properties in hadronic variables in a
strong-coupling perturbation approach. We shall call this a
``bottom-up" approach. The advantage of the latter approach is
that one can resort to standard nuclear physics information using
phenomenological Lagrangians at tree order with more or less known
coupling constants. This approach has been found to be successful
in reproducing the medium-dependent $\rho$-meson properties
deduced from the CERN CERES dilepton experiments. In these
experiments. most of the $\rho$-mesons come from densities less
than nuclear matter density. In the low-density regime, the
phenomenological approach is justified by {\it construction} in a
well-prescribed way. The disadvantage of this approach is that as
density increases beyond nuclear matter density, the possible
vacuum change makes the tree order calculation unreliable because
of the strong coupling or more precisely a large anomalous
dimension which signals that one is fluctuating at a wrong ground
state and that it is preferable to shift to a different vacuum
even though no phase transitions are involved. This aspect was
associated above with differences in topological vacua.

In \cite{BLRRW,bKRBR,KRBR}, the top-down BR scaling and the
bottom-up Rapp-Wambach approach were combined into a unified
description. We will follow this description. We consider the
most important component of the Rapp-Wambach theory, i.e., the
introduction of the ``$\rho$-sobar" which is the excitation of
the $I=1$, $J=3/2$ $N^\star (1520)$ coupled together with the
nucleon-hole to the quantum numbers of the $\rho$. There are of
course other $N^\star$-hole states of the same quantum numbers
which would come in but here we shall focus on the dominant
component only. At zero density this isobar-hole state is at 580
MeV. The phenomenological Lagrangian coupling the ``elementary
$\rho$" and the isobar-hole $\hat{\rho}$ that we shall
use~\footnote{The non-covariant form of the Lagrangian is not
unique as pointed out in \cite{riskabrown}.} is
 \be
\calL_{\rho N^\star
N}^{s-wave}=\frac{f_\rho}{m_{\rho}}\psi_{N^*}^\dagger (q_0\vec
s\cdot\vec \rho_a - \rho_a^0\vec s\cdot \vec q) \tau_a \psi_N  +
~  h.c. \  , \label{LL1520} \ee where $f^2_{\rho N^\star
N}/4\pi\simeq 5.5$ from fits to photodisintegration data. Here
($\vec{q}$, $q_0$) are the $\rho$-meson momenta. we shall keep
only the $q_0$ term in our dilepton discussion. The two branches
of the $\rho$ spectral function can be located by solving
self-consistently the real part of the $\rho$-meson dispersion
relation (taking $\vec{q}=0$)
 \be
q_0^2=m_\rho^2 +{\rm Re}\Sigma_{\rho N^\star N} (q_0).
 \ee
The natural width of the \N1520 is supplemented at $\rho=\rho_0$
by an additional medium-dependent width of 250 MeV.

Including also the backward-going graph, the \N1520$N^{-1}$
excitation contributes to the self-energy of the $\rho$-meson at
nuclear matter density $n_0$ (in this section we denote density by
$n$ to avoid confusion with the $\rho$-meson))
 \be
\Sigma_{\rho N^\star N} (q_0)=f^2_{\rho N^\star N}\frac 83
\frac{q_0^2}{m_\rho^2} \frac{n_0}{4} \left(\frac{(\Delta
E)^2}{(q_0+i\Gamma_{tot}/2)^2-(\Delta E)^2}\right),\label{SE}
 \ee
where $\Delta E\simeq 1520-940=580$ MeV and
$\Gamma_{tot}=\Gamma_0 +\Gamma_{med}$ is the total width of the
\N1520. If we neglect any in-medium correction to the width of
the \N1520 we find two solutions, located at $q_0^-\simeq 540$
MeV and $q_0^+\simeq 895$ MeV. As $n$ increases, $\Sigma$ will
become larger and initially the $q_0$ of the lower branch will go
down. However, this downward movement will be halted once $q_0$
has decreased substantially. We note that already at $n=0$, the
\N1520 decays $\sim 15$ \% of the time to $N \rho$ indicating
that there is already mixing there. The mixing becomes stronger
with increasing $n$.
\begin{figure}
\vskip -2cm \centerline{\epsfig{file=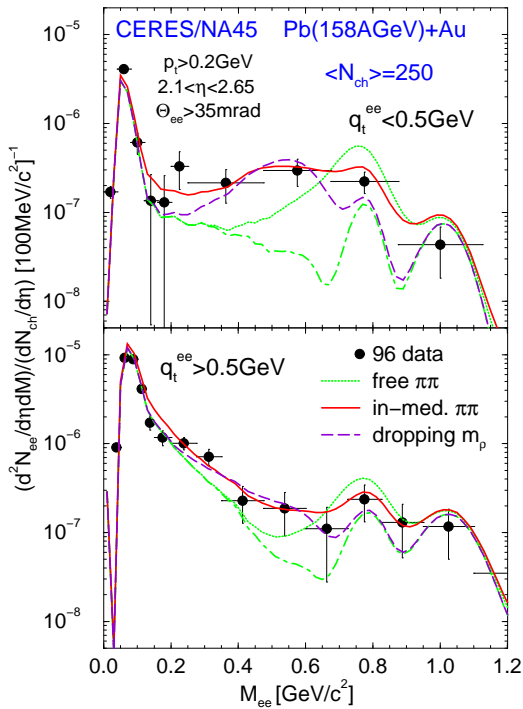,width=8cm}}
\caption{\small Dilepton data from central Pb (158 GeV/u)+Au
collisions compared with Rapp/Wambach (labelled as ``in-medium
$\pi\pi$") and Brown/Rho (labelled as ``dropping $m_\rho$") for
$q_t^{ee}<0.5$ GeV (upper) and $q_t^{ee}>0.5$ GeV (lower).
}\label{dilepton}
\end{figure}

 As noted, the Rapp-Wambach theory works well for the low
densities $n\lsim n_0$ important for the CERES experiments. We
show in Fig.\ref{dilepton} fits to the CERES spectra with both
Rapp-Wambach and Brown-Rho theories. In the BR approach, however,
the vacuum is ``sliding," that is, for each new density, one has a
new ``vacuum" to fluctuate around. Recalling the skyrmion
structure for given baryon number, we observe that ``vacua" are
solitonic in nature and cannot be connected by perturbation
theory. Thus, the vacuum at $n\neq 0$ has no continuous connection
with that at $n=0$. This means that with the Lagrangian
(\ref{LL1520}), the Rapp-Wambach theory can never approach the BR
theory as density increases. One can see from (\ref{SE}) that the
self-energy will stop bringing down the energy of the lower branch
of the $\rho$ as $(q_0/m_\rho)^2$ becomes smaller. This situation
is somewhat like the usual Walecka mean field theory where the
nucleon self-energy has a scalar density
$\rho_s=\psi^\dagger\gamma_0\psi$ in it. As $\rho_s$ goes to zero,
the self-energy which involves $\rho_s$ linearly does likewise, so
the nucleon mass can go to zero only at $n=\infty$. However with
the introduction of negative energy states in Nambu-Jona-Lasinio
theory, the mass of the constituent quark goes smoothly to zero at
a density $n_c\sim 3n_0$~\cite{BBR}. As argued in
\cite{bKRBR,KRBR}, this can happen with (\ref{SE}) {\it only if}
we replace there the factor $(q_0/m_\rho)^2$ by $\sim \epsilon
(q_0/m^\star_\rho)^2$. The unknown factor $\epsilon$ can be
adjusted to give the same critical density as in NJL model. It is
plausible that this change-over is related to the requirement in
the Harada-Yamawaki scheme which is not present in few-order
perturbation theory that the phase transition coincide with
Georgi's vector-limit fixed point. As mentioned, the Harada-Sasaki
work~\cite{harada-sasaki} in temperature indeed suggests that {\it
the Wilsonian matching with QCD with hidden gauge invariance
requires the parametric $\rho$ mass to have an ``intrinsic" $T/n$
dependence.} Once one accepts this, what follows turns out to be
quite reasonable: with $\epsilon=1$, $m_\rho^\star$ drops to zero
at the same density as in the NJL model, $n_c\sim 3n_0$.

\subsection{Distinguishing Rapp/Wambach and Brown/Rho}
\itt
 The modified sobar picture obtained by replacing $m_\rho$ by
$m_\rho^\star$ as described above still differs in detail from the
BR scaling picture of 1991 even though they describe more or less
equally well the $m_\rho^\star\rightarrow 0$ limit, because at
any density less than $n_c$ there are two branches of the $\rho$
quantum number. However as $n\rightarrow n_c$ the width of the
lower branch goes to zero and all of the spectral strength goes
into it so we do recover the original BR scaling. We will
encounter this feature once more in Section \ref{complementarity}
from the consideration of hidden local symmetry and
color-flavor-locking in QCD.

At densities $n\lsim n_0$, the lower $\rho$ branch most likely
becomes very wide because of the large number of open channels
that it couples to, which has led Rapp and Wambach to interpret
their fit to the dilepton data from the broadening of the $\rho$,
the lower parts of the spectral function contributing importantly
to the dilepton production because of the greater Boltzmann
factors.

So how do we distinguish between Rapp/Wambach and Brown/Rho;
equivalently between broadening and movement towards chiral
restoration? Formally we do not violate gauge invariance or any
other invariance by using $m_\rho^\star$ rather than $m_\rho$. It
is really up to nature to decide.

In a recent paper, Alam, Sarkar, Hatsuda, Nayak and
Sinha~\cite{hatsuda-photon} have suggested that nature has
decided for Brown/Rho~\footnote{In fact Alam et al suggest two
scenarios that can describe the photons, one being the hadronic
description with BR scaling and the other one in which
quark-gluon-plasma (QGP) is initially produced at a temperature
$T_i=196$ MeV. One has a mixed phase of QGP and hadronic matter
which persists down to a transition to hadronic matter at
$T_c=160$ MeV. In a later paper~\cite{sarkar}, Sarkar et al find
that for relatively central collisions, the dilepton yield can be
described by either of the two scenarios, BR scaling or one in
which the QGP is originally produced as above. These authors find
that ``as of yet, it has not been possible to explain the observed
low-mass enhancement of dileptons measured in the Pb+Au collisions
as well as in the S+Au collisions at the CERN SPS in a scenario
which does not incorporate in-medium effects in the vector meson
mass."

As noted earlier, we believe that two descriptions, BR scaling
and the one in which the QGP is initially produced in a mixed
phase, to be ``dual" as long as the number of degrees of freedom
in the former is constrained to equal that in the latter. The
mixed phase in the scenario with the QGP formation could,
however, be just the part of a second-order phase transition in
the BR scaling scenario. In the latter the energy is used up,
going down in temperature through the phase transition, in
building up the scalar field energy as the hadrons gain back
their masses. Most of the entropy decrease will come from the
higher mass hadrons disappearing as their mass is increased, due
to decreasing Boltzmann factors, rather than through expansion.
It seems therefore reasonable that the various freeze-out species
exhibit thermal equilibration at the chiral restoration
temperature, the one at which hadron masses go to zero in the
chiral limit in the BR scaling scenario, since the system spends
a long time there.

Although the dropping masses with BR scaling and production of
QGP in mixed phase at $T\sim 200$ MeV may be dual around the
phase transition, it is excluded that the latter scenario
describe the dropping masses found in $\sim 300$ MeV inelastic
electron scattering, which we described in the previous section,
or the many indirect evidences we described earlier in our
review. In other words, the duality is useful only at higher
scales, near and above the chiral restoration transition. One
should remember the B\'eg-Shei theorem, that the nature of
symmetry realization (i.e., Wigner-Weyl vs. Nambu-Goldstone) is
irrelevant to discussing the short-distance
symmetry~\cite{beg-shei}.}. This work is extended in Alam et
al~\cite{alametal}. In these papers, the calculations were based
on mesonic processes such as $\pi\rho\rightarrow \pi\gamma$,
$\pi\pi\rightarrow \rho\gamma$, etc. leading to thermal
production of photons. The observation then was that when
increasing the width of the $\rho$ meson to $\sim 1$ GeV in order
to take into account medium effects, the photon yield decreases
as compared to the case using the free $\rho$ width, thus
underestimating the spectra. However these calculations neglect
the contributions to the photons coming from some of the
processes that increase the width, e.g., the $\rho\rightarrow
N^\star (1520) N^{-1}$ excitation (the collective
``$\rho$-sobar") which provides the lower-mass $\rho$ excitation
in the Rapp/Wambach calculation. This contributes to the photon
through the $N^\star (1520)\rightarrow N\gamma$ decay.
Calculations in progress (by C. Gale, M.A. Hadasz and R. Rapp)
show that these baryonic processes contribute importantly so the
conclusions of \cite{hatsuda-photon, alametal} may be premature.

The bottom-up approach of Rapp and Wambach should be reliable at
low density since it is based on phenomenological Lagrangians
constructed in standard nuclear physics with known properties. Our
principal argument is that this picture should cede to the
Brown-Rho picture at higher densities replacing $m_\rho$ by
$m_\rho^\star$ in scaling the Lagrangian required to be
consistent with Harada-Yamawaki's vector manifestation. Of course
much more detailed experiments will be needed to confirm this
scenario.

%

 \section{``BROAD-BAND EQUILIBRATION" OF\\
 STRANGENESS IN HEAVY-ION
COLLISIONS}\label{strangeness}
\setcounter{equation}{0} 
\renewcommand{\theequation}{\mbox{\ref{strangeness}.\arabic{equation}}}
 \subsection{Kaons and Chiral Symmetry}
 \itt
The strange quark is massive compared with the light (u, d)
quarks, with $m_s\sim 150$ MeV. This gives the kaon a mass of
$\sim 1/2$ GeV. This is a manifestation of both explicit and
spontaneous chiral symmetry breaking. Unlike the previous cases
where the decondensing of the condensate $\la\bar{q}q\ra$ -- that
is, restoring the spontaneously broken chiral symmetry -- enters
directly into the phenomena we were looking at, here it is the
``rotating away" of the explicit chiral symmetry breaking
intricately tied in with the decondensing of $\la \bar{q}q\ra$
that is principally manifested in the behavior of kaons in dense
medium. This section deals with how this aspect is probed in the
kaon production experiments at GSI. Based on work done
recently~\cite{broadband}, we will develop, in conjunction with BR
scaling, the notion of ``broadband equilibration" in heavy-ion
processes and suggest the vector decoupling in dense medium.
\subsection{Equilibration vs. Dropping Kaon Mass}
 \itt
 Following work by Hagedorn~\cite{hagedorn} on production of
 anti-$^3$He, Cleymans et al~\cite{cleymans} have shown that for low
 temperatures, such as found in systems produced at GSI,
 strangeness production is strongly suppressed. The abundance of
 $K^+$ mesons, in systems assumed to be equilibrated, is given by
 \cite{cleymans2},
 \be
 n_{\tiny{K^+}}\sim e^{-E_{\tiny{K^+}}/T}\ \
 V\left\{g_{\bar K}\int\frac{d^3p}{(2\pi)^3}
 e^{-E_{\bar K}/T}+g_\Lambda\int \frac{d^3p}{(2\pi)^3}
 e^{-(E_\Lambda-\mu_B)/T}\right\}.\label{K+}
 \ee
 Here the $g$'s are the degeneracies. Because strangeness must be
 conserved in the interaction volume $V$, assumed to be that of
 the equilibrated system for each $K^+$ which is produced, a
particle of ``negative strangeness"~\footnote{By ``negative
strangeness" we are referring to the negatively charged strange
quark flavor. The positively charged anti-strange quark will be
referred to as ``positive strangeness."} containing ${s}$, say,
$\bar K$ or $\Lambda$, must also be produced, bringing in the
$\bar K$ or $\Lambda$ phase space and Boltzmann factors. The $K^+$
production is very small at GSI energies because of the low
temperatures which give small Boltzmann factors for the $\bar K$
and $\Lambda$ in addition to the small Boltzmann factor for the
\Kp. Note the linear dependence on interaction volume which
follows from the necessity to include $\bar K$ or $\Lambda$ phase
space.

In an extensive and careful analysis, Cleymans, Oeschler and
Redlich~\cite{cleymans2} show that measured particle multiplicity
ratios $\pi^+/p$, $\pi^-/\pi^+$, $d/p$, $K^+/\pi^+$, and
$K^+/K^-$ -- but not $\eta/\pi^0$ -- in central Au-Au and Ni-Ni
collisions at (0.8-2.0)A GeV are explained in terms of a thermal
model with a common freeze-out temperature and chemical
potential, if collective flow is included in the description. In
other words, a scenario in which the kaons and anti-kaons are
equilibrated appears to work well. This result is puzzling in
view of a recent study by Bratkovskaya et al~\cite{brat} that
shows that the $K^+$ mesons in the energy range considered would
take a time of $\sim 40$ fm/c to equilibrate. We remark that this
is roughly consistent with the estimate for higher energies in the
classic paper by Koch, M\"uller and Rafelski~\cite{koch} that
strangeness equilibration should take $\sim 80$ fm/c. Such
estimates have been applied at CERN energies and the fact that
emergent particle abundances are described by Boltzmann factors
with a common temperature $\sim 165$ MeV~\cite{peter0} has been
used as part of an argument that the quark/gluon plasma has been
observed.

We interpret the result of \cite{cleymans2} as follows. Since
free-space masses are used for the hadrons involved, Cleymans et
al~\cite{cleymans2} are forced to employ a $\mu_B$ substantially
less than the nucleon mass $m_N$ in order to cut down $\Lambda$
production as compared with $K^-$ production, the sum of the two
being equal to $K^+$ production. This brings them to a diffuse
system with density of only $\sim \rho_0/4$ at chemical
freeze-out. But this is much too low a density for equilibration.

We shall first show how this situation can be improved by
replacing the $K^-$ mass by the \Km energy at rest $\wkm
\equiv\omega_-(k=0) < m_K$. (The explicit formula for
$\omega_\pm$ is given later, see eq. (\ref{energy}).) In doing
this, we first have to reproduce the $K^+$ to $K^-$ ratio found
in the Ni + Ni experiments~\cite{menzel}:
 \be
n_{K^+}/n_{K^-}\simeq 30.\label{ratio}
 \ee
Cleymans et al reproduce the earlier smaller ratio of $21\pm 9$
with $\mu_B=750$ MeV and $T=70$ MeV. How this or rather
(\ref{ratio}) comes out is easy to see. The ratio of the second
term on the RHS of eq.(\ref{K+}) to the first term is roughly the
ratio of the exponential factors multiplied by the phase space
volume
 \be
R=\frac{g_\Lambda}{g_{\bar K}} \left(\frac{\bar p_\Lambda}{\bar
p_{\bar K}}\right)^3
\frac{e^{-(E_\Lambda-\mu_B)/T}}{e^{-E_{\tiny{K^-}}/T}}\approx
\left(\frac{m_\Lambda}{m_{\bar K}}\right)^{3/2}
\frac{e^{-(m_\Lambda-\mu_B)/T}}{e^{-m_{\tiny{K^-}}/T}}\approx 21
 \label{R} \ee
where we have used $g_\Lambda\approx g_{\bar K}\simeq 2$, $(\bar
p)^2/2 m\simeq \frac 32 T$, $E_\Lambda=1115$ MeV and
$E_{\tiny{\bar K}}=495$ MeV. Inclusion of the $\Sigma$ and $\Xi$
hyperons would roughly increase this number by 50\% with the
result that the ratio of $K^-$ to $\Lambda$, $\Sigma$, $\Xi$
production is~\footnote{In order to reproduce this result with
$\mu_B=750$ MeV and $T=70$ Mev within our approximation, we have
assumed only the $\Sigma^-$ and $\Sigma^0$ hyperons to
equilibrate with the $\Lambda$. This may be correct because the
$\Sigma^+$ and $\Xi$ couple more weakly. Inclusion of the latter
could change our result slightly. Probably they should be
included in analysis of the AGS experiments at higher energies
where they would be more copiously produced.}
 \be
\frac{n_{K^-}}{n_{\Lambda+ \Sigma+ \Xi}}\simeq 1/32.
 \ee
Since a $K^+$ must be produced to accompany each particle of one
unit of strangeness (to conserve strangeness flavor), we then have
 \be
n_{K^+}/n_{K^-}\sim 33.
 \ee
This is consistent with the empirical ratio (\ref{ratio}). It
should be noted that had we set $\mu_B$ equal to $m_N$, we would
have had the $K^+$ to $K^-$ ratio to be $\sim 280$ because it
costs so much less energy to make a $\Lambda$ (or $\Sigma$)
rather than $K^-$ in this case. In other words the chemical
potential $\mu_B$ is forced to lie well below $m_N$ in order to
penalize the hyperon production relative to that of the $K^-$'s.

One can see from Fig.5 in Li and Brown~\cite{LB} that without
medium effects in the $K^-$ mass, the $K^+/K^-$ ratio is $\sim
100$, whereas the medium effect decreases the ratio to about 23.
This suggests how to correctly redo the Cleymans et al's analysis,
namely, by introducing the dropping $K^-$ mass into it.

In the next subsection we show that positive strangeness
production takes place chiefly at densities greater than
$2\rho_0$. As the fireball expands to lower densities the amount
of positive strangeness remains roughly constant. The number of
\Kp's is such as to be in equilibrium ratio $K^+/\pi^+$ with the
equilibrated number density of pions at $T=70$ MeV, $n_\pi\approx
0.37\, T_{197}^3\, {\rm fm}^{-3}$. Only in this sense do the \Kp's
equilibrate. It will be noted, however, that with the $T$ of 70
MeV, $\mu_B$ is chosen so that the empirical number of hyperons
are produced. Since the number of \Kp is one greater than that of
the hyperons (including the \Km in the negative strangeness), it
will also be the apparent equilibrated ratio for this $\mu_B$ and
$T$. Thus, with these two latter parameters the $\pi$, \Kp and
\Km can be put into apparent equilibration. It is fairly easy for
the nucleons to equilibrate with the pions at the given
temperature because of the strong interaction.

It is amusing to note that the ``equilibrated ratio" of $\sim 30$
for the $n_{K^+}/n_{K^-}$ holds over a large range of densities
for $T=70$ MeV, once density-dependent \Km masses are introduced,
in that the ratio $R$ of (\ref{R}) is insensitive to density.
(Remember that because of the small number of \Km's, the number
of \Kp's must be nearly equal to the number of hyperons,
$\Lambda$, $\Sigma^-$ and $\Xi$, in order to conserve
strangeness.) This insensitivity results because $\wkm $
decreases with density at roughly the same rate as $\mu_B$
increases. We can write $R$ of (\ref{R}), neglecting possible
changes in $T$ and $m_{\bar{K}}$ in our lowest approximation,
as~\footnote{To be fully consistent, we would also have to
consider the medium modifications of the $\Lambda$ and \Kp
properties. These and other improvements are left out for our
rough calculation.}
 \be
R=\left(\frac{m_\Lambda}{m_{\bar{K}}}\right)^{3/2} e^{(\mu_B+\wkm
)/T}\, e^{-m_\Lambda/T}.\label{Rp}
 \ee
As will be further stressed later, the most important point in
our arguments is that {\it the $\mu_B+\wkm $ is nearly constant
with density}. This is because whereas $\mu_B$ increases from 860
MeV to 905 MeV as $\rho$ goes
from $1.2\rho_0$ to $2.1\rho_0$, $\wkm$ decreases from 380 MeV to
332 MeV, the sum $\mu_B+m^\star_{\tiny{K^-}}$ decreasing very
slightly from 1240 MeV to 1237 MeV. Indeed, even at
$\rho=\rho_0/4$, $\mu_B+m^\star_{\tiny{K^-}}\sim 1218$ MeV, not
much smaller.

We believe that the temperature will change only little in the
region of dropping masses because in a consistent evolution
(which we do not carry out here) the scalar field energy
$m_\sigma^2\sigma^2/2$ in a mean field theory plays the role of an
effective bag constant. In ref.\cite{BBR} (which was reviewed in
Section \ref{PT}) this is phrased in terms of a modified Walecka
theory, \be B_{eff}=\frac12 m_\sigma^2\sigma^2\Rightarrow\frac12
m_\sigma^2 (M_N/g_{\sigma NN})^2, \ee the $\sigma$ going to
$M_N/g_{\sigma NN}$ as the nucleon effective mass goes to zero
with chiral restoration. Most of the energy with compression to
higher densities goes with this effective bag constant, estimated
\cite{BBR} to be $\sim 240$ MeV/fm$^3$, rather than heat, mocking
up the behavior of a mixed phase with constant temperature.
Moreover at $\rho=2\rho_0$ where $m_N^\star$ may be $\sim 0.5
m_N$, only about 25\% of the bag constant $B$ may have been
achieved, so there may be some increase in temperature. We shall,
at the same level of accuracy, have to replace $m_{\bar{K}}$ in
the prefactor of eq.(\ref{Rp}) by $\wkm$. We adjust the increase
in temperature so that it exactly compensates for the decrease in
prefactor so that the \Kp/\Km ratio is kept the same, as required
by experiment~\cite{menzel}. We then find that the temperature at
$\rho=2\rho_0$ must be increased from 70 to 95 MeV. This is
roughly the change given by that in inverse slopes of \Km and \Kp
transverse momentum distributions found in going from low
multiplicities to high multiplicities~\cite{menzel}.

In any case, we see that $R$ will depend but little on density.
This near cancellation of changes in the factors is fortunate
because the $K^- +p \leftrightarrow \Lambda$ reaction, operating
in the negative strangeness sector, is much stronger than the
positive strangeness reactions, so the former should equilibrate
to densities well below $2\rho_0$ and we can see that the
``apparent equilibration" might extend all the way down to
$\rho\sim \rho_0/4$.

The near constancy of $R$ with density also explains the fact the
\Kp/\Km ratio does not vary with centrality~\footnote{We are
grateful to Helmut Oeschler for pointing this out to us.}.
Although $R$ is the ratio of $\Lambda$'s to \Km's, both of which
are in the negative strangeness sector, nonetheless, the number
of \Kp's must be equal to the sum of the two and since the
$\Lambda$'s are much more abundant than the \Km's, $R$
essentially  represents the \Kp/\Km ratio.

\begin{figure} \centerline{\epsfig{file=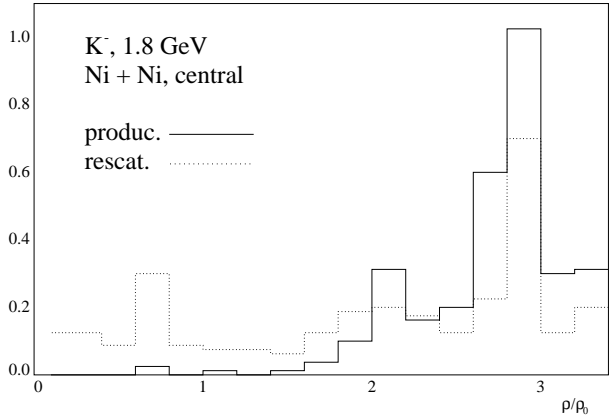,width=8cm}}
\caption{\small Calculations by Bratkovskaya and Cassing (private
communication) which show the density of origin and that of the
last interaction of the $K^-$ mesons.}\label{nini}
\end{figure}
Detailed transport results of Bratkovskaya and Cassing (see
Fig.\ref{nini}) show the last scattering of the detected \Km to be
spread over all densities from $\rho_0/2$ to $3\rho_0$, somewhat
more of the last scatterings to come from the higher density.
This seems difficult to reconcile with a scenario of the \Km
numbers being decided at one definite density and temperature,
but given our picture of dropping masses, one can see that the
\Kp/\Km ratio depends little on density, that is, on $\mu_B$ at
which the \Km last interacts. In any case we understand from our
above argument that the apparent density of equilibration can be
chosen to be very low in a thermal description and still get more
or less correct \Kp/\Km ratio.
\subsection{The Equilibrium \Kp/$\pi^+$ Ratio}
 \itt
We show here how our argument that gives a correct $K^+/K^-$ ratio
can reproduce the $K^+/\pi^+$ ratio. Let us leave $T=70$ MeV and
choose $\rho\sim 2\rho_0$ as educated guesses. We are thereby
increasing the equilibration density by a factor $\sim 8$. We
then calculate the baryon density for this $\mu_B$ and $T=70$ MeV
and find $\rho=2\rho_0$ which checks the consistency.

 According to Brown et al~\cite{BKWX} the $K^+$ production under
 these conditions will come chiefly from $BB\rightarrow N\Lambda
 K$, with excited baryon states giving most of the production.
 {}From the solid curve for $\rho/\rho_0=2$ in fig.9 of \cite{BKWX}
 we find $\langle \sigma v\rangle\sim 2\times 10^{-3}\ {\rm mb}=
 2\times 10^{-4}\ {\rm fm}^2$. The rate equation reads
  \be
\delta \Psi_K=\frac 12 (\sigma_{BB}^{BYK} v_{BB}) n_B^2\simeq
\frac 12 (2\times 10^{-4}\ {\rm fm}^2)\rho_B^2=dn/dt=\frac
19\times 10^{-4}\ {\rm fm}^{-4}\label{A1}
 \ee
where $B$, $Y$ and $K$ stand, respectively, for baryon, hyperon
and kaon. For this, we have taken $\langle \sigma_{BB}^{BYK}
v_{BB}\rangle$ from fig.9 of \cite{BKWX} and $\rho=2\rho_0=\frac
13\ {\rm fm}^{-3}$. Choosing a time $t=10\ {\rm fm}/c$ we obtain
 \be
n_{\tiny{K^+}}\simeq\delta \Psi_K\ t= \frac 29\times 10^{-3}\ {\rm
fm}^{-3}.\label{A2}
 \ee
Now equilibrated pions have a density
 \be
n_\pi=0.37 (T/197{\rm MeV})^3\ {\rm fm}^{-3}=0.016\ {\rm
fm}^{-3}\label{A3}
 \ee
for $T=70$ MeV. From (\ref{A2}) and (\ref{A3}) we get
 \be
n_{\tiny{K^+}}/n_{\pi^+}\simeq 0.0069\label{A4}
 \ee
 which is slightly below the ``equilibrated value" of 0.0084 of Table 1 of
 Cleymans et al~\cite{cleymans2}. Production of \Kp by pions may
 increase our number by $\sim 25$\%~\cite{ko2}.

Our discussion of \Kp production is in general agreement with
earlier works by Ko and collaborators~\cite{ko2,ko1}. In fact, if
applied at the quark level, the vector mean field is conserved
through the production process in heavy-ion collisions, so it
affects only the strangeness condensation in which there is time
for strangeness non-conservation. These earlier works establish
that at the GSI energies the \Kp content remains roughly constant
once the fireball has expanded to $\sim 2\rho_0$, so that in this
sense one can consider this as a chemical freeze-out density.

It should be noted that in the papers \cite{cassing,BKWX,ko2,ko1}
and others, the net potential -- scalar plus vector -- on the
\Kp-meson is slightly attractive at $\rho\sim 2\rho_0$ even
though the repulsive vector interaction is not decoupled. A hint
for such change-over was noted already at nuclear matter density
by Friedman, Gal and Mares~\cite{friedman}. Since in our top-down
description the vector interaction can be thought of as applied
to the quark (matter) field in the \Kp, the total vector field on
the initial components of a collision is then the same as on the
final ones, so the vector mean fields have effectively no effect
on the threshold energy of that process. Our proposal here is
that the vector mean field on the \Kp should be below the values
used by the workers in \cite{cassing,BKWX,ko2,ko1} due to the
decoupling. However, in comparison with \cite{cassing}, our total
potential on the \Kp at $\rho=2\rho_0$ is $\sim -85$ MeV, as
compared with their $\sim -30$ MeV. We have not redone the
calculation of \cite{BKWX} to take into account this difference.

\subsection{The Top-Down Scenario of $K^\pm$ Production}
 \itt
Brown and Rho~\cite{BR96} discussed fluctuations in the kaon
 sector in terms of a simple Lagrangian
 \be
 \delta {\cal
L}_{KN}=\frac{-6i}{8{f^\star}^2}(\overline{N}\gamma_0
N)\overline{K}\del_t K +
\frac{\Sigma_{KN}}{{f^\star}^2}(\overline{N}N)\overline{K}K+\cdots\equiv
{\cal L}_\omega +{\cal L}_\sigma+\cdots\label{kaonL} \ee
 It was suggested there that at high densities, the constituent
quark or quasi-quark  description can be used with the
$\omega$-meson coupling to the kaon viewed as a {\it matter
field} (rather than as a Goldstone boson). Such a description
suggests that the $\omega$ coupling to the kaon which has one
non-strange quark is 1/3 of the $\omega$ coupling to the nucleon
which has three non-strange quarks. The ${\cal L}_\omega$ in the
Lagrangian was obtained by integrating out the $\omega$-meson as
in the baryon sector. We may therefore replace it by the
interaction
 \be
V_{K^\pm}\approx \pm \frac 13 V_N.
 \ee
In isospin asymmetric matter, we shall have to include also the
$\rho$-meson exchange~\cite{BR96} with the vector-meson coupling
treated in the top-down approach.

For the top-down scenario, we should replace the chiral
Lagrangian (\ref{kaonL}) by one in which the ``heavy" degrees of
freedom figure explicitly. This means that
$\frac{1}{2{f^\star}^2}$ in the first term of (\ref{kaonL})
should be replaced by ${g^\star}^2/{m_\omega^\star}^2$ and
$\frac{\Sigma_{KN}}{{f^\star}^2}$ in the second term by $\frac 23
m_K \frac{{g_\sigma^\star}^2}{{m_\sigma^\star}^2}$ assuming that
both $\omega$ and $\sigma$ are still massive. We will argue below
that while the $\omega$ mass drops, the ratio
${g^\star}^2/{m_\omega^\star}^2$ should stay constant or more
likely decrease with density and that beyond certain density
above nuclear matter, the vector fields should decouple. On the
other hand, $g_\sigma$ is not scaled in the mean field that we are
working with; the motivation for this is given in Brown, Buballa
and Rho~\cite{BBR} who construct the chiral restoration
transition in the mean field in the Nambu-Jona-Lasinio model. Thus
 \be
\frac{\Sigma_{KN}}{{f^\star}^2}\approx \frac 23 m_K
\frac{g_\sigma^2}{{m_\sigma^\star}^2}.\label{sigmatop}
 \ee
In this framework, $m_\sigma^\star$ is the order parameter for
chiral restoration which drops {\`a la}\, BR scaling~\cite{BR91}:
 \be
\frac{m_\sigma^\star}{m_\sigma}\equiv \Phi (\rho)\simeq
\frac{1}{1+y\rho/\rho_0}
 \ee
with $y\simeq 0.28$, at least for $\rho\lsim
\rho_0$~\footnote{$y$ may well be different from this value for
$\rho>\rho_0$. In fact the denominator of $\Phi (\rho)$ could
even be significantly modified from this linear form. At present
there is no way to calculate this quantity from first
principles.}.

\begin{figure} \centerline{\epsfig{file=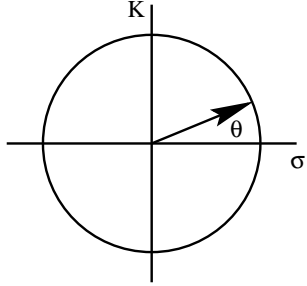,width=4cm}}
\caption{\small Projection onto the $\sigma, K$ plane. The
angular variable $\theta$ represents fluctuation toward kaon mean
field.\label{Ksigma}}
\end{figure}

Once the vector is decoupled, a simple way to calculate the
in-medium kaon effective mass, equivalent to using the
${\calL}_\sigma$, is to consider the kaon as fluctuation about
the ``$\sigma$"-axis in the V-spin formalism~\cite{BKR87} as
depicted pictorially in Fig.\ref{Ksigma}. The Hamiltonian for
explicit chiral symmetry breaking is
 \be
H_{\chi SB}&=&\Sigma_{KN}\langle\overline{N}N\rangle \cos (\theta)
+\frac 12 m_K^2 {f^\star}^2 \sin^2 (\theta)\nonumber\\
&\simeq&\Sigma_{KN}\langle\overline{N}N\rangle
(1-\frac{\theta^2}{2})+ \frac 12 m_K^2 {f^\star}^2 \theta^2
 \ee
where the last expression is obtained for small fluctuation
$\theta$. Dropping the term independent of $\theta$, we find
 \be
{m_K^\star}^2=m_K^2\left(1-\frac{\Sigma_{KN}\langle\overline{N}N\rangle}{{f^\star}^2
m_K^2}\right).\label{mlow}
 \ee
Using eq. (\ref{sigmatop}) we obtain
 \be
{m_K^\star}^2=m_K^2\left(1-\frac 23
\frac{g_\sigma^2\langle\overline{N}N\rangle} {{m_\sigma^\star}^2
m_K}\right).\label{mhigh}
 \ee
In accord with Brown and Rho~\cite{BR96} we are proposing that
eq.(\ref{mlow}) should be used for low densities, in the
Goldstone description of the $K^\pm$, and that we should switch
over to eq.(\ref{mhigh}) for higher densities. It is possible that
the $m_K$ appearing in (\ref{mhigh}) should be replaced by
$m_K^\star$ for self-consistency but the dropping of
${m_\sigma^\star}^2$ makes the $m_K^\star$ of (\ref{mhigh})
decrease more rapidly than that of (\ref{mlow}) so that
eq.(\ref{mlow}) with $\langle\overline{N}N\rangle$ set equal to
the vector density $\rho$~\footnote{The correction to this
approximation which may become important as the nucleon mass
drops comes as a ``1/m" correction in the heavy-fermion chiral
perturbation theory (as for the ``range term" mentioned below)
and can be taken into account systematically. It can even be
treated fully relativistically using a special regularization
scheme being developed in the field. Our approximation does not
warrant the full account of such terms, so we will not include
this correction here.}, a much used formula valid to linear order
in density
 \be
{m_K^\star}^2\approx m_K^2\left(1-\frac{\rho\Sigma_{KN}}{f^2
m_K^2}\right)\label{mlowp}
 \ee
obviously gives too slow a decrease of $m_K^\star$ with density.

Although the above are our chief points, there are two further
points to remark. One, even without scaling, our vector
interaction on the kaon is still too large. Two, more
importantly, there is reason to believe in the large
$\Sigma_{KN}$ term,
 \be
\Sigma_{KN}\sim 400\ \ {\rm MeV}.\label{value}
 \ee
This comes from scaling of the pion sigma term
 \be
\Sigma_{KN}\equiv \frac{(m_u+m_s)\langle
N|\bar{u}u+\bar{s}s|N\rangle}{(m_u+m_d)\langle
N|\bar{u}u+\bar{d}d|N\rangle}\Sigma_{\pi N}.
 \ee
Taking $m_s\sim 150$ MeV, $m_u+m_d\sim 12$ MeV, $\Sigma_{\pi
N}=46$ MeV and $\langle N|\bar{s}s|N\rangle\sim \frac 13 \langle
N|\bar{d}d|N\rangle$ from lattice calculations~\cite{liu}, one
arrives at (\ref{value}).

Other authors, in adjusting the $\Sigma$ term to fit the
kaon-nucleon scattering amplitudes, have obtained a somewhat
smaller $\Sigma_{KN}$. This can be understood in the chiral
perturbation calculation of C.-H. Lee~\cite{chl} where the only
significant effect of higher chiral order terms can be summarized
in the ``range term"\footnote{As mentioned, in the language of
heavy-baryon chiral perturbation theory, this corresponds to a
``1/m" correction term.}; namely $\Sigma_{KN}$ is to be replaced
by an effective $\Sigma$,
 \be
(\Sigma_{KN})_{eff}=(1-0.37\omega_K^{\star
2}/m_K^2)\Sigma_{KN}.\label{sigmaeff}
 \ee
It should be pointed out that although the $\Sigma_{KN}$ is
important at low densities,
$\omega_K$ decreases with $m_K^\star$, this ``range-term"
correction becomes less important at higher densities. This
effect -- which is easy to implement -- is included in the
realistic calculations.
\subsection{Partial Decoupling of the Vector Interaction}
 \subsubsection{\it Theoretical and experimental evidences}\label{3.1}
\itt
 In Section \ref{how}, a general argument for the vector decoupling
in hot/dense matter was presented based on Wilsonian matching with
QCD of HLS. Here we supply more specific theoretical and empirical
reasons why we believe that the vector interaction should decouple
at high density.
 \ben
 \item
We first give the theoretical arguments why the vector coupling
$g_\omega^\star$ should drop with density.
 \bitem
 \item The first is the observation by Song et al~\cite{song}
that describing nuclear matter in terms of chiral Lagrangian in
the mean field requires the ratio $g_\omega^\star/m_\omega^\star$
to at least be roughly constant or even decreasing as a function
of density. In fact to quantitatively account for non-linear
terms in a mean-field effective Lagrangian, a dropping ratio is
definitely favored~\footnote{Since the non-linear terms -- though
treated in the mean field -- are fluctuation effects in the
effective field theory approach, this represents a quantum
correction to the BR scaling.}. For instance, as discussed in
\cite{song}, the in-medium behavior of the $\omega$-meson field is
encoded in the ``FTS1" version of the non-linear theories of
ref.\cite{FTS1}. In fact, because of the attractive quartic
$\omega$ term in the FTS1 theory, the authors of \cite{FTS1} have
(for the parameter $\eta=-1/2$ favored by experiments)
${g_\omega^\star}^2/{m_\omega^\star}^2\simeq
0.8{g_\omega}^2/{m_\omega}^2$ as modification of the quadratic
term when rewritten in our notation . In other words, their
vector mean field contains a partial decoupling already at
$\rho\approx \rho_0$ although they do not explicitly scale
$g_\omega$ as we do.

Historically, Walecka-type mean field theories with only
quadratic interactions (i.e., linear Walecka model) gave
compression modulus $K\sim 500$ MeV, about double the empirical
value. This is cured in nonlinear effective field theories like
FTS1 by higher-dimension non-renormalizable terms which
effectively decrease the growth in repulsion in density. As
suggested in \cite{song}, an effective chiral Lagrangian with BR
scaling can do the same (by the increase in magnitude of the
effective scalar field with density) but more economically and
efficiently.

This decrease of ${g_\omega^\star}/{m_\omega^\star}$ as density
increases toward nuclear matter density contrasts with the
increase of $g^\star_{\rho NN}/m^\star_\rho$ within the same
density range discussed in Section \ref{finite}. This can be
explained as follows. In free space $g_\omega\equiv g_{\omega
NN}=3 g_{\rho NN}$ whereas near chiral restoration, we expect (as
in the case of the quark number susceptibility mentioned below)
that $g_{\omega NN}^\star\simeq g^\star_{\rho NN}$. This implies
that at low density, the $\omega$ coupling must fall faster than
the $\rho$ coupling by ``shedding off" the factor of 3.
\item Implementing baryons into the HLS Lagrangian, Kim and
Lee~\cite{kl} showed explicitly that both $g_V^\star$ and
$m_V^\star$ (where $V$ stands for hidden gauge bosons) are
predicted to fall very rapidly with baryon chemical
potential~\cite{kl}~\footnote{Kim and Lee found that even the
ratio $g_V^\star/m_V^\star$ fell rapidly. This is at variance with
what one would expect if the vector manifestation \`a la
Harada-Yamawaki were realized. As noted, this may be due to the
incompleteness of the renormalization group analysis of
\cite{kl}.}. The main agent for this behavior is found in
\cite{kl} to be the pionic one-loop contribution linked to chiral
symmetry which is lacking in the mean-field treatment for BR
scaling~\cite{BR91}. The Kim-Lee arguments were made for the
$\rho$ which has a simple interpretation in terms of hidden gauge
symmetry but it will apply to the $\omega$ if the nonet symmetry
is assumed to hold in nuclear medium.
 \item Finally, close
to chiral restoration in temperature, there is clear evidence from
QCD for an equally rapid drop, specifically, from the quark number
susceptibility that can be measured on the lattice~\cite{BRPR}.
The lattice calculation of the quark number susceptibility dealt
with quarks and the large drop in the (isoscalar) vector mean
field was found to be due chiefly to the change-over from hadrons
to quarks as the chiral restoration temperature is approached
from below. The factor of 9 in the ratio $g_{\omega
NN}^2/g_{\omega QQ}^2$ (where $Q$ is the constituent quark)
should disappear in the change-over.
Now since the electroweak properties of a constituent quark
(quasiquark) are expected to be the same as those of a bare Dirac
particle with $g_A=1$ and no anomalous magnetic moment (i.e., the
QCD quark)~\cite{weinbergquark} with possible corrections that
are suppressed as $1/N_c$~\cite{peris}, there will be continuity
between before and after the chiral transition. This is very much
in accordance with the ``Cheshire-Cat picture" developed
elsewhere~\cite{NRZ}. In fact, it is possible to give a dynamical
(hadronic) interpretation of the above scenario. For instance in
the picture of \cite{KRBR} sketched in Section \ref{heavyion},
this may be understood as the ``elementary" $\omega$ strength
moving downwards into the ``nuclear" $\omega$, the $[N^\star
(1520) N^{-1}]^{J=1,I=0}$ isobar-hole state involving a
single-quark spin flip~\cite{riskabrown}. The mechanism being
intrinsic in the change-over of the degrees of freedom, we expect
the same phenomenon to hold in density as well as in temperature.
The upshot of this line of argument is that the suppression of the
vector coupling is inevitable as density approaches the critical
density for chiral transition. In Section
\ref{complementarity},we will have an additional support for this
from color-isospin locking in QCD both in low density and in high
density.
 \eitem

We believe that the different behavior of vector and scalar mean
fields, the latter to be discussed below, follows from their
different roles in QCD. With the vector this is made clearer in
the lattice calculations of the quark number susceptibility which
involves the vector interactions. In Brown and Rho~\cite{BRPR},
it is shown that as the description changes from hadronic to
quark/gluonic at $T\sim T_c$, the critical temperature for chiral
restoration, the vector interaction drops by an order of
magnitude, much faster than the logarithmic decrease due to
asymptotic freedom. We expect a similar feature in density,
somewhat like in the renormalization-group analysis for the
isovector vector meson $\rho$ of Kim and Lee~\cite{kl}. The
scalar interaction, on the other hand, brings about chiral
restoration and must become more and more important with
increasing density as the phase transition point is approached.

\item
{}From the empirical side, the most direct indication of the
decoupling of the vector interaction is from the baryon
flow~\cite{Bflow} which is particularly sensitive to the vector
interaction. The authors of \cite{Bflow} find a form factor of
the form
 \be
 f_V (\Bp)= \frac{\Lambda_V^2-\frac 16 \Bp^2}{\Lambda_V^2+\Bp^2}
 \ee
with $\Lambda_V=$0.9 GeV is required to understand the baryon
flow. Connecting momenta with distances, one finds that this
represents a cutoff at
 \be
R_{cutoff}\sim \frac{\sqrt{6}}{\Lambda_V}\sim 0.5\ \ {\rm fm}.
 \ee

Furthermore it is well known that the vector mean field of the
Walecka model must be modified, its increase as $E/m_N$ removed,
at a scale of $\sim 1$ GeV. The reason for this is presumably that
inside of $R\sim 0.5$ fm, the finite size of the solitonic
nucleon must be taken into account. A repulsion still results,
but it is scalar in nature as found in \cite{vento} and for which
there are direct physical indications~\cite{holinde}.
 \een
\subsubsection{\it Kaons at GSI} \itt
 The \Kp and \Km energies in the top-down scenario are given
by
 \be
\omega_{\pm}=\pm \frac{\omega_{\pm}}{m_K}V_K
+\sqrt{k^2+{m_K^\star}^2}\label{energy}
 \ee
with $V_K$ given in (\ref{vectordc}) below. Although at high
densities it will decouple, the term linear in $V_K$ that figures
in the range correction in $(\Sigma_{KN})_{eff}$ will give a
slightly different effective mass for \Kp and \Km before
decoupling. Although the large distance vector mean field must
arise from vector meson exchange, this must be cut off at a
reasonably large distance, say, $\sim$ 0.5 fm as indicated by the
baryon-flow mentioned above.

For the GSI experiments with temperature $\sim 70$ MeV,
the nucleon and kaon momenta are $|\Bp_N|\sim 444$ MeV and
$|\Bp_K|\sim 322$ MeV, respectively, and
 \be
{f_V}(p)\sim 0.82.
 \ee
We therefore propose to use \be
 \frac{V_K}{m_K}\approx \frac 13 f_V^2 V_N (p=0)/m_K
 \sim 0.15\label{vectordc}
 \ee
This is small. Furthermore we assume it to be constant above
$\rho_0$. This assumption amounts to taking the vector coupling to
drop as $\sim 1/\rho$.

The above arguments could be quantified by a specific model. For
example, as alluded to above, the low-lying $\rho$- and
$\omega$-excitations in the bottom-up model can be built up as
$N^\star$-hole excitations~\cite{KRBR}. At higher densities,
these provide the low-mass strength. One might attempt to
calculate the coupling constants to these excitations in the
constituent-quark (or quasi-quark) model, which as we have
suggested would be expected to be more applicable at densities
near chiral restoration. Riska and Brown~\cite{riskabrown} find
the quark model couplings to be a factor $\sim 2$ lower than the
hadronic ones~\cite{frimanetal}.

\subsection{Schematic Model}\label{schematic}
 \subsubsection{\it First try with the simplest form}
 \itt
 On the basis of our above considerations, a first try in
transport calculation might use the vector potential with the
Song scaling~\cite{song} as
$g_\omega^\star/m_\omega^\star=constant$ and the effective mass
 \be
{m_K^\star}^2\approx
m_K^2\left(1-\frac{\rho(\Sigma_{KN})_{eff}}{f^2
m_K^2}\right)\label{mlowpp}
 \ee
with $\Sigma_{KN}=400$ MeV and $(\Sigma_{KN})_{eff}$ given by eq.
(\ref{sigmaeff}). While as argued above the vector coupling will
decouple at very high densities, as $\omega_K$ drops, the vector
potential will become less important even at moderate densities
since the factor $\omega_K/m_K$ comes into the coupling of the
vector potential to the kaon.

Our schematic model (\ref{mlowpp}) gives roughly the same mass as
used by Li and Brown~\cite{LB} to predict kaon and antikaon
subthreshold production at GSI. For $\rho\sim 3\rho_0$ it gives
$m_{K^-}^\star\sim 230$ MeV, less than half the bare mass. We
predict somewhat fewer $K^-$-mesons because the (attractive)
vector interaction is largely reduced if not decoupled. Cassing et
al~\cite{cassing} have employed an $m_{K^-}^\star$ somewhat lower
than that given by eq.(\ref{mlowpp}) to describe a lot of data.

The experimental data verify that our description is quite good
up to the densities probed, i.e., $\sim 3\ \rho_0$. In order to
go higher in density, we switch to our top-down description which
through eq.(\ref{sigmatop}) involves
$g_\sigma^2/{m_\sigma^\star}^2$. Although the scalar interaction
could have roughly the same form factor as the vector, cutting it
off at $\sim 0.5$ fm as mentioned above,  we believe that this
will be countered by the dropping scalar mass $m_\sigma^\star$
which must go to zero at chiral restoration (viewed as an order
parameter). Treating the scalar interaction linearly as a
fluctuation (as in (\ref{sigmatop})) cannot be expected to be
valid all the way to chiral restoration but approaching the
latter the $\sigma$-particle becomes the ``dilaton" in the sense
of Weinberg's ``mended symmetry"~\cite{weinberg-salam,beane} with
mass going to zero (in the chiral limit) together with the pion .

At high densities at which the vector interaction decouples, the
$K^+$ and $K^-$ will experience nearly the same very strong
attractive interactions. This can be minimally expressed through
the effective mass $m_K^\star$. At low densities where the vector
potential not only comes into play but slightly predominates over
the scalar potential, the $K^+$ will have a small repulsive
interaction with nucleons. It is this interaction, extrapolated
without medium effects by Bratkovskaya et al~\cite{brat} which
gives the long equilibration time of 40 fm/c. However, clearly
the medium effects will change this by an order of magnitude.
\subsubsection{\it Implications on kaon condensation and maximum
neutron-star mass}
 \itt
While in heavy-ion processes, we expect that taking
$m_\sigma^\star$ to zero (or nearly zero in the real world) is a
relevant limiting process, we do not have to take $m_\sigma^\star$
to zero for kaon condensation in neutron stars, since the
$K^-$-mass $m_K^\star$ must be brought down only to the electron
chemical potential $\mu_e\simeq E_F (e)$, the approximate equality
holding because the electrons are highly degenerate. Arguments
based on effective chiral Lagrangians that are consistent with
low-energy phenomenology in kaon-nucleon and kaon-nuclear systems
typically give the critical density in neutron-star matter of
$\rho_c\sim 3\rho_0$~\cite{BTKR,BLRT,chl}. We should however
mention that it has been suggested that the electron chemical
potential $\mu_e$ could be kept low by replacing electrons plus
neutrons by $\Sigma^-$ hyperons (or more generally by exploiting
Pauli exclusion principle with hyperon introduction) in neutron
stars~\cite{glendenning}. In this case, the $\mu_e$ might never
meet $m_K^\star$.

Hyperon introduction may or may not take place, but even if it
does, the scenario will be more subtle than considered presently.
To see what can happen, let us consider what one could expect
from a naive extrapolation to the relevant density, i.e.,
$\rho\sim 3\rho_0$, based on the {\it best} available nuclear
physics. The replacement of neutron plus electron will take place
if the vector mean field felt by the neutron is still high at that
density. The threshold for that would be
 \be
E_F^n +V_N +\mu_e\simeq M_{\Sigma^-} + \frac 23 V_N
+S_{\Sigma^-}\label{thr}
 \ee
where $E_F^n$ is the Fermi energy of the neutron, $M_{\Sigma^-}$
the bare mass of the $\Sigma^-$ and the $S_{\Sigma^-}$ the scalar
potential energy felt by the $\Sigma^-$. Here we are simply
assuming that the two non-strange quarks of the ${\Sigma^-}$
experience $2/3$ of the vector mean field felt by the neutron.
Extrapolating the FTS1 theory~\cite{FTS1}\footnote{There is
nothing that would suggest that the effective Lagrangian valid up
to $\rho\sim \rho_0$ will continue to be valid at $\rho\sim
3\rho_0$ without addition of higher mass-dimension operators,
particularly if the chiral critical point is nearby. So this
exercise can be taken only as indicative.} and taking into
account in $V_N$ the effect of the $\rho$-meson using vector
dominance , we find $E_F^n+V_N\sim 1064$ MeV at $\rho\approx
3\rho_0$. From the extended BPAL 32 equation of state with
compression modulus 240 MeV~\cite{prakash}, the electron chemical
potential comes out to be $\mu_e\simeq 214$ MeV. So the left-hand
side of (\ref{thr}) is $E_F^n +V_N +\mu_e\sim 1278$ MeV. For the
right-hand side, we use the scalar potential energy for the
$\Sigma^-$ at $\rho\approx 3\rho_0$ estimated by Brown, Lee and
Rapp~\cite{BLR} to find that $M_{\Sigma^-} + \frac 23 V_N
+S_{\Sigma^-}\sim 1240$ MeV. The replacement of neutron plus
electron by $\Sigma^-$ looks favored but only slightly.

What is the possible scenario on the maximum neutron star mass if
we continue assuming that the calculation we made here can be
trusted?  A plausible scenario would be as follows. \Km
condensation supposedly occurs at about the same density and both
the hyperonic excitation (in the form of $\Sigma N^{-1}$ -- where
$N^{-1}$ stands for the nucleon hole -- component of the
``kaesobar"~\cite{BLR}) and \Km condensation would occur at
$T\sim 50$ MeV relevant to the neutron-star matter. Now if as is
likely the temperature is greater than the difference in energies
between the two possible phases, although the hyperons will be
more important initially than the kaons, all of the phases will
enter more or less equally in constructing the free energy of the
system. In going to higher density the distribution between the
different phases will change in order to minimize the free
energy. Then it is clear that dropping from one minimum to
another, the derivative of the free energy with density -- which
is just the pressure -- will decrease as compared with the
pressure from any single phase. This would imply that the maximum
neutron star mass calculated with either hyperonic excitation or
kaon condensation alone must be greater than the neutron star
mass calculated with inclusion of both.

The story will be quite different if the vector field decouples.
We showed in Section \ref{3.1} that the isoscalar vector mean
field must drop by a factor $\gsim 9$ in the change-over from
nucleons to quasiquarks as variables as one approaches the chiral
restoration density. Hyperons will disappear during this drop. It
is then highly likely that the kaon will condense {\it before}
chiral restoration and that the kaon condensed phase will persist
through the relevant range of densities which determine the
maximum neutron star mass.

\subsection{Discussions}
  \itt
In this section, we have given arguments to support that not only
the kaon mass but also its coupling to vector mesons should drop
in matter with density. In particular, with the introduction of
medium effects the apparent equilibration found in strangeness
production at GSI can be increased from the baryon number density
of $\sim\frac 14\rho_0$ up to the much more reasonable $\sim
2\rho_0$. These properties do not appear at first sight to be
connected to BR scaling {\it per se}. But they must be connected
in an albeit indirect way.

{}From the baryon flow analysis we have direct indications that
the
 vector interaction decouples from the nucleon at a three-momentum
 of $|\Bp|\sim 0.9$ GeV/c or at roughly 0.5 $m_N c$. In colliding
 heavy ions this is reached at a kinetic energy per nucleon of
 $\sim \frac 18 m_N c^2$ which means a temperature of 78 MeV when
 equated to $\frac 32 T$. This is just the temperature for
chemical freeze-out at GSI energies. We have given several
theoretical arguments why the vector coupling should drop rapidly
with density.

Once the vector mean field, which acts with opposite signs on the
$K^+$ and $K^-$ mesons is decoupled, these mesons will feel the
same highly attractive scalar meson field. Their masses will fall
down sharply; e.g., from eq.(\ref{mlowpp}) with proposed
parameters,
 \be
 \frac{m_K^\star}{m_K}\sim 0.5
 \ee
at $\rho\approx 3\rho_0$ and possibly further because of the
dropping $m_\sigma^\star$. The differing slopes of $K^+$ and
$K^-$  with kinetic energy will then develop after chemical
freeze-out, as suggested by Li and Brown~\cite{LB}.

In this section we have focused on the phenomenon at GSI
energies. Here the chief role that the dropping \Km-mass played
was to keep the combination $\mu_B+ \wkm$ nearly constant so that
low freeze-out density in the thermal equilibration scenario
became irrelevant for the \Kp/\Km ratio. We suggest that the same
scenario applies to AGS physics, where the freeze-out density in
the thermal equilibration picture comes out to be $\sim
0.35\rho_0$~\cite{peter}.
In fact, there is no discernible dependence on centrality in the
\Km/\Kp ratios measured at $4A$ GeV, $6A$ GeV, $8A$ GeV and
$10.8A$ GeV~\cite{dunlop}. From this it follows either that the
ratio of produced \Km to hyperons is nearly independent of
density or that the negative strangeness equilibrates down to a
lower freeze-out density and then disperses. Given the relative
weakness of the strange interactions, we believe the former to be
the case. In fact, we suggest that near constancy with
multiplicity of the \Km/\Kp ratio found experimentally be used to
determine temperature dependence of $\wkm$ in the region of
temperatures reached in AGS physics. As was done for GSI
energies, the temperature can be obtained from the inverse slopes
of the kaon and antikaon distributions of $p_\perp$ and then the
temperature dependence of $\wkm$ can be added to the density
dependence as in \cite{BKWX}, in such a way that $\mu_B+\wkm$
stays roughly constant as function of density. At least this can
be done in the low-density regime considered in \cite{peter} when
the approximation of a Boltzmann gas is accurate enough to
calculate $\mu_B$. Our ``broad-band equilibration"; i.e., the
production of the same, apparently equilibrated ratio of \Km
-mesons to hyperons over a broad band of densities, avoids
complications in the way in which the \Km\ degree of freedom is
mixed into other degrees of freedom at low
density~\cite{Ramos:2000ku}. Most of the \Km -production will
take place at the higher densities, as shown in Fig. \ref{nini},
where the degrees of freedom other than \Km\ have been sent up to
higher energies by the Pauli principle.

Unless the electron chemical potential in dense neutron star
matter is prevented from increasing with density (as might happen
if the repulsive $\Sigma$-nuclear interaction turns to attraction
at $\rho> \rho_0$), kaon condensation will take place {\it before}
chiral restoration. This has several implications at and beyond
chiral restoration. For instance, its presence would have
influence on the conjectured color superconductivity at high
density, in particular regarding its possible coexistence with
Overhauser effects, skyrmion crystal and other phases with
interesting effects on neutron star cooling.

The phenomenon of vector decoupling, if confirmed to be correct,
will have several important spin-off consequences. The first is
that it will provide a refutation of the recent
claim~\cite{pandh} that in the mean-field theory with a
kaon-nuclear potential given by the vector-exchange
(Weinberg-Tomozawa) term -- both argued to be valid at high
density -- kaon condensation would be pushed up to a much higher
density than that relevant in neutron-star matter. Our chief point
against that argument is that the vector decoupling and the
different role of scalar fields in QCD (e.g., BR scaling)
described in this article cannot be accessed by the mean field
reasoning used in \cite{pandh} or by any other standard nuclear
physics potential models. The second consequence that could be of
a potential importance to the interpretation of heavy-ion
experiments is that if the vector coupling rapidly diminishes
with density, the strong-coupling perturbation calculation of the
vector response functions used in terms of ``melting" vector
mesons to explain~\cite{rapp}, e.g., the CERES dilepton data must
break down rapidly as density increases. This could provide a
specific mechanism for the quasiparticle description of BR
scaling found to be successful in nuclear matter~\cite{migdal}).
It is of course difficult to be quantitative as to precisely
where this can happen.

Finally we should stress that the scenario presented in this
section -- which is anchored on Brown-Rho scaling-- should not be
considered as an {\it alternative} to  a possible quark-gluon
scenario currently favored by the heavy-ion community. It is once
again more likely a sort of complementarity of the same physics
along the line that the Cheshire-Cat Principle~\cite{NRZ} embodies
that would continue to apply at higher energies.

 \section{EFFECTIVE FIELD THEORIES FOR DENSE QCD}\label{denseqcd}
 \setcounter{equation}{0} 
\renewcommand{\theequation}{\mbox{\ref{denseqcd}.\arabic{equation}}}
 \itt
Thus far we have been climbing up in density in what may be aptly
called ``bottom-up" approach. In this section, we will start at
some asymptotic density and come down to the density regime
relevant to dense compact stars and possibly in certain regimes of
heavy-ion collisions. This will be a ``top-down" approach
exploiting once more how chiral symmetry in nature manifests
itself through (Nambu-)Goldstone mode. There is an intense
activity on this matter, comprehensively reviewed in the recent
literature~\cite{raja-wilcz}, so we will not go into most of the
topics. Here we will focus on the most intriguing aspect of dense
QCD that reflects an intriguing repeat of the Cheshire-Cat
phenomenon. We wish to highlight the seemingly continuous role of
chiral symmetry from low-density strong-coupling regime to high
density weak-coupling regime in the form of a skyrmion replica
that maps high-density excitations to low-density excitations. In
view of the rapid development on this matter, our discussion will
necessarily be incomplete and primitive but the general feature
of the Cheshire Cat phenomenon is likely to survive the test of
the time.
\subsection{Color-Flavor Locking for $N_F=3$}
\subsubsection{\it Instability at high density} \itt Asymptotic freedom
of QCD suggests that at very high density with
$\mu\rightarrow\infty$, the QCD gauge coupling becomes very weak
and so interactions disappear. However if there is an attraction
in a certain channel, now matter how weak, the same
renormalizationg group consideration discussed above in
connection with Landau Fermi liquid state~\cite{shankar} tells us
that the system will be unstable against a phase change. Indeed in
the weak coupling limit, one gluon exchange that survives
asymptotic freedom has an attraction in the channel that involves
diquarks sitting on the opposite side of the Fermi sphere,
namely, the BCS channel and this inevitably leads to the Cooper
pair condensation as in superconductivity. That this must occur
at asymptotic density is without doubt as dictated by the
renormalization group argument. Whether or not -- and in what
manner -- this will occur in a physically relevant regime of
density cannot be addressed by the asymptotic QCD and it appears
most plausible that the scenario is much richer in various
different ways than just the Cooper pairing type as one
approaches top-down to the density regime of relevance. This
matter is amply discussed in the literature~\cite{raja-wilcz} but
the situation is totally unclear. It is not our aim to dwell on
the multitude of possibilities. We will focus on one particular
aspect of this phenomenon, that is, the situation where color and
flavor of QCD get locked, i.e., the color-flavor-locked (CFL)
phase which renders the discussion simple and transparent.
\subsubsection{\it Symmetry breaking and excitations}
\itt We shall discuss first the case of $N_c=N_F=3$ which is
relevant at asymptotic density, returning to the case of $N_c=3$,
$N_F=2$ in the next subsection.

Gluon exchange between two quarks is attractive in the color
$\bar{\bf 3}$ channel and repulsive in the ${\bf 6}$ channel. So
one expects diquarks would condense in the $\bar{\bf 3}$ channel.
We shall ignore a small mixing to the ${\bf 6}$ channel which is
allowed by symmetry. The diquark condensate would be in the form
\be
 \la q_{L\alpha}^{ia} q_{L\beta}^{jb}\ra=-\la q_{R\alpha}^{ia}
q_{R\beta}^{jb}\ra=\kappa
\epsilon^{ij}\epsilon^{abI}\epsilon_{\alpha\beta
I}\label{condensate} \ee where $\kappa$ is a constant, $i,j$ are
$SL(2,C)$ indices, $a,b$ are color indices and $\alpha,\beta$ are
flavor indices. This condensation breaks the symmetry
 \be
G\rightarrow H
 \ee
with
 \be
G&=& SU(3)_c\times SU(3)_L\times SU(3)_R\times U(1)_B,\label{G}\\
H&=& SU(3)_{c+L+R}\times Z_2.\label{H}
 \ee
At superdensity, instantons are suppressed, but the anomaly never
disappears, so the $U(1)_A$ symmetry is never truly restored. We
shall not be concerned with it here. The color is completely
broken and since the vectorial color is locked to both $L$ and
$R$, chiral symmetry is also broken so the resulting invariance
group is $SU(3)_{C+L+R}$. Since $\la q_{L\alpha}^{a}
q_{L\beta}^{b}\bar{q}^{\gamma}_{Ra} \bar{q}^{\delta}_{Rb}\ra\sim
\kappa^2 \epsilon_{\alpha\beta
\Gamma}\epsilon^{\gamma\delta\Gamma}$ is gauge invariant, it can
be used as an order parameter but while it breaks chiral symmetry,
it leaves $Z_2$ invariant, thus its appearance in $H$.

The spontaneous symmetry breaking induces Goldstone bosons: eight
from the breaking of the color $SU(3)_C$, eight from chiral
symmetry breaking and one from $U(1)_B\rightarrow Z_2$. We shall
forget the last one since it does not concern us here. The
striking feature we would like to focus on here is the uncanny
resemblance of the spectrum in the CFL phase to the that of low
density in terms of a hidden local symmetry~\cite{bandoetal}. The
scalar Goldstones are eaten up by the gluons thereby the vector
bosons becoming massive. The octet of the massive gluons in the
CFL are the analogs to the (octet) massive light-quark vector
mesons in the low-density vacuum. The octet pseudoscalar
Goldstones in the CFL are the analog to the octet pions in the
low density vacuum, the dynamics of which can be described by an
analogous chiral Lagrangian. The quarks in the system are gapped
-- and hence massive -- and bear the quantum numbers of the octet
baryon of the zero-density vacuum. We shall show below that this
can be understood by looking at the solitons in the chiral
Lagrangian~\cite{hongetal}. Since the soliton corresponds to a
gapped colored quark, we call it ``qualiton" in analogy to the
qualiton considered by Kaplan for the constituent
quark~\cite{kaplan}. (The breaking of $U(1)_B$ leads to
superfluidity which resembles the same in nuclear matter, another
uncanny correspondence.) The one-to-one matching between the low
density spectrum in terms of hadronic variables and the high
density spectrum in terms of quark-gluon variable is referred to
as ``quark-hadron continuity."~\cite{continuity}. We will
identify this as belonging to the class of the Cheshire Cat
phenomenon in the strong interaction physics~\cite{NRZ}.
\subsubsection{\it Chiral Lagrangians and qualitons}
\itt As suggested in \cite{hongetal,casalbuoni}, the dynamics of
the surviving Goldstone modes can be described by an effective
chiral field theory as in the zero-density situation. We
introduce the chiral effective field~\footnote{One has to be
careful with a singularity in the product of two fermion fields
at one point. See \cite{hongetal} for a more proper definition.}
 \be
 \xi_{La\alpha} (x)\sim
 \epsilon^{ij}\epsilon_{abc}\epsilon_{\alpha\beta\gamma}q_{Li}^{b\beta}
 (-\vb_F,x) q_{Lj}^{c\gamma} (\vb_F,x)
\ee corresponding to the map $SU(3)_c\times SU(3)_L/SU(3)_{C+L}$
and similarly for $R$,
 \be
\xi_{Ra\alpha} (x)\sim
 \epsilon^{ij}\epsilon_{abc}\epsilon_{\alpha\beta\gamma}q_{Ri}^{b\beta}
 (-\vb_F,x) q_{Rj}^{c\gamma} (\vb_F,x)
\ee corresponding to the map $SU(3)_c\times SU(3)_R/SU(3)_{c+R}$.

Under an $SU(3)_c\times SU(3)_L\times SU(3)_R$ transformation by
unitary matrices $(g_c,g_L,g_R)$, $\xi_L$ transforms as
$\xi_L\mapsto g_c^*\xi_Lg_L^{\dagger}$ and $\xi_R$ transforms as
$\xi_R\mapsto g_c^*\xi_Rg_R^{\dagger}$. In the ground state of the
CFL superconductor, $\xi_L$ and $\xi_R$ take the same constant
value. QCD symmetries with (\ref{condensate}) imply
\begin{equation}
\left<{\xi_L}_{a\alpha}\right>=-\left<{\xi_R}_{a\alpha}\right>=\kappa\,
\delta_{a\alpha}.
\end{equation}
The Goldstone bosons are the low-lying excitations of the
condensate, given as unitary matrices $\xi_L(x)=g_c^T(x)g_L(x)$
and $\xi_R(x)=g_c^T(x)g_R(x)$. For the present decomposition, we
note the extra invariance under the (hidden) local transformation
$g_{c+L+R}(x)$ -- which is an analog to the hidden gauge transform
$h(x)$ of \cite{bandoetal} -- within the diagonal $SU(3)_{c+L+R}$
through \be g_c^T (x) \rightarrow g_c^T (x) g_{c+L+R}^{\dagger}
(x) \qquad\qquad g_{L,R} (x)\rightarrow g_{c+L+R}(x) g_{L,R} (x)
\label{local} \ee Hence, the spontaneous breaking of
$SU(3)_c\times \left(SU(3)_L\times SU(3)_R\right)\rightarrow
SU(3)_{c+L+R}$ can be realized non-linearly through the use of
$\xi_{L,R}(x)$ or linearly through the use of $g_{c,L,R}(x)$ with
the addition of an octet vector gauge field transforming
inhomogeneously under local $g_{c+L+R}(x)$. This can be identified
with the hidden local symmetry~\cite{bandoetal}~\footnote{We
comment as a side remark on a technical detail which is somewhat
outside of the scope of this review but may be helpful to those
interested in subtleties involved in hidden-gauge-symmetry aspects
of the problem.

As mentioned, when the color is completely broken, the octet
gluons become massive by Higgs mechanism with the mass
proportional to the chemical potential $\mu$. These were found to
have the same quantum numbers as those of the light-quark vector
mesons present at zero density. The hidden gauge bosons excited
at high density  discussed here belong to the unbroken diagonal
subgroup $SU(3)_{c+L+R}$. How are these vectors related to the
massive gluons?

In \cite{RSWZ}, the hidden gauge bosons $\in SU(3)_{c+L+R}$ were
identified with the spin-1 two-quark bound states with the
properties that match with the Georgi vector limit at some
non-asymptotic density. At the leading order in $1/\mu$, these
states seem to have nothing to do with the Higgsed gluons: The
overlap is zero. But this cannot be correct. In nature, the
low-lying vector excitations at large density must be a coherent
mixture of the two but at the leading order, they must be
equivalent. This is analogous to the complementary description of
the pion as a Goldstone mode and as a zero-mass bound state of a
quark and antiquark. Indeed this complementarity has also been
verified in the pion channel in dense QCD in terms of bound
diquarks ~\cite{RWZ,RSWZ}. To see this aspect in the vector
channel, we follow the CCWZ (Callan-Coleman-Wess-Zumino)
formalism~\cite{CCWZ}. Consider the symmetry group
$G=SU(3)_L\times SU(3)_R\times SU(3)_c$ broken down to
$H=SU(3)_{L+R+c}$. Using a notation slightly different from what
is used in this subsection, let the $G$ transformation be
represented in the block diagonal form $g={\rm diag} (L,R,C)$
corresponding to the left, right and color transformations. Let
the generator of the diagonal subgroup be given by $h={\rm diag}
(U,U,U)$. Now we would like to write $\Xi (x)\in G$, the {\it
rotation matrix that transforms the standard vacuum configuration
to the local field configuration}. One choice consistent with the
CCWZ prescription is
 \ba
\Xi (x)={\rm diag}\, (\xi_L (x), \ \xi_R (x), \ 1)\nonumber
 \ea
which transforms as
 \ba
g\Xi h^{-1}= {\rm diag}\, (L\xi_L U^{-1}, \ R\xi_R U^{-1}, \
CU^{-1}).\nonumber
 \ea
It is clear that we are required to take
 \ba
CU^{-1}=1\nonumber
 \ea
which means that $U=C$ and that $\xi_{L,R}$ transform as
 \ba
\xi_L&\rightarrow& L\xi_L C^{-1},\nonumber\\
\xi_R&\rightarrow& R\xi_R C^{-1}.\nonumber
 \ea
Now $U$ is a non-linear function of $L$, $R$, $\xi_L$ and
$\xi_R$, and so must be $C$. In HGS, this is the group that is
gauged. Therefore the ``hidden gauge symmetry" can be equated to
the color symmetry of QCD as was done by Casalbuoni and
Gatto~\cite{casalbuoni}. This can be seen also in Eq.(\ref{2x})
given below.}. Clearly the composites carry color-flavor in the
unbroken subgroup, with a mass of the order of the
superconducting gap.

The current associated with $\det \xi_L$ ($\det \xi_R$) is the
left (right)-handed $U(1)_L$ ($U(1)_R$) baryon number current.
Because of the $U(1)$ axial anomaly, the field $\det \xi_L/ \det
\xi_R$ is massive due to instantons. We decouple the massive field
from the low energy effective action by imposing $\det \xi_L/ \det
\xi_R=1$. The Goldstone boson associated with spontaneously broken
$U(1)_B$ symmetry is described by $\det \xi_L\cdot\det \xi_R$ and
responsible for baryon superfluidity. But since it is not
directly relevant for our problem, we further choose $\det
\xi_L\cdot\det \xi_R=1$ to isolate the Goldstone bosons resulting
from the spontaneous breaking of chiral symmetry, from the massive
ones eaten up by the gluons. We now parameterize the unitary
matrices $\xi_{L,R}$ as
\begin{equation}
\xi_L(x)=\exp\left(i \pi_L^AT^A/f\right), \quad
\xi_R(x)=\exp\left(i\pi_R^AT^A/f\right),
\end{equation}
where $T^A$ are $SU(3)$ generators, normalized as ${\rm
Tr}~(T^AT^B)=\delta^{AB}/2$. The Goldstone bosons $\pi^A$
transform nonlinearly under $SU(3)_c\times SU(3)_L\times SU(3)_R$
but linearly under the unbroken symmetry group $SU(3)_{c+L+R}$.

The effective Lagrangian for the CFL phase at asymptotic
densities follows by integrating out the `hard' quark modes at
the edge of the Fermi surface~\cite{HONG}. The effective
Lagrangian for $U_{L,R}$ is a standard non-linear sigma model in
D=4 dimensions including the interaction of Goldstone bosons with
colored but ``screened" gluons $G$. Expanding in powers of
derivative, the effective Lagrangian for the Goldstone bosons is
then
\begin{eqnarray}
{\cal L} =&+& \frac {f_T^2}4 \,\,{\rm Tr} (\partial_0
\xi_L\partial_0 \xi_L^{\dagger}) -\frac {f_S^2}4 \,\,{\rm Tr}
(\partial_i \xi_L\partial_i \xi_L^{\dagger}) + g_s G\cdot J_L +
(L\rightarrow R)+\cdots \label{2x}
\end{eqnarray}
Here the ellipsis stands for higher derivative terms including the
Wess-Zumino-Witten term~\cite{witten1} that is needed to account
for anomalies. The vectorial coupling to the gluon field $G_\mu$
via the color current $J_\mu$ indicates that the left-right chiral
symmetry is also locked to each other. In the presence of a
chemical potential, the $SL(2,C)$ breaks down to $O(3)$, so the
``decay constant" $f$ has two different values for the time and
space components. The temporal and spatial decay constants
$f_{T,S}$ are fixed by the `hard' modes at the Fermi surface.
Their exact values are determined by the dynamics at the Fermi
surface. They go like $f_S\sim f_T\sim \mu$.

The effective Lagrangian (\ref{2x}) in the CFL phase bears much
in common with Skyrme-type Lagrangians~\cite{NRZ}, with the
screened color mediated interaction analogous to the exchange of
massive vector mesons. Indeed in the CFL phase the `screened'
gluons and the `Higgsed' gauge composites are the analog of the
massive vector mesons in the low density phase. In the CFL phase
the WZW term is needed to enforce the correct flavor anomaly
structure.

The `Higgsed' gluons are very massive in the large $\mu$ limit,
so can be integrated out from the Lagrangian (\ref{2x}). This is
very much like the hidden gauge symmetry theory wherein the
vector mesons can be integrated out to give rise to a nonlinear
sigma model. The theory then can have a soliton as suggested by
Hong, Rho and Zahed~\cite{hongetal} if there are higher
derivative terms that stabilize it. The qualiton will be colored.
However the pair-condensed ground state is colored also, so
combined with the background, the excitation will be effectively
color-singlet. In other words, the chiral field from which the
soliton arises can be considered as color-singlet fluctuation.
This means that one can rewrite the effective field theory in
terms of the {\it color-singlet field}
 \be
 U^j_i\equiv \xi_{Lai} \xi_R^{*aj}
\ee which transforms $U\rightarrow g_L U g_R^\dagger$. The
Lagrangian (\ref{2x}) can then be rewritten in the form familiar
from low density
 \be
 \calL = \frac{f_T^2}{4}\Tr(\del_0 U\del_0
 U^\dagger)-\frac{f_S^2}{4}\Tr(\del_iU\del_i U^\dagger)
 +{\cal O} (\del^4) +{\cal O} ({\cal M}^2)+\cdots\label{Leff}
\ee The resulting theory is essentially identical to the familiar
nonlinear sigma model giving rise to the Skyrme
soliton~\cite{NRZ}, except that the mass terms which we did not
specify here should be quadratic in the quark mass matrix ${\cal
M}$ because of the $Z_2$ invariance. The qualiton can be
collective-quantized in a way paralleling the low-density
skyrmions. This was worked out recently in \cite{honghong} in
which the authors argue that since a qualiton is a quark on top
of the condensed state, it would correspond to the quasiparticle
of the particle-hole complex and the qualiton mass that is
obtained as a soliton solution must therefore correspond to
$2\Delta$ plus the ground-state (or condensation) energy where
$\Delta$ is the gap.
\subsection{Complementarity of Hidden Gauge Symmetry and
Color Gauge Symmetry and BR Scaling}\label{complementarity}
 \itt
The $N_c=3$, $N_F=2$ case is a bit more intricate and involved
than the $N_c=N_F=3$ CFL  discussed above. It turns out, however,
that color-flavor locking in this case~\cite{berges-wett} is more
relevant in the density regime closer to nature, both in high
density and in low density. It can also be mapped as we shall
argue to hidden local symmetry discussed in the previous sections.
 \subsubsection{\it Quark number susceptibility}
 \itt
It was observed in \cite{BRPR,hatsuda} that the quark number
susceptibility $\chi_\pm=(\del/\del\mu_u\pm \del/\del\mu_d)
(\rho_u\pm \rho_d)$ where $\rho_{u,d}$ and $\mu_{u,d}$ are
respectively $u,d$-quark number density and chemical potential
measured on lattice as a function of temperature~\cite{lattice}
exhibited a smooth change-over from a flavor-gauge symmetry or
hidden gauge symmetry to QCD color-gauge symmetry at the chiral
phase transition critical temperature $T_c$ . It was
suggested~\cite{BRPR} that at the phase transition, the flavor
gauge symmetry -- which is induced and hence not fundamental --
gets {\it converted directly} to the color gauge symmetry --
which is fundamental, implying that they could be related in an
intricate way. In this subsection, following \cite{BR2001}, we
suggest how this can be realized in terms of color-flavor-locked
(CFL) quark-antiquark and diquark condensates and ``vector
manifestation" (VM) of chiral symmetry explained in Section
\ref{how}. It will be seen how BR scaling can be fit into the
general scheme that results from these developments: Its proof
will be the most convincing case for the validity of BR scaling.

Our argument relies on two recent remarkable developments that
come from seemingly unrelated sectors. One is the suggestion by
Harada and Yamawaki~\cite{VM} that the phase transition from the
Nambu-Goldstone phase to the Wigner-Weyl phase involves ``vector
manifestation" of chiral symmetry, that is, at the phase
transition, the longitudinal components of the light-quark vector
mesons (i.e., the triplet $\rho$ in the 2-flavor case) and the
triplet pions ($\pi^a$) come together becoming massless in the
chiral limit, the vectors decoupling \`a la Georgi's vector
limit~\cite{georgi} but with the pion decay constant $f_\pi$
vanishing at that point. The other important development is the
proposal by Berges and Wetterich~\cite{wetterich,berges-wett} that
color and flavor get completely locked in the Nambu-Goldstone
phase (a) for three flavors ($N_f=3$)~\cite{wetterich} by the
quark-antiquark condensate in the color-octet ($\bf{8}$) channel
 \be \chi=\la\bar{q}_\alpha^a\sum_{i=1}^3
(\tau_i)_{\alpha\beta} (\lambda_i)^{ab} q_\beta^b\ra\label{chi}
 \ee
and (2) for two flavors ($N_F=2$)~\cite{berges-wett} by
(\ref{chi}) together with the diquark condensate in the
color-antitriplet ($\bar{\bf 3}$) channel
 \be
\Delta=\la q_\alpha^a (\tau_2)_{\alpha\beta} (\lambda_2)^{ab}
q_\beta^b\ra.\label{delta}
 \ee
In (\ref{chi}) and (\ref{delta}), the indices $\alpha, \beta$
denote the flavors and $a, b$ the colors. In the Berges-Wetterich
scenario, the chiral phase transition and deconfinement occur
through the melting of the condensates $\chi$ and $\Delta$.
 \subsubsection{\it Color-isospin locking}
 \itt
The scenario for color-flavor-locking (CFL) is a bit different
depending on the number of flavors $N_F$. For our purpose, the
case of two flavors is more relevant, so we shall focus on this
case. In \cite{berges-wett}, Berges and Wetterich argue that both
the $\chi$ and $\Delta$ condensates can be nonzero in the vacuum,
thereby completely breaking the color. Should one of the
condensates turn out to be zero, then color would be only
partially broken (see \cite{raja-wilcz} for a simple explanation
for this and references). This pattern of color breaking and
color-flavor locking renders the octet gluons and six quarks
massive by the Higgs mechanism and generates three Goldstone
pions due to the broken chiral symmetry. All of the excitations
are integer-charged. Among the eight massive gluons, three of
them are identified with the isotriplet $\rho$'s with the mass
 \be
m_\rho=\kappa g_c \chi\label{cfl-rho}
 \ee
where $\kappa$ is an unknown constant and $g_c$ the color gauge
coupling. The fourth vector meson is identified with the
isosinglet $\omega$ with the mass
 \be
m_\omega=\kappa^\prime g_c \Delta\label{cfl-omega}
 \ee
where $\kappa^\prime$ is another constant. The remaining four
vector mesons turn out to have exotic quantum numbers and are
presumably heavy. We assume that they decouple from the
low-energy regime. As for the fermions, there are two baryons
with the quantum numbers of the proton and neutron with their
masses proportional to the scalar condensate $\phi$,
 \be
 \phi=\la \bar{q}_\alpha^a q_\alpha^a\ra.
 \ee
The four remaining fermions are also of exotic quantum numbers
with zero baryon number and heavier, so we assume that they also
decouple from the low-energy sector. What concerns us here is
therefore the three pions, the proton and neutron, the
$\rho$-mesons and the $\omega$-meson.
\subsubsection{\it Implications of the lattice measurements of
quark number susceptibility}
 \itt
It was argued in \cite{BR96} that the ``measured" singlet and
non-singlet QNS's~\cite{lattice} indicate that both the $\rho$ and
$\omega$ couplings vanish at the transition temperature $T_c$.
This meant that the $\omega NN$ coupling which is $\sim 3$ times
the $\rho NN$ coupling at zero temperature became equal to the
latter at near the critical temperature. This also meant that the
both vector mesons became massless  and decoupled. Viewed from the
CFL point of view, it follows from (\ref{cfl-rho}) and
(\ref{cfl-omega}) that the condensates $\chi$ and $\Delta$ ``melt"
at that point. This is consistent with the observation by
Wetterich~\cite{wetterich} that for three-flavor QCD, the phase
transition -- which is both chiral and deconfining -- occurs at
$T_c$ with the melting of the color-octet condensate $\chi$. The
transition is first-order for $N_F=3$ in agreement with lattice
calculations, so the vector meson mass does not go to zero
smoothly but makes a jump from a finite value to zero. We expect
however that in the case of $N_F=2$ the transition will be
second-order with the vector mass dropping to zero continuously.

Noting that both the CFL formulas
(\ref{cfl-rho})-(\ref{cfl-omega}) and the vector manifestation
result (\ref{ksrf}) are of the Higgsed type, we invoke the lattice
results to arrive at
 \be
ag_V f_\pi\approx \kappa g_c\chi \approx \kappa^\prime
g_c\Delta.\label{crossover}
 \ee
We admit that this relation follows neither from the
group-theoretical considerations of Berges and
Wetterich~\cite{berges-wett,wetterich} nor from the vector
manifestation of HLS~\cite{VM}. We are proposing that this is
indicated by the lattice data, at least near the phase transition
point. It holds empirically at zero temperature and zero density,
so we are led to assume that it holds at least approximately from
$T=0$ up to $T=T_c$~\footnote{This would imply that the nonet
symmetry for $N_F=3$ or the quartet symmetry $N_F=2$ is a good
symmetry not only at $T=0$ but also for $T\neq 0$. Why this
symmetry should hold at any non-zero temperature or density is not
obvious either in CFL QCD or in HLS effective theory. In
discussing color-flavor locking in two-flavor QCD, Berges and
Wetterich~\cite{berges-wett} entertain among others the
possibility that the two condensates $\chi$ and $\Delta$ could be
different at the critical point. As for the HLS theory, at
one-loop order, the RG flows are expected to be different for the
$\rho$ and $\omega$ properties, so it is not obvious that the
$\rho$ and $\omega$ mesons would reach the chiral restoration
with the Georgi vector limit at the same temperature. Nonetheless
if our interpretation of QNS is correct, it seems most plausible
and appealing that the nonet or quartet symmetry does hold at the
phase transition. Proving this conjecture remains as a
theoretical challenge.}.  It should be noted that the vanishing of
the hidden gauge coupling $g_V$ corresponds to the vanishing of
the condensates $\chi$ and $\Delta$ with the color gauge coupling
remaining non-vanishing, $g_c\neq 0$.

Next, we have shown in \cite{BR96} that above the chiral
transition temperature $T\gsim T_c$, the QNS's can be well
described by perturbative gluon exchange with a gluon coupling
constant $\frac{g_c^2}{4\pi}\approx 0.19$ and argued that the
flavor gauge symmetry {\it cedes} to the fundamental QCD gauge
symmetry. Now the HLS theory is moot on what it could be beyond
the chiral restoration point since the theory essentially
terminates at $T_c$. We propose that this is where the
color-flavor locking of \cite{wetterich,berges-wett} phrased in
the QCD variables takes over by supplying a logical language for
crossing-over from below $T_c$ to above $T_c$. Indeed
(\ref{crossover}) describes the {\it relay} that must take place
in terms of the hidden flavor gauge coupling $g_V$ on one side and
the color gauge coupling $g_c$ on the other side. Now above
$T_c$, the color and flavor must unlock, with the gluons becoming
massless and releasing the scalar Goldstones. The dynamics of
quarks and gluons in this regime will then be given by hot QCD in
the proper sense. The way the two condensates melt as temperature
is increased is a dynamical issue which seems to be difficult to
address unambiguously within the present scheme. It will have to
be up to lattice measurements to settle this issue. Our chief
point here is that their melting is intricately connected.
 \subsubsection{\it Link to BR scaling and Landau parameter
$F_1$}
 \itt
The situation appears to be quite different in dense medium.
Since there is no guidance from lattice as it is impossible at
present to put density on lattice except for unphysical cases of
two colors or adjoint quarks, we shall simply assume that the
above scenario holds in density up to $n=n_c$~\footnote{In this
subsection, we again denote density by $n$ reserving $\rho$ for
the vector meson.}. Certain models indicate that the phase
structure near chiral restoration could be quite involved and
complex. As suggested by Sch\"afer and Wilczek~\cite{continuity},
an intriguing possibility is that the three-flavor color-flavor
locking operative at asymptotic density continues all the way
down to the ``chiral transition density" ($n_c$) in which case
there will be no real phase change since there will then be a
one-to-one mapping between hadrons and quark/gluons, e.g., in the
sense of ``hadron-quark continuity." However the non-negligible
strange-quark mass is likely to spoil the ideal three-flavor
consideration. One possible alternative scenario is that viewed
from ``bottom-up," one gets into the phase where $\chi=\sigma=0$
and $\Delta\neq 0$ corresponding to the two-flavor color
superconducting (2csc) phase~\cite{raja-wilcz}. Unless $\Delta$
goes to zero at $\rho_c$, this would mean that the $\rho$ mesons
become massless but the $\omega$ meson remains massive. One
cannot say that this is inconsistent with the vector
manifestation since the HLS does not require that $U(2)$ symmetry
hold at the chiral restoration point or in medium in general. But
this seems unlikely. On the other hand, it is highly plausible
that both $\chi$ and $\Delta$ approach zero (or near zero if it
is first-order) from below $n_c$ and then $\Delta$ picks up a
non-zero value at or above $n_c$ in which case we will preserve
the mass formula (\ref{ksrf}) as one approaches $n_c$. This is
the scenario that we favor.

Among the scaling relations implied by BR
scaling~\cite{BR91,BR96}, the one most often discussed in the
literature is the dropping of the $\rho$-meson mass in medium.
This relation has been extensively discussed recently in
connection with the CERN-CERES data on dilepton production in
heavy-ion collisions. The simplest explanation for the observed
dilepton enhancement at an invariant mass $\sim 400$ MeV is to
invoke BR scaling for the excitations relevant in the
process~\cite{LKB}. It turns out however that this explanation is
not unique. One could explain it equally well if the $\rho$ meson
``melted" in dense medium with a broadened width~\cite{rapp}.
Since the process is essentially governed by a Boltzmann factor,
all that is needed is the shift downward of the $\rho$ strength
function: the expanding width simply does the job as needed for
the dilepton yield. If one calculates the current-current
correlation function in low-order perturbation theory with a
phenomenological Lagrangian, it is clear, because of the strong
coupling of the $\rho$ meson with the medium, that the meson will
develop a large width in medium and ``melt" at higher density.
The upshot of the dilepton experiments then is that they cannot
distinguish the variety of scenarios that probe average
properties of hadrons in the baryon density regime -- which is
rather dilute -- encountered in the experiments.

In the vector manifestation scenario, the width should become
narrower, decreasing like $\sim g_V^2$. Then the vector mesons
become more a quasiparticle at high density than at lower
density. {\it This is the underlying picture of BR scaling}.

Thus far, we have made a link between the color-flavor-locked
condensates and hidden gauge symmetry. We can go even one step
further and via BR scaling, make an intriguing connection between
QCD ``vacuum" properties and many-body nuclear interactions. This
connection was discussed in Section \ref{fermiliquid}. Simply
put, it comes about because nuclear matter owes its stability to
a Fermi-liquid fixed point~\cite{shankar}. Certain interesting
nuclear properties were found to be calculable in terms of the
Fermi-liquid fixed point parameters~\cite{frsall}. Among others,
it was shown that the Landau parameter $F_1$ -- which is a
component of quasiparticle interactions -- can be expressed in
terms of the BR scaling factor $\Phi (n)\equiv m_\rho^\star
(n)/m_\rho (0)$
 \ba
\tilde{F}_1=3(1-\Phi^{-1}) +\tilde{F}_1 (\pi)
 \ea
where $\tilde{F}_1 (\pi)$ is the contribution from the pion which
is completely fixed by chiral dynamics. One of the most
remarkable prediction was Eq.(\ref{chptd}) for the anomalous
gyromagnetic ratio $\delta g_l$ in nuclei
 \be
\delta g_l=\frac{4}{9}\left[\Phi^{-1}-1-\frac 12\tilde{F}_1
(\pi)\right]\tau_3.\label{deltagl}
 \ee
At nuclear matter density $n=n_0$, we have $\tilde{F}_1
(\pi)|_{n=n_0}=-0.153$. Note that (\ref{deltagl}) depends on only
one parameter, $\Phi$. This parameter can be extracted from
various sources and all give about the same value, $\Phi
(n_0)\approx 0.78$. Given that this is not very accurately
determined, it is probably a better strategy to determine $\Phi$
from the data on $\delta g_l$. In any event, given $\Phi$ at
nuclear matter density, Eq.(\ref{deltagl}) makes a simple
prediction,
 \be
\delta g_l=0.23\tau_3
 \ee
which should be compared with the measurement in the Pb
region~\cite{schumacher},
 \be
 \delta g_l^p=0.23\pm 0.03.
 \ee
We believe this to be a good agreement within the theoretical
uncertainty involved. It is intriguing that what is an intrinsic
QCD quantity can be related to what appears to be a standard
many-body nuclear interaction summarized in the Landau
quasiparticle parameter $F_1$.
\subsection{Kaon Condensation: {\it Encore} Cheshire Cat}
\itt
 In addition to the ``continuity" in excitations and chiral phase
transitions between the hadronic phase and the quark phase, meson
condensations can occur in high density matter with $n >n_c$ and
in hadronic matter with $n<n_c$ where $n_c$ is the presumed
critical density for  the so-called ``chiral restoration" --
whatever that may be. Pion condensation is unlikely in the
density regime that is relevant for laboratory or astrophysical
observations, but the negatively charged kaon $K^-$ can condense
at a relatively low density as well as at an asymptotically high
density providing yet another support for ``continuity" or
Cheshire Cat. For completeness we briefly describe this
phenomenon although the story is by no means final.\vskip 0.3cm

$\bullet$ {\bf $K^-$ Condensation in the Hadronic Sector}\vskip
0.3cm

Since this matter was discussed already in Section
\ref{strangeness}, we shall simply summarize the pertinent feature
in a slightly different language. As originally pointed out by
Kaplan and Nelson~\cite{Knelson} and reinterpreted by Brown et
al~\cite{BKR87}, the S-wave condensation of $K^-$'s is driven by
``rotating away" of the kaon mass associated with both the
explicitly broken chiral symmetry and spontaneously broken chiral
symmetry. As is widely discussed, hyperons can also participate
through P-wave coupling with the kaons but near the condensation
transition, the P-wave coupling would be
``irrelevant"~\footnote{This presumably is related to Migdal
theorem. See Polchinski in Ref.\cite{shankar} for a related
discussion in condensed matter physics.} in contrast to the S-wave
that has to do with ``relevant" terms~\cite{LRS} and hence may be
ignored in this qualitative discussion. The condensation therefore
occurs when the kaon mass is eaten up by attractive interactions.
Negative kaons have attractive interactions with nucleons by
exchanging the vector mesons $\rho$ and $\omega$ as well as scalar
mesons, e.g., the $\sigma$ meson. The vector exchange is dictated
by the vector current conservation and is given by the
Weinberg-Tomozawa term and the $\sigma$ exchange by the
$\Sigma_{KN}$ -- the KN sigma term. When many-body interactions
through many-Fermi contact terms in the Lagrangian are implemented
through BR scaling, one then has a Lagrangian of the form
(\ref{kaonL}). One can generate this effective Lagrangian in
chiral perturbation theory as reviewed e.g. by Lee~\cite{chl}. The
condensation occurs when these relevant terms drive the system to
instability~\cite{LRS}. In neutron star matter, the kaon energy
$\omega_K$ need not go all the way to zero. It suffices to drop to
the electron chemical potential $\mu_e$. Since the $\mu_e$
increases as a function of density in nonrelativistic nuclear
systems and the $\omega_K$ must fall, the crossing is bound to
occur at some density. However exactly at what density it will
occur will depend upon details of the dynamics and this is still a
controversial issue~\cite{pandh}. In Section \ref{strangeness}, we
have proposed that the critical density is rather low,
$\rho_c^K\sim 3\rho_0$.\vskip 0.3cm

$\bullet$ {\bf $K^-$ Condensation in the CFL Phase}\vskip 0.3cm

The effective chiral Lagrangian discussed above, (\ref{Leff}),
implemented with mass terms predicts~\cite{SS,RWZ,HLM,MT,BBS} at
high density where the color-flavor locking sets in that the kaon
mass is of the form
 \ba
m_{K^\pm}=cm_d (m_u +m_s)
 \ea
where $c$ is a constant that can be evaluated in the
weak-coupling QCD, $c=\frac{3\Delta^2}{\pi^2 f_\pi^2}$ where
$\Delta$ is the pairing gap and the subscripts u, d and s stand
for up, down and strange quarks respectively. Only the pairing
gap $\Delta$ appears in the mass formula despite that an
antiquark is involved. This is because the Majorana mass of the
antiquark is equal to the gap~\cite{hong-majorana}. Because the
Goldstone bosons in this phase are of $q^2\bar{q}^2$
configuration, the kaon is less massive than the pion which has
the mass
 \ba
m_{\pi^\pm}=cm_s(m_u+m_d).
 \ea
It is found numerically that for $\mu\sim 500$ -- 1000 MeV, the
kaon mass ranges from $\sim 5$ MeV to $\sim 1$ MeV. The reason
for this small mass is easy to understand. First of all, because
of the $Z_2$ symmetry, the Goldstone mass is quadratic in the
quark mass and secondly, the mass is proportional to
$(\Delta/\mu)^2$ that goes to zero as $\mu\rightarrow \infty$.

For large density, one can ignore the attractive kaon-quark
interaction analogous to the kaon-nuclear since the interaction is
suppressed by $(1/\mu)^2$. Considering the Fermi sea filled with
non-interacting massless up, down and strange quarks, the electron
chemical potential $\mu_e$ is found in this case to drop as $\sim
m_s/k_F$. This contrasts with the increasing $\mu_e$ found in the
nonrelativistic hadronic system. Sch\"afer~\cite{TS} found that
for a reasonable range of parameters involved for the CFL phase,
the kaon mass is {\it always} less than the electron chemical
potential. However this does not lead to a $K^-$ condensation of
the type seen in the hadronic sector since as argued by Rajagopal
and Wilczek~\cite{neutral}, the CFL phase is charge-neutral more
or less independently of the mass of the strange quark and hence
no electrons need to be present in the phase.

Kaon condensation $can$ however occur in superdense matter, not
just because of the small mass of the kaons (which helps) but
because of the presence of the chiral symmetry breaking mass term.
Though perhaps different in character, this is analogous to the
case of kaon condensation in the hadronic sector where the phase
change is essentially effectuated by ``rotating away" the large
$\Sigma$ term arising from the chiral symmetry breaking due to the
strange-quark mass. Indeed in superdense regime, a term of the
form
 \be -M^2/2\mu \ee
where $M$ is the quark-mass matrix plays the role of a flavor
chemical potential that provides an effective attraction favoring
neutral kaon condensation~\cite{densekaon}.

Although interesting purely from the theoretical point of view,
the physical relevance of this observation to the physics of
compacts stars is yet to be established. Whereas the phenomenon
lends itself to a simple and elegant analysis for asymptotic
densities, the relevant process for the formation and structure of
compact stars requires instead a ``bottom-up" approach starting
from low density at the stage of supernovae explosion and climbing
up to high but non-asymptotic density as the matter is compressed
in the interior of compact stars. This must involve, along the
way, such hadronic phases as kaon (or pion) condensation, hyperon
presence etc. for which QCD is intractable and it is not clear
that the simple picture based on QCD at asymptotic density is
directly relevant to what actually takes place in the stellar
matter. Working out the change-over from hadronic variables to QCD
variables necessary for a realistic description of the process --
which we believe involves a ``Cheshire cat" mechanism -- remains a
challenge for nuclear theorists in the sense discussed in this
review.

\subsection*{Acknowledgments}
\itt We are grateful for discussions with and comments from Bengt
Friman, Masayasu Harada, Deog-Ki Hong, Youngman Kim, Kuniharu
Kubodera, Shoji Nagamiya and Koichi Yamawaki. This work was
initiated while both of us were visiting Korea Institute for
Advanced Study in June 2000 and completed when one of us (MR) was
spending the Spring 2001 at Seoul National University and Yonsei
University. We acknowledge the generous hospitality of the host
institutes. The work of GEB was supported by the US Department of
Energy under Grant No. DE-FG02-88 ER40388 and that of MR in part
by the Brain Korea 21 in 2001.

\end{document}